\documentclass{aa}  

\usepackage{natbib}
\usepackage{graphicx}
\usepackage{txfonts}
\usepackage[colorlinks,allcolors=blue]{hyperref}
\usepackage{multirow}
\usepackage{xspace}
\usepackage{scalerel}
\usepackage[dvipsnames,usenames,table]{xcolor}
\usepackage[normalem]{ulem}
\usepackage{booktabs}
\usepackage{cleveref} 
\usepackage{enumitem}

\newcommand{\citenop}[1]{\citeauthor{#1} \citeyear{#1}}
\newcommand{\orcit}[1]{\protect\href{https://orcid.org/#1}{\protect\includegraphics[width=8pt]{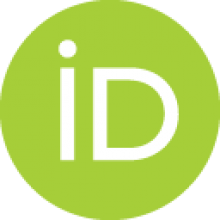}}}

\defcitealias{liberatori2026ggscs}{L26}

\defcitealias{trabucchi2021longperiods}{T21}

\makeatletter
\renewcommand*\aa@pageof{, page \thepage{} of \pageref*{LastPage}}
\makeatother

\begin{document}

    \title{Light curve spectral energy distribution fitting of 13 carbon-rich Mira variables stars in the Small Magellanic Cloud}
    \titlerunning{Running title}
    
    \authorrunning{Liberatori et al.}
    \author{
            A.~Liberatori\orcit{0009-0002-2078-7966}\inst{\ref{inst_nkua}, \ref{inst_pd}, \ref{inst_oapd}}
            \fnmsep\thanks{
                Corresponding author: A. Liberatori
                (\href{mailto:alessio.liberatori.astro@gmail.com}{\tt alessio.liberatori.astro@gmail.com})
            }
        \and
            M.~Trabucchi\orcit{0000-0002-1429-2388}\inst{\ref{inst_pd},\ref{inst_oapd},\ref{inst_gen}}
        \and
        D.~Hatzidimitriou\inst{\ref{inst_nkua}}
        \and
        M.A.T.~Groenewegen\inst{\ref{inst_belg}}
        \and
        A.~Bressan\inst{\ref{inst_oapd}, \ref{inst_sissa}, \ref{inst_ch}}
        \and
        G.~Pastorelli\inst{\ref{inst_oapd}, \ref{inst_pd}}
        \and
        L.~Girardi\inst{\ref{inst_oapd}}
        \and
        K.~Shepherd\inst{\ref{inst_pd}}
        \and
        B.~Aringer\inst{\ref{inst_at}}
        \and
        A.~Udalski\inst{\ref{inst_pl}}
        \and 
        A.~Nunnari\inst{\ref{inst_torvergata},\ref{inst_inaf_oar}}
        }

    \institute{
            National and Kapodistrian University of Athens, 
            Panepistimiopolis, 
            15784 Zografos, 
            Athens, 
            Greece\label{inst_nkua}
        \and
            Dipartimento di Fisica e Astronomia
            ``Galileo Galilei'',
            Universit\`a di Padova,
            Vicolo dell'Osservatorio 3,
            I-35122 Padova,
            Italy\label{inst_pd}
        \and
            INAF-Osservatorio Astronomico di Padova,
            Vicolo dell'Osservatorio 5,
            I-35122 Padova,
            Italy\label{inst_oapd}
        \and
            Department of Astronomy,
            University of Geneva,
            Ch. Pegasi 51,
            CH-1290 Versoix,
            Switzerland\label{inst_gen}
        \and
            Koninklijke Sterrenwacht van Belgi\"e, 
            Ringlaan 3, 
            B-1180 Brussels,
            Belgium\label{inst_belg}
        \and
            SISSA, 
            Via Bonomea 265, 
            I-34136 Trieste, 
            Italy\label{inst_sissa}
        \and
            Purple Mountain Observatory, 
            Chinese Academy of Sciences, 
            Nanjing 210023, 
            People’s Republic of China\label{inst_ch}
        \and
            Department of Astrophysics, 
            University of Vienna, 
            Türkenschanzstraße 17, 
            1180 Wien, 
            Austria\label{inst_at}
        \and
            Astronomical Observatory, 
            University of Warsaw, 
            Al. Ujazdowskie 4, 
            00-478 Warszawa,
            Poland\label{inst_pl}
        \and 
            Dipartimento di Fisica, 
            Università di Roma Tor Vergata, 
            via della Ricerca Scientifica 1, 
            00133 Rome, Italy\label{inst_torvergata}
        \and
            INAF – Osservatorio Astronomico di Roma, 
            Via Frascati 33, 
            00078 Monte Porzio Catone, Italy\label{inst_inaf_oar}
        }

    \date{\today}
\abstract
{
    Carbon-rich Mira variables are highly dynamic stars characterized by large-amplitude radial pulsations and intense mass-loss. Standard spectral energy distribution (SED) fitting often relies on time-averaged observations, which flatten the intrinsic variability and provide only an approximation of the stellar and circumstellar parameters.
}
{
    We aim to overcome the limitations of static SED modeling by dynamically tracing the physical parameters of a selected sample of 13 carbon-rich Mira variables in the Small Magellanic Cloud (SMC) throughout their entire pulsation cycle.
}
{
    We combine time-series photometry from \textit{Gaia}, WISE, VMC, 2MASS, and OGLE to reconstruct simultaneous multi-band light curves. We extract independent SEDs at 14 distinct phases across the pulsation cycle and model them using the ATHENA-C grid.
}
{
    Our phase-resolved approach allows us to coherently trace the cyclic variations of effective temperature ($T_{\rm eff}$), optical depth ($\tau_{0.55}$), and bolometric luminosity. We found peak-to-peak $T_{\rm eff}$ variations ranging from $\sim 100$~K to over $1000$~K. A strong anti-correlation between $\tau_{0.55}$ and bolometric luminosity is observed, consistent with the dust condensation zone being pushed outward at maximum brightness, thus making the star's circumstellar environment optically thinner. Additionally, we recover the classical Period-Luminosity relation and derive the amplitude of the radius variation, along with the radial expansion velocities, which we find to be consistent with independent measurements.
}
{
    Phase-resolved SED fitting reveals the true dynamic ranges of atmospheric and circumstellar parameters, which traditional static fits fail to capture. Although the determination of these parameters is limited by the coarse sampling of the hydrostatic model grid and the gaps in parameter space where converged models are unavailable, explicitly accounting for time-dependent behavior provides a much more physically realistic picture of the interplay between stellar pulsation and dust formation. This highlights the importance of time-resolved methodologies for studying large-amplitude variables.
}

\maketitle

\section{Introduction}
\label{sec:Introduction}

Carbon-rich Mira variables represent an advanced evolutionary stage for stars with initial masses in the range of $1 \lesssim M \lesssim 4~M_{\odot}$ on the Asymptotic Giant Branch (AGB) \citep{diCriscienzo2016studying}. 
These are intrinsically dynamic stars, characterized by large-amplitude radial pulsations with periods spanning several hundred days \citep{wood1977miras}. 
During the later stages of the AGB, these stars undergo recurrent thermal pulses (TPs) driven by thermal instabilities in the helium-burning shell, leading to cyclic variations in the two-shell burning structure. 
These TPs trigger the third dredge-up (TDU), a deep convective mixing process that transports newly synthesized carbon ($^{12}$C) to the stellar surface. 
Because carbon readily binds with available oxygen to form highly stable CO molecules, a critical threshold is reached when the atmospheric carbon abundance exceeds that of oxygen. 
A carbon star is characterized by a surface abundance $\rm C/O > 1$.

This surface chemical change fundamentally alters the atmospheric opacity and directly dictates the stellar mass-loss mechanism. 
The strong shock waves induced by the large-amplitude pulsations elevate and compress the cool outer layers of the atmosphere, creating a dense, supersaturated environment optimal for molecular chemistry and dust condensation. 
In these carbon-rich environments, the condensing dust is primarily composed of highly opaque amorphous carbon (amC) with a minor fraction of silicon carbide (SiC). 
The intense radiation pressure from the central star acting on these opaque grains, coupled with the outward momentum from pulsational shocks, drives a massive stellar wind. 
This pulsation-enhanced outflow gradually strips the star of its envelope, strongly enriches the interstellar medium, and forms a dense circumstellar shell with high optical depth \citep{Hofner2018Review}. 
Depending on the specific evolutionary state and metallicity of the star, the resulting mass-loss rates typically range from roughly $10^{-11}$ to $10^{-4}~M_{\odot}~\text{yr}^{-1}$. 
As a result, they play a crucial role in the chemical evolution of galaxies.

A highly effective approach to investigate the physical properties of these stars is through spectral energy distribution (SED) fitting. 
By comparing multi-band photometric observations with appropriate synthetic models, this method allows to reliably constrain key stellar and circumstellar quantities, such as effective temperatures, luminosities, mass-loss rates, and dust properties. 
Over the past decade, numerous studies have successfully employed this technique across the optical and infrared regimes to characterize large samples of carbon stars (e.g.,~\citenop{groenewegen2009luminosities}; \citenop{riebel2012mass}; \citenop{rau2017adventure}; \citenop{nanni2018estimating}; \citenop{groenewegen2018luminosities}; \citenop{nanni2019mass};  \citenop{groenewegen2022wise}; \citenop{liberatori2026ggscs}).

Despite the pronounced dynamic nature of these stars, the vast majority of studies in the literature based on SED fitting adopt a fundamentally static approach. 
Often, these works rely on single-epoch photometric observations or time-averaged magnitudes, comparing them against grids of hydrostatic atmospheric models. 
Although this approach has thus far been a necessary standard for analyzing large data samples, it flattens the intrinsic variability of the star, yielding stellar and circumstellar parameters that represent an approximate average snapshot of the true physical state of the star.
The recent advent of multiple surveys providing high-precision photometric time series, such as \textit{Gaia} \citep{gaia2016gaia}, the All-Sky Release Catalog \citep[AllWISE;][]{cutri2021vizier} of the Wide-field Infrared Survey Explorer \citep[WISE;][]{Writhg_etal_2010}, and VMC \citep{cioni2011vmc}, now allows to partially overcome the limitations of static approaches.

A recent notable work towards the direction of dynamical state reconstruction is \cite{iwanek2021lcurves}. 
The authors simultaneously modeled multi-band light curves of Magellanic Cloud Miras to reconstruct time-series SEDs throughout their pulsation cycles. 
However, their primary goal was not a faithful, detailed physical reconstruction of the stellar atmosphere and dust parameters, but rather to study variability amplitude ratios and phase lags, and to derive synthetic period-luminosity relations across various photometric surveys.

In this work, we present a study based on phase-resolved SED fitting. 
We analyze a selected sample of 13 carbon-rich Mira variables in the Small Magellanic Cloud (SMC), for which we reconstructed simultaneous multi-band light curves from the optical to the mid-infrared. 

Instead of deriving a single set of time-averaged parameters, we model the SED at multiple phases throughout the entire pulsation cycle, which allows us to coherently and dynamically trace the actual variation of the effective temperature, optical depth, and bolometric luminosity as the star pulsates and the circumstellar envelope reacts.

This paper is organized as follows. Section~\ref{sec:Data} describes the multi-band dataset and the selected sample. 
Section~\ref{sec:Methods} outlines our light curve analysis, the phase-resolved fitting methodology, and the synthetic models used for the SED fitting. 
We present our findings in Sect.~\ref{sec:Results} and discuss their physical implications and limitations in Sect.~\ref{sec:Discussion}. 
Conclusions are summarized in Sect.~\ref{sec:Conclusions}.

\section{Data}
\label{sec:Data}

As discussed in \citet{liberatori2026ggscs} (hereafter \citetalias{liberatori2026ggscs}), the literature contains numerous, often overlapping, catalogs of AGB carbon star candidates. Selecting sources from heterogeneous catalogs yields samples with diverse metallicities and stellar properties, complicating homogeneous analyses. We therefore adopt as our primary sample the SMC AGB carbon stars analysed by \citetalias{liberatori2026ggscs}, which is based on the \textit{Gaia} Golden Sample of carbon stars \citep{Creevey2023gaiagoldensample}.

Reconstructing the SED at multiple pulsation phases requires simultaneous optical and near-infrared coverage. We combine time-series data from three facilities: \textit{Gaia} ($BP$, $G$, and $RP$ bands) for the optical range; WISE (AllWISE+NEOWISE, $W1$ and $W2$ bands) for the mid-infrared, where dust shell emission becomes significant; and the VMC survey ($K$ band) for the near-infrared. We exclude the VMC $Y$ and $J$ bands due to insufficient temporal coverage for reliable light curve modeling.
We supplement the dataset with OGLE $I$-band photometry. Its extensive, well-sampled time series significantly improves period determination and light curve fitting. However, we exclude the $I$ band from the SED fitting because its response closely mimics the \textit{Gaia} $RP$ filter, adding no independent spectral information while potentially introducing noise.

Every target has a counterpart in the 2MASS point source catalogue (the Two Micron All Sky Survey; \citealt{skrutskie2006two}). 
Given the similar response of the 2MASS $K_s$ and VMC $K$ filters, we treat the 2MASS single-epoch measurement as an additional VMC $K$ data point. To ensure reliable phase assignment, we strictly select stars where the 2MASS observation epoch falls within the OGLE temporal baseline. 

We also apply the same extinction correction to the photometric data as done in \citetalias{liberatori2026ggscs}, by utilizing the \cite{lallement2022dustmap} 3D dust map and a distance of 62.44 kpc \citep{graczyk2020distance}.

To define our final sample, we first applied strict automated quality cuts to ensure reliable phase-resolved analysis. We required regular temporal coverage across all photometric bands, the absence of strong outliers, and a physically motivated constraint: the optical variability amplitude must exceed the infrared one, as expected for Mira-type pulsators.

Then, we proceeded with a rigorous visual inspection of the individual multi-band light curves. For this pilot study, we deliberately discarded sources exhibiting complex pulsational behaviors, such as multiple competing sub-periods or strong secondary modulations. Instead, we manually selected a final, robust sample of 13 Mira variables that display an exceptionally clean, single dominant period. 

We emphasize that this work is designed as a pilot study to demonstrate the potential of phase-resolved SED modeling. Consequently, our goal was not to extract a statistically complete subset representative of every Mira sub-class, but rather to construct a high-fidelity benchmark sample tailored for methodological validation.

\section{Methods}
\label{sec:Methods}
The overall workflow requires extracting phase-resolved photometry from multi-band light curve modeling, followed by independent SED fitting at distinct pulsation phases.

\subsection{Multi-band Light Curve Modeling}
\begin{figure*}
    \centering
    \includegraphics[width=0.95\linewidth]{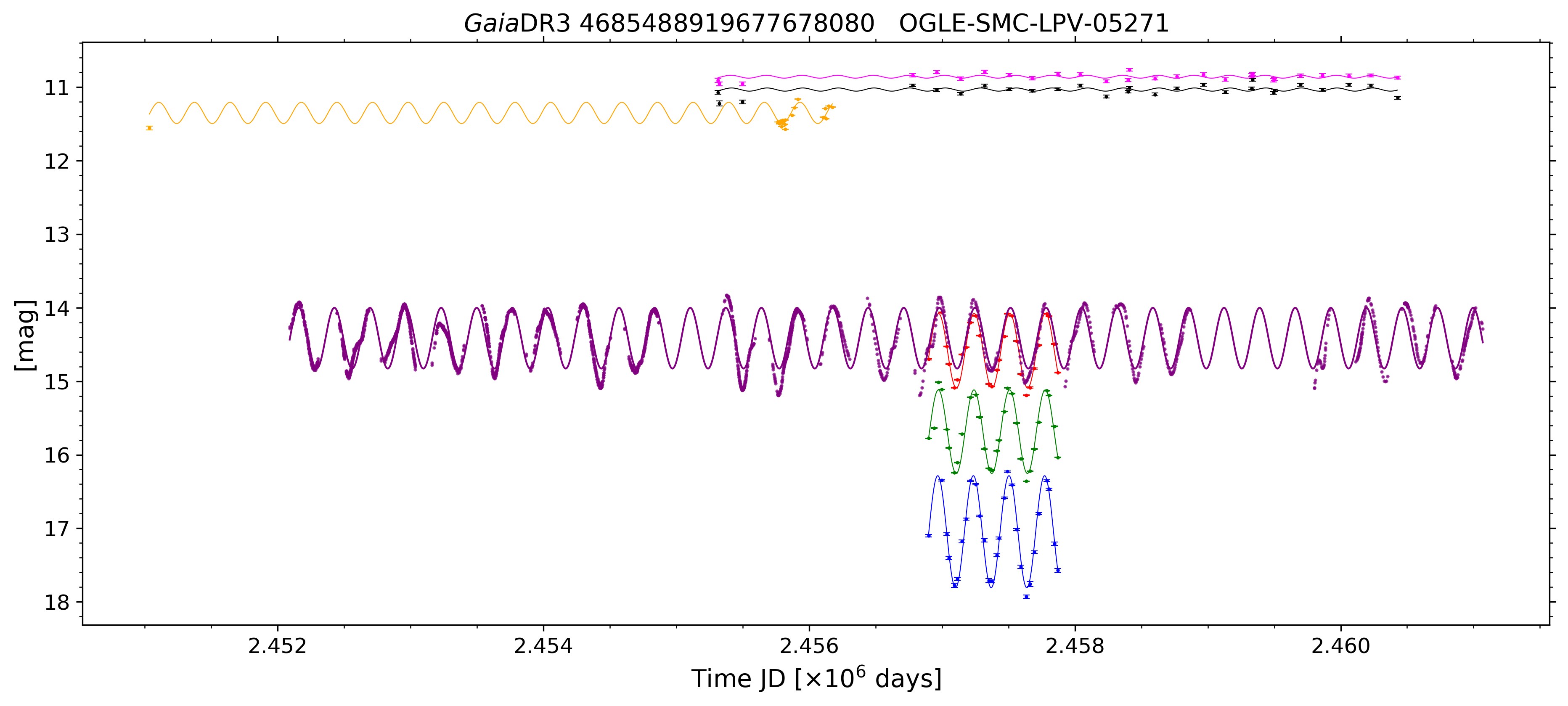}
    \caption{Time series multi band data with their best fit models from our custom pipeline; each color represents a different photometric band. In particular, blue for $Gaia$ BP, red for $Gaia$ RP, green for $Gaia$ G, purple fo OGLE I, yellow for VMC K + 2MASS K, black for WISE W1 and magenta for WISE W2.}
    \label{fig:timeseries_example}
\end{figure*}

We developed a customized Python pipeline based on the multiband Lomb-Scargle periodogram (MLSP) method \citep{VanderPlas_Ivezic_2015}, using the \texttt{astropy} implementation \texttt{LombScargleMultiBand}, to simultaneously model multi-band light curves, enforcing a single periodicity across all photometric bands\footnote{The core routines of this pipeline are publicly available at \url{https://github.com/aliberatori/LightCurveFit}}.
This ensures the identification of the dominant pulsation period for each variable star while preserving the relative phase information between filters. Figure~\ref{fig:timeseries_example} shows the time-series data and the corresponding best-fit light curve models for a representative star.

\subsection{VMC $K$-band Saturation Correction}
Observations of bright AGB stars in the VMC $K$ band are affected by known saturation effects \citep[e.g.][]{VMC_XXXVII}. 
We correct for this by using the MLSP model to predict the unsaturated K-band magnitude at each observation epoch.  

\begin{figure}
    \centering
    \includegraphics[width=0.95\linewidth]{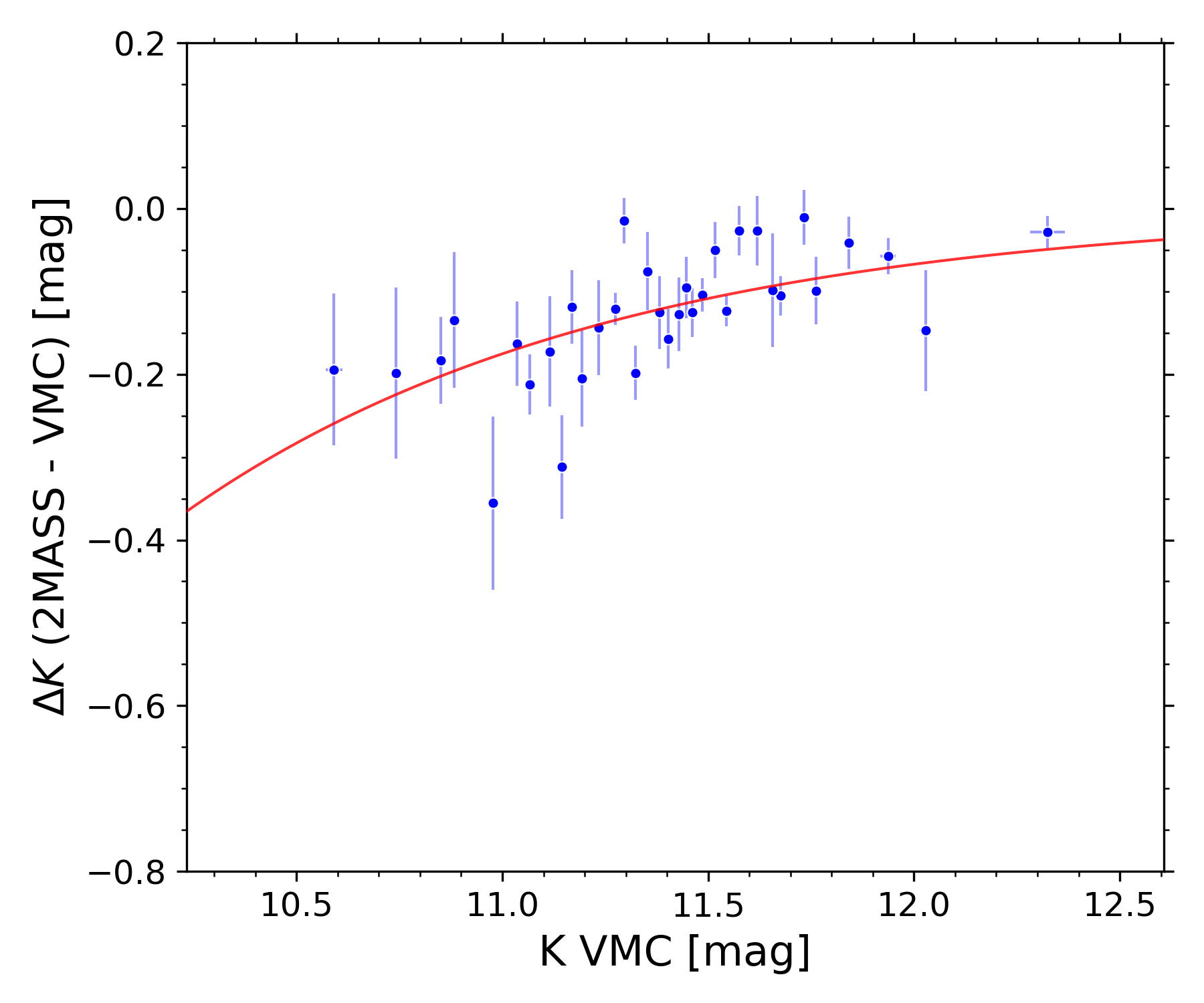}
    \caption{Difference between the $K$ magnitude from VMC and from 2MASS for all the SMC stars studied in \citetalias{liberatori2026ggscs} as a function of the VMC K mean magnitude as obtained by the fitted light curve best model for each star. The red line represents the fit, the blue points the stars binned in subgroups and the errorbars represent the magnitudes spanned by each star's subgroup in their time series VMC data.}
    \label{fig:deltaK_vs_Kmag_fit_all_stars}
\end{figure}

Specifically, we first determine the best-fit lightcurve model for the VMC $K$-band time series. 
We then evaluate this model at the epoch of the corresponding 2MASS K-band observation (which is unaffected by saturation), and compute the offset. 
These offsets provide empirical estimates of the saturation effect for individual sources.
We modeled this effect with an empirical exponential relation:
\begin{equation}
    \Delta K = -\exp[-a(K_{\mathrm{K_{VMC}}} - b)]; \ \ a = 0.9579; \ b=9.1825 \ ,
\end{equation}
and applied this correction to all VMC $K$ magnitudes before finalizing the global light curve fitting. 
Consequently, VMC and 2MASS $K$ measurements were treated as a single homogeneous dataset. Figure \ref{fig:deltaK_vs_Kmag_fit_all_stars} shows the non-linear fit used for this correction.

\subsection{Phase-Resolved SED Extraction}
For each star, we reconstructed the light curves in all filters from the best-fit multi-band model and sampled them at 14 uniformly distributed phases across the pulsation cycle. 
At each phase, we extracted the corresponding magnitudes for the SED fitting, excluding OGLE data (used solely for period determination). This yielded 14 phase-resolved SEDs per star.
The observed bolometric luminosity at each phase was estimated via a three-dimensional spline interpolation of the multi-band photometry, serving as a key input for the SED fitting.

\subsection{Synthetic SED Models and Fitting Priors}
\label{subsect:phase-resolved_sedfit}
For the SED fitting, we employ version 2.0 of the ATHENA-C model grid\footnote{\url{https://zenodo.org/records/18418537}}, based on the grid presented by \citetalias{liberatori2026ggscs}. 
This updated release includes minor numerical corrections to the stellar masses and C/O abundance ratios, ensuring that they accurately reflect the physical properties of the underlying COMARCS atmospheric models \citep{aringer2016synthetic, aringer2019carbon}. 
To improve computational usability, the dataset has also been migrated into a single, highly efficient Apache Parquet file, consolidating the physical parameters, synthetic photometry, and full SEDs into one tabular structure.
The atmospheric models are coupled with the DUSTY code \citep{ivezic1997dusty, ivezic1999dusty, nenkova2000dusty}. 
The grid provides robust coverage of the relevant parameter space—incorporating fully validated numerical values for stellar mass and C/O ratio parameters—spanning effective temperatures ($T_{\rm eff}$) from $2500$ to $5000\,$K, optical depths at 0.55 $\rm \mu m$ ($\tau_{0.55}$) from $0.0001$ to $10$, masses from $1$ to $3\,M_\odot$, and C/O ratios from $1.05$ to $2.00$. The dust density distribution follows a $\rho \propto r^{-2}$ power law, assuming a composition of 100\% amorphous carbon utilizing \citet{Zubko1996} optical constants. 
The inner boundary of the dust shell is strictly determined by the dust sublimation temperature, which we fixed at $T_{\rm sub} = 1200$~K, a standard value for amorphous carbon grains.
Additional technical details about the models grid can be found in \citepalias{liberatori2026ggscs}, Sect.~4.1.

At each of the 14 phase slices, an independent SED fit is performed using this grid to derive $T_\mathrm{eff}$, stellar mass ($M$), optical depth ($\tau_{0.55}$), and surface C/O ratio. We modeled all bands with their measured inter-band phase lags (see Sect. \ref{subsect:light_curve_fit} for further physical explanation of the phase lag).

Physically, $M$ and the C/O ratio evolve on timescales vastly exceeding a single pulsation cycle. Any apparent phase-dependent variation is therefore an artifact of fitting degeneracies rather than genuine stellar evolution. 
Consequently, we searched for solutions where $M$ and C/O remain constant across all 14 phases. 
However, fixing these parameters introduces a practical limitation: the COMARCS grid does not provide uniform $T_{\rm eff}$ coverage for all $M$ and C/O combinations, which can artificially suppress the recovered temperature variation if these parameters are arbitrarily constrained. 
This non-uniformity of the grid is due to missing converging solutions in specific regions of the parameter space. 
While these gaps could theoretically be filled via interpolation, we chose to strictly rely on the originally computed models. Implementing a multidimensional interpolation scheme would introduce an additional, uncontrolled variable into our fitting procedure, which goes beyond the scope of this work. 
To balance physical consistency and model flexibility, we select the best-fit $M$ and C/O following this methodology: for every available $M$ and C/O combination in the COMARCS grid, we compute the mean $\chi^2$ averaged over all 14 phases. The combination yielding the lowest mean $\chi^2$ is adopted. This criterion is purely statistical.

\subsection{Uncertainties}
\label{subsect:uncertainties}
Rigorous uncertainty estimation for the reconstructed multi-band magnitudes is critical, as these errors govern the predictive intervals of the physical parameters. Since the \texttt{LombScargleMultiBand} algorithm does not provide uncertainties for Fourier parameters, we evaluated the $1\sigma$ errors on the global period, amplitudes, and phases via non-linear least-squares optimization. 
By utilizing the \texttt{curve\_fit} algorithm by \texttt{SciPy} over the concatenated time-series, using the Lomb-Scargle parameters as initial guesses, we extracted the covariance matrix to obtain standard parametric errors across all bands.
However, the magnitude uncertainties at each phase slice must account for both fit precision and intrinsic, phase-dependent physical variations. 
We defined a comprehensive error budget combining three sources in quadrature:
\begin{equation}
    \sigma_{\mathrm{tot}}(\phi) = \sqrt{\sigma_{\mathrm{model}}^2(\phi) + \sigma_{\mathrm{scatter}}^2(\phi) + \bar{\sigma}_{\mathrm{phot}}^2} \ ,
\end{equation}
where $\phi$ represents the pulsation phase, $\sigma_{\mathrm{model}}(\phi)$ is the analytical parametric model uncertainty from the covariance matrix, $\bar{\sigma}_{\mathrm{phot}}$ is the mean photometric error of the band, and $\sigma_{\mathrm{scatter}}(\phi)$ is the local empirical scatter, the standard deviation of photometric residuals within a $\Delta\phi = \pm 0.05$ window. 
The empirical scatter ($\sim 5\%$ of the magnitude) highly dominates both the model and photometric errors ($\sim 0.5\%$), rendering any statistical interdependence between model error and scatter negligible. 

This localized approach ensures that uncertainties dynamically adapt to the star's physical state. 
For instance, most sources exhibit larger cycle-to-cycle variations near minimum brightness compared to maximum light. 
By computing scatter within narrow phase windows, our methodology naturally accounts for this phase-dependent noise, providing a robust, physics-driven error budget for the SED fit.
To rigorously propagate these photometric uncertainties into the derived physical parameters, we employed a Monte Carlo randomization approach at each of the 14 phase slices. Specifically, for every phase, the reconstructed magnitudes were randomly perturbed 150 times, drawing from a Gaussian distribution with a standard deviation equal to our estimated $\sigma_{\mathrm{tot}}(\phi)$. The SED fitting procedure was then independently repeated for each of these 150 perturbed multi-band sets. This allowed us to construct a posterior probability distribution for each fitted physical parameter. The final adopted values for the stellar and circumstellar parameters are defined as the median of these distributions, while the corresponding $1\sigma$ uncertainties are reported as the 16th and 84th percentiles in the final results.

Global pulsation period uncertainties were determined using a dual approach. Statistical precision was extracted from the covariance matrix. To estimate systematic uncertainties from epoch sampling and inter-survey alignment, we merged \textit{Gaia} $G_\mathrm{RP}$ and OGLE $I$ observations, which have highly similar spectral responses, into a single synthetic pseudo-band. 
Re-evaluating the model with this dataset forces strict phase coherence, naturally perturbing the converged period. 
The absolute difference between this perturbed period and our nominal period was adopted as the systematic uncertainty.

While full Monte Carlo analysis would be more exhaustive, our SED fitting is relatively insensitive to such small period variations, rendering the computational expense unnecessary. 
The final period uncertainty is the sum in quadrature of the formal covariance error and the systematic difference.

\section{Results}
\label{sec:Results}
\begin{figure*}
    \centering
    \includegraphics[width=0.45\linewidth]{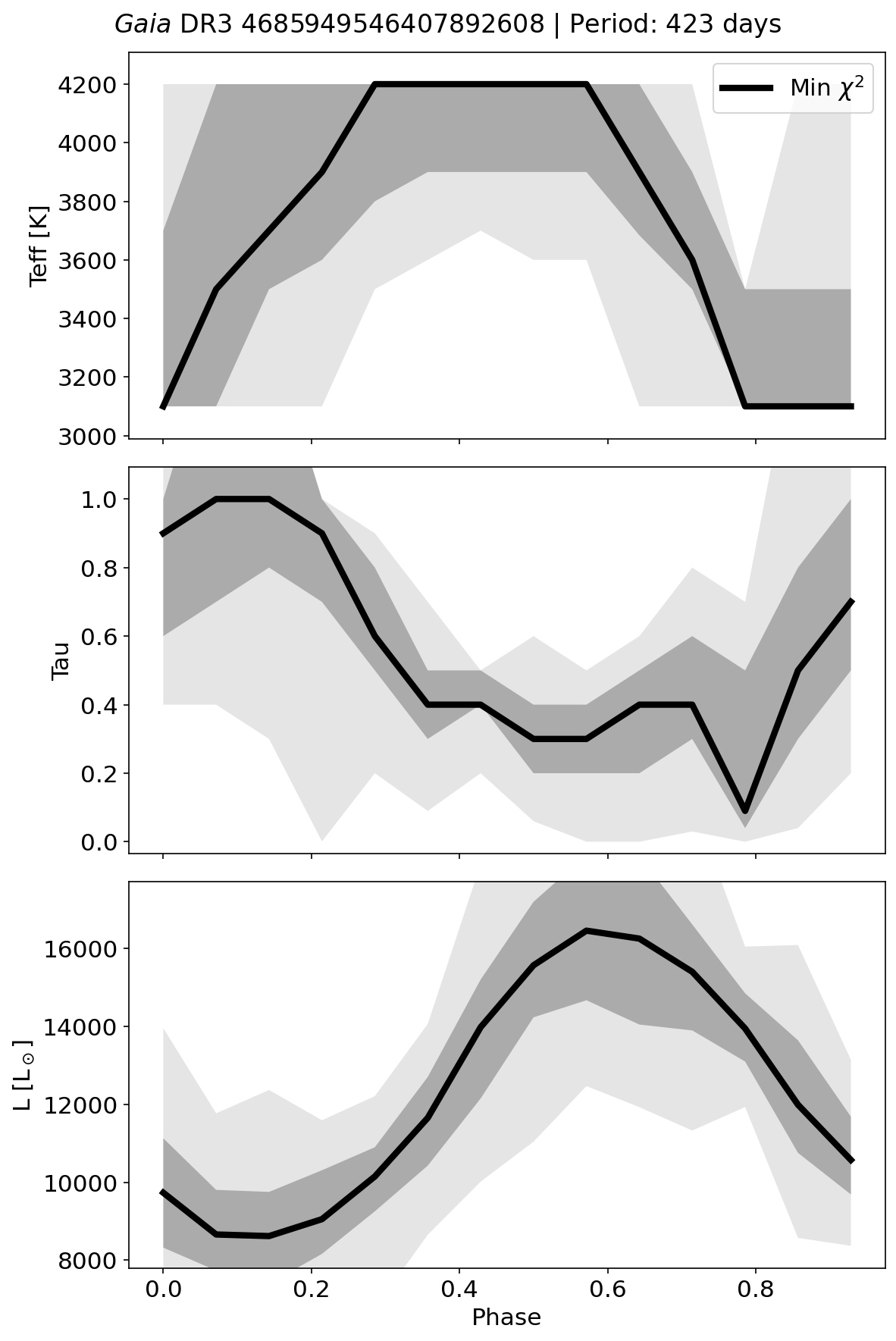}
    \includegraphics[width=0.45\linewidth]{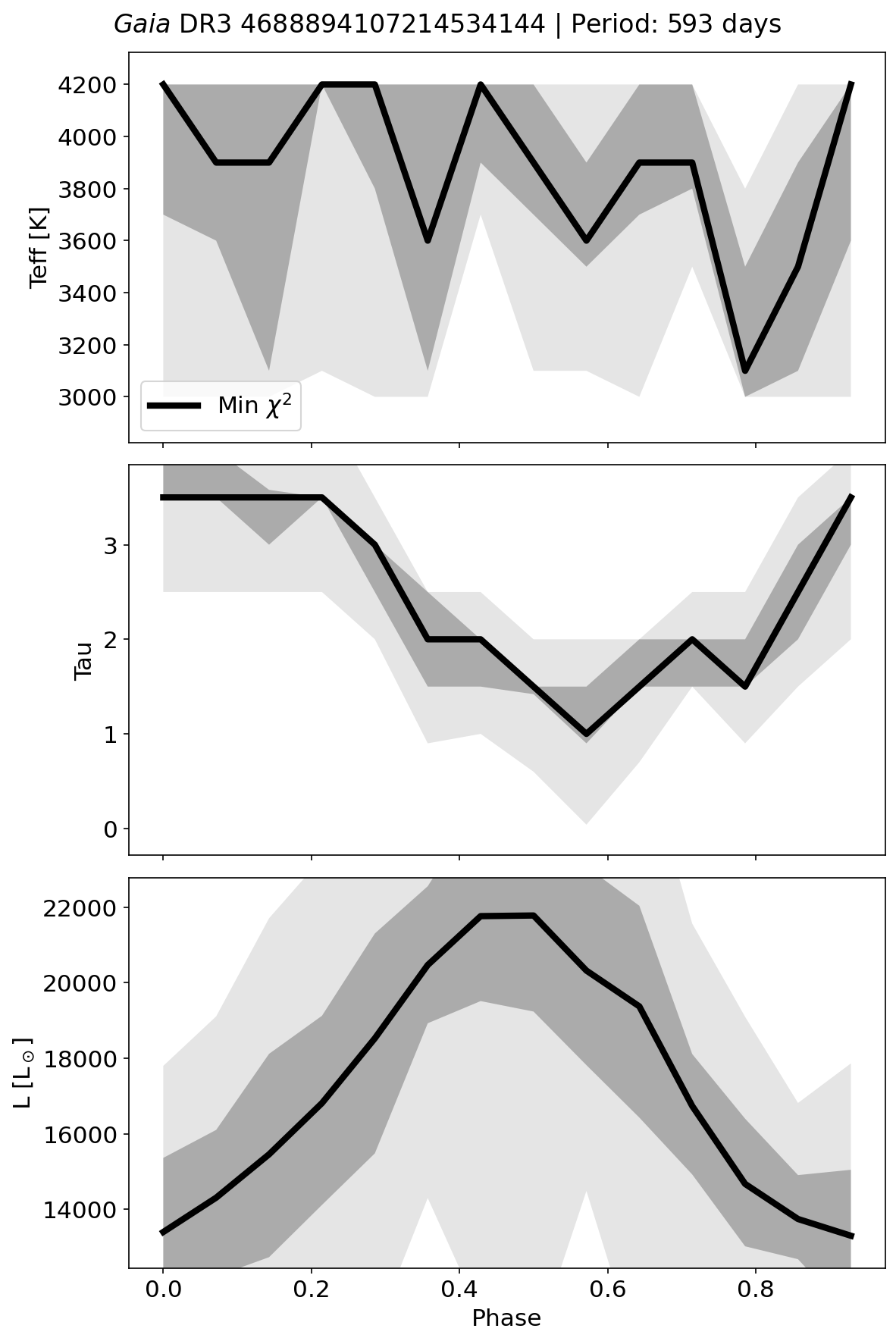}
    \caption{Diagnostic plot for two stars showing the evolution of three parameters with pulsation phase. Black continuous line defines the results from the phase-resolved SED fit. Shaded areas represent the 0th, 16th, 84th and 100th percentile of each parameter at each pulsation phase. Dark gray shades represent the 16th and 84th percentile of the distribution, while light gray represent the 0th and 100th.}
    \label{fig:diagnostic}
\end{figure*}

\begin{figure*}
    \centering
    \includegraphics[width=0.48\linewidth]{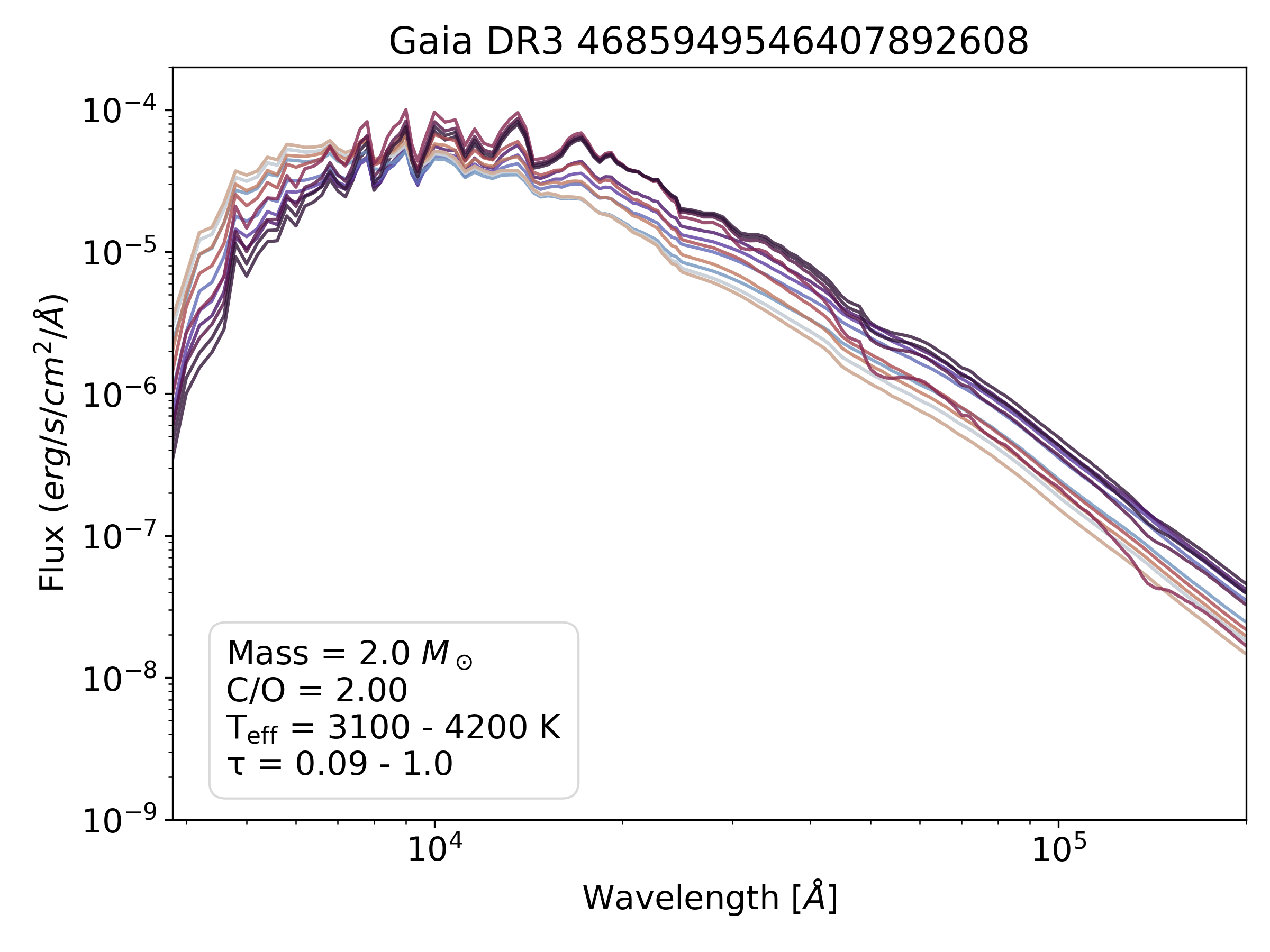}
    \includegraphics[width=0.48\linewidth]{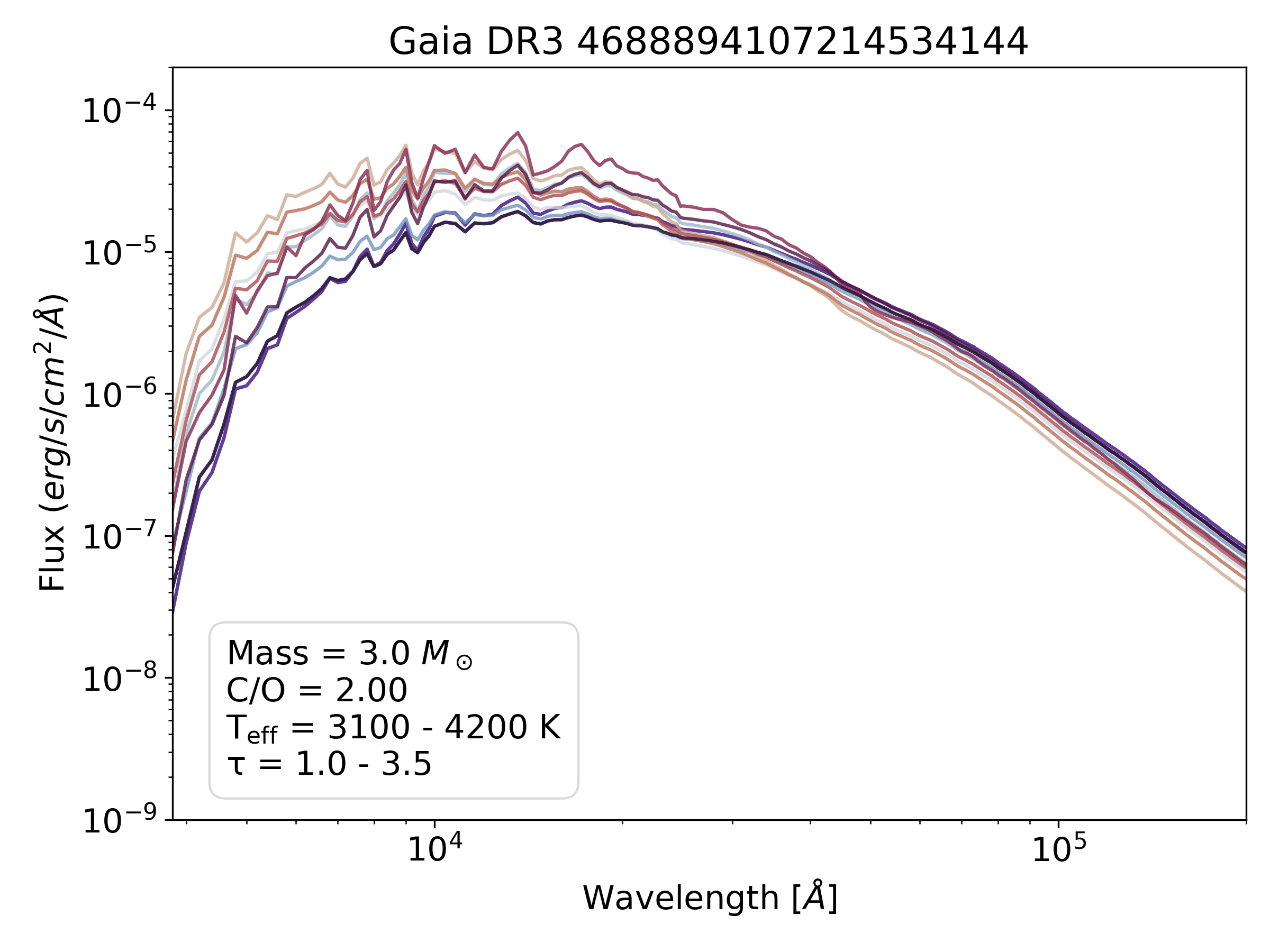}
    \caption{Best-fit SED models over the pulsation cycle for two stars with different masses, temperatures, and optical depth ranges. Colors indicate the pulsation phases. The two stars are the same star of Fig.~\ref{fig:diagnostic}.}
    \label{fig:sed}
\end{figure*}

This section presents the main results of our light curve and SED fitting. 
Table~\ref{tab:results_table} provides the values obtained for the various parameters from the light curve fit and the SED fitting, with the corresponding uncertainties.
Figure~\ref{fig:diagnostic} illustrates the phase-dependent evolution of the derived parameters for two stars, while Fig.~\ref{fig:sed} shows the variation of the best-fit SED over the pulsation cycle for the same two stars.
These two types of figures are presented, for each star our sample, in Appendix~\ref{appendix:diagnostic_plots}.

\subsection{Light curve fit}
\label{subsect:light_curve_fit}
In this subsection we present and discuss the three parameters derived from the light curve fit, namely the period, the pulsation amplitude and the pulsation phase of each photometric band.
The 13 selected stars exhibit periods between 198 and 424~days, typical for this class of variables. A notable exception is \textit{Gaia} DR3 4688894107214534144, which has a significantly longer period of $P = 593 \pm 4$~days, and the largest optical amplitude.

\begin{figure}
    \centering
    \includegraphics[width=0.95\linewidth]{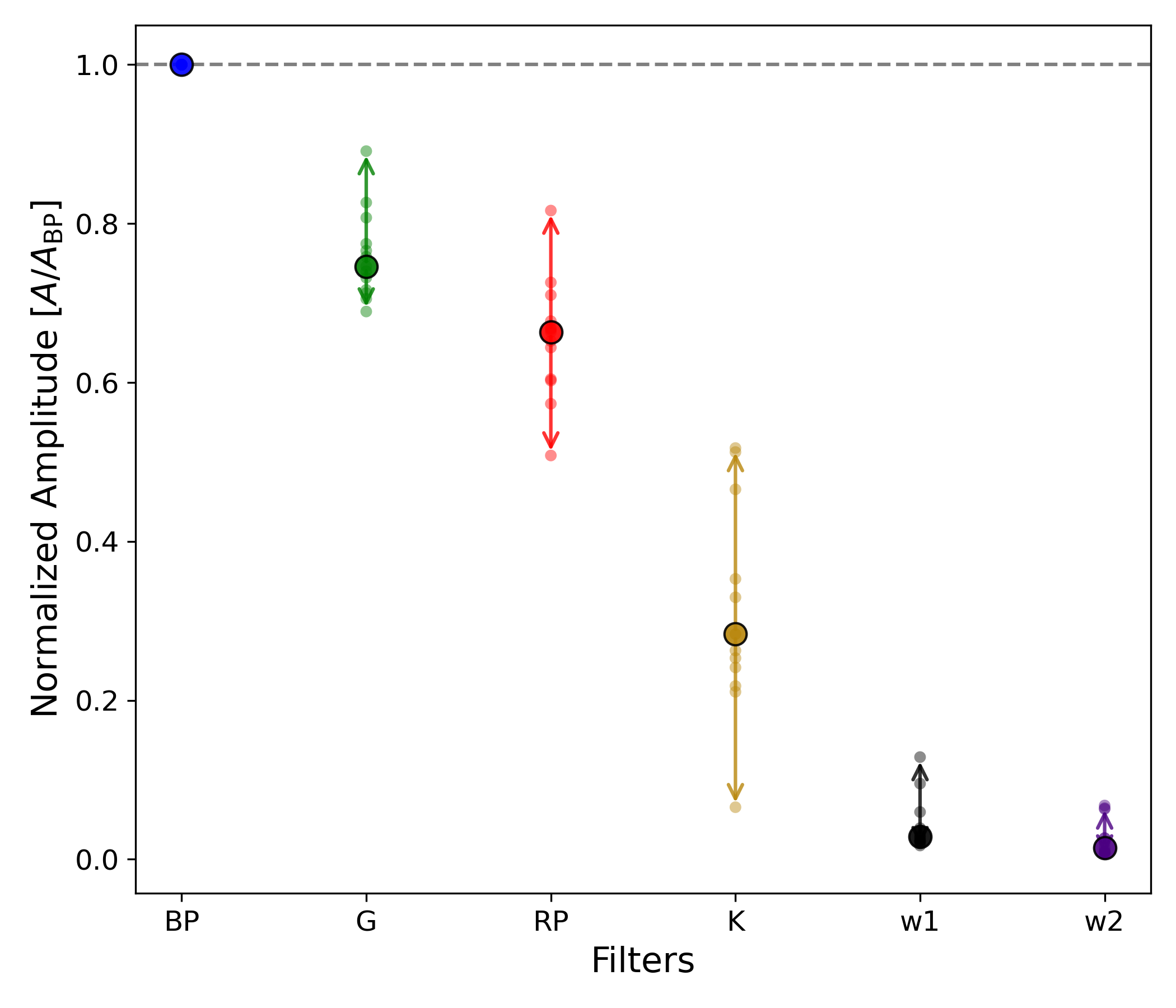}
    \caption{Pulsation amplitude values normalized to the amplitude in the \textit{Gaia} band $BP$. Large dots represent the average value of the amplitude for each filter, while smaller dots are the single measures. The vertical lines are useful to visualize the entire range of values spanned by each filter.}
    \label{fig:amplitudes}
\end{figure}

\begin{figure}[h]
    \centering
    \includegraphics[width=0.95\linewidth]{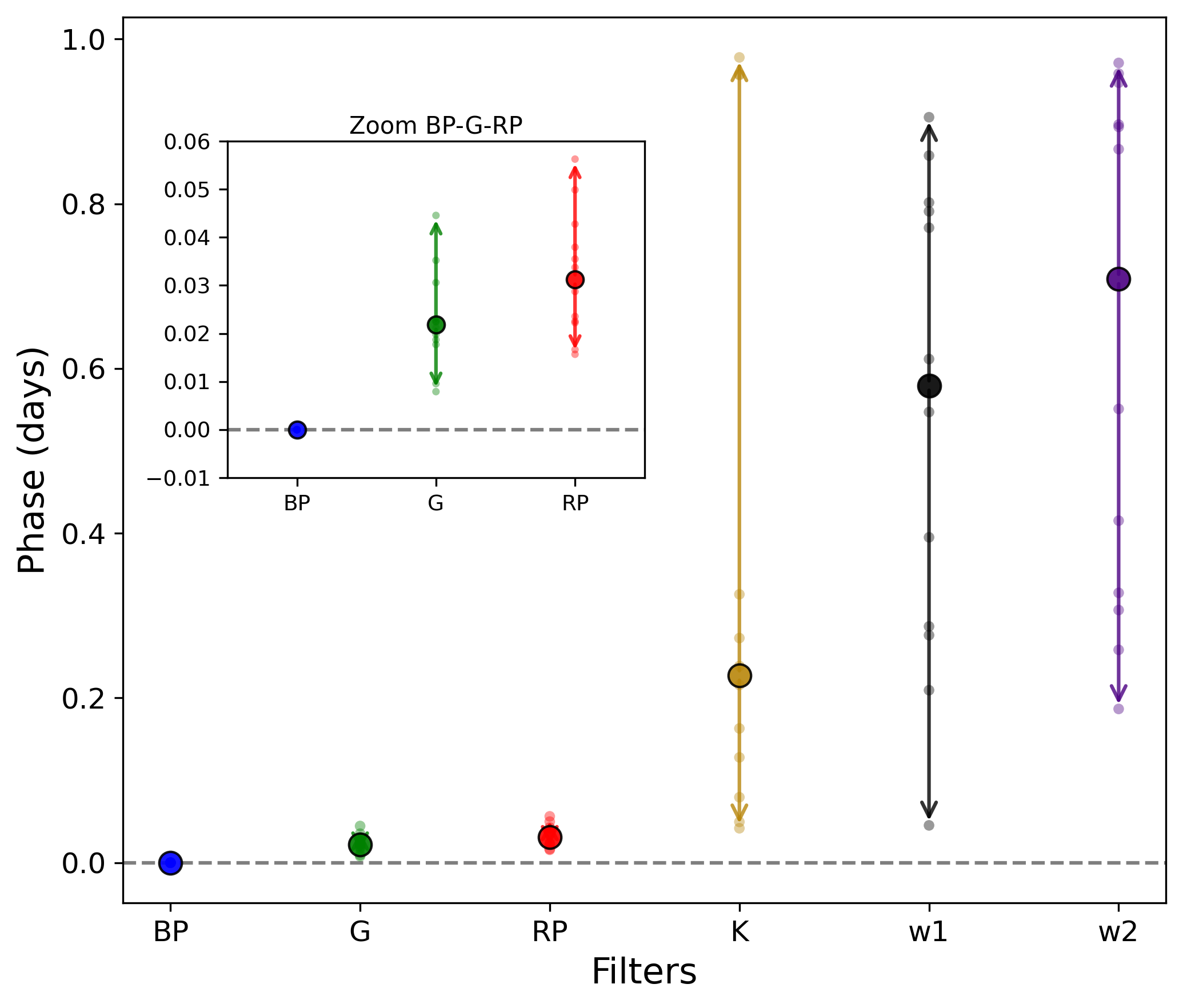}
    \caption{Pulsation phases across the photometric filters used for SED fitting. Colors correspond to the different photometric bands employed in the light curve fit. Large dots represent the average value of the amplitude for each filter, while smaller dots are the single measures. The vertical lines are useful to visualize the entire range of values spanned by each filter. The small window inside is a zoom of the $Gaia$ filters.}
    \label{fig:folded_bands_phase_notsynchro}
\end{figure}

Furthermore, our multi-band modeling reveals a wavelength dependence in both the amplitude and the phase of the pulsation. 
Figure~\ref{fig:amplitudes} highlights a systematic and drastic decrease in pulsation amplitude from the optical to the infrared (IR) bands. 
At the same time, Figure~\ref{fig:folded_bands_phase_notsynchro} reveals a progressive phase shift (delay) toward longer wavelengths relative to the bluest filter (\textit{Gaia} $G_\mathrm{BP}$).
This is a known feature of Mira variables, already documented in foundational literature \citep[e.g.,][]{lockwood1971lcurvesmiras, smith2006phaselags}.

These two observational features are physically consistent and jointly trace the interplay between the pulsating stellar atmosphere and its circumstellar environment. 
This amplitude dampening at longer wavelengths primarily originates in the photosphere. Given the typical effective temperatures of our sample ($T_{\rm eff} \approx 3000$~K), the optical bands sample the steep, short-wavelength side of the Planck function, where the emitted flux is highly sensitive to even small pulsational variations in $T_{\rm eff}$. 
This intrinsic sensitivity is further amplified by molecular blanketing (e.g., $\mathrm{C}_2$, $\mathrm{CN}$), which strongly suppresses the optical flux at minimum light when the star is at its coolest. Conversely, the IR bands sample the long-wavelength side of the Planck curve, where the thermal emission is inherently much less responsive to temperature changes.
Moreover, the circumstellar dust absorbs this variable optical flux and thermally re-emits it in the IR. 
This reprocessing dampens the IR amplitude, while the light-travel time to the dust shell naturally introduces the observed optical-IR phase lag.

While the optical-to-IR phase lag is clearly evident, the trend becomes highly scattered when moving further into the mid-IR (e.g., the WISE $W1$ and $W2$ bands). 
As seen in Fig.~\ref{fig:folded_bands_phase_notsynchro}, the measured phases for these filters span nearly the entire $0$ to $1$ range. 
This large spread is primarily an observational limitation rather than a physical effect. 
In our multiband modeling approach, the global period is fixed across all bands, but the amplitude and phase remain strictly independent variables. Consequently, the algorithm requires sufficient time-series sampling to properly constrain the phase parameter. While the dense, simultaneous \textit{Gaia} observations successfully anchor the relative phase shifts in the optical, the much sparser time-series data in the mid-IR fail to robustly constrain this independent phase, leading to the observed spread.
Coupled with the complex atmospheric opacity stratification driven by dynamical structural effects, this data sparsity makes it exceptionally challenging to pinpoint a precise phase lag in the IR. 
Indeed, \cite{iwanek2021lcurves} similarly found no strong evidence for a clean systematic phase lag trend strictly among the IR bands in C-rich Miras.

Moreover, the WISE observing cadence of approximately one year is close to the pulsation periods of many of the stars in our sample. 
As a result, the WISE observations may repeatedly sample similar pulsation phases, leading to systematically underestimated variability amplitudes. 
The reported WISE amplitudes for these objects should therefore be interpreted as lower limits. \\

\begin{figure}[!h]
    \centering
    \includegraphics[width=1.0\linewidth]{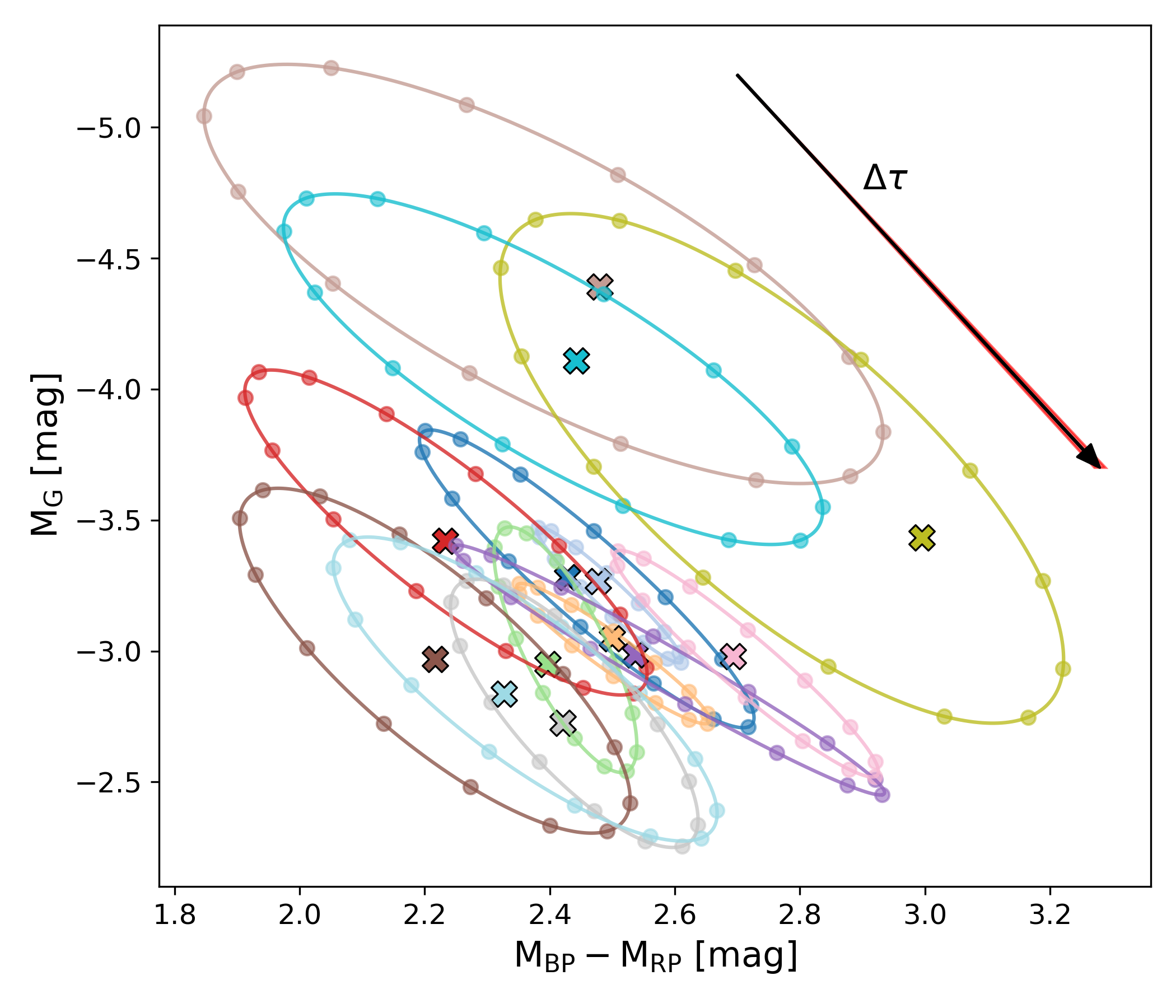}
    \caption{Color-magnitude diagram evolution of the 13 analyzed stars. Each color represents a different star. Solid lines trace the entire pulsation cycle, while circles indicate the phases sampled for SED fitting. The central crosses mark the static color and magnitude utilized by \citetalias{liberatori2026ggscs} for time-averaged SED fitting. The black arrow represents the average reddening vector for all the models considered, tracing the expected displacement due to increasing circumstellar dust. The red shaded region highlights the total area spanned by all variations within the models parameters.}
    \label{fig:cmd_evolution}
\end{figure}

To further illustrate the interplay between stellar pulsation and the circumstellar environment, Figure~\ref{fig:cmd_evolution} presents the phase-resolved magnitude and color evolution of our targets in the \textit{Gaia} Color-Magnitude Diagram (CMD). 
The cyclic variations of the stars form distinct, elongated loops. 

To understand the physical driver of these variations, we compare their orientation against the theoretical circumstellar reddening vector. As defined by \cite{marigo2013MNRAS}, this vector traces the exact displacement a star would undergo in the CMD strictly due to an increase in circumstellar dust, assuming a standard carbon-rich composition of 85\% amorphous carbon and 15\% SiC for various $T_{\rm eff}$ values between 2650~K and 3400~K.

Notably, the major axes of these ellipses are closely aligned with the reddening vector. 
This alignment provides direct photometric evidence that the observed color and magnitude variations across the pulsation cycle are not solely due to changes in the underlying stellar effective temperature, but are heavily driven by the periodic variation of the circumstellar optical depth. Consequently, the more closely the major axis of the pulsational loop parallels the reddening vector, the more strongly the star's evolution in the CMD is dominated by dust opacity variations.
A dedicated simulation demonstrating how pure temperature variations fail to reproduce this alignment, which is instead perfectly matched when dynamic dust condensation is included, is presented and discussed in Appendix \ref{fig:appendix_cmd_sim}.

\subsection{Effective temperature}
Overall, the analyzed sample exhibits a mean effective temperature of $\sim 3300$~K, consistent with literature values. 
Across a single pulsation cycle, as presented in Fig.~\ref{fig:diagnostic}, peak-to-peak $T_{\rm eff}$ variations range from $\sim 100$~K to over $1000$~K in the most extreme cases, depending on the star.  
The average $T_{\rm eff}$ excursion within a pulsation cycle is $\overline{\Delta T_{\rm eff}} \approx 590$~K.
Furthermore, the formal uncertainty on $T_{\rm eff}$ is highly phase-dependent. On average, this uncertainty increases from roughly $200\,$K at maximum brightness to $375\,$K at minimum light. This behavior is physically expected: fainter phases naturally suffer from larger intrinsic photometric errors, while enhanced dust formation more heavily obscures the stellar photosphere, degrading the precision of the surface temperature inference. 
We caution that these formal uncertainties are also inherently limited by the finite parameter spacing of the model grid.

\begin{figure}[!h]
    \centering
    \includegraphics[width=1\linewidth]{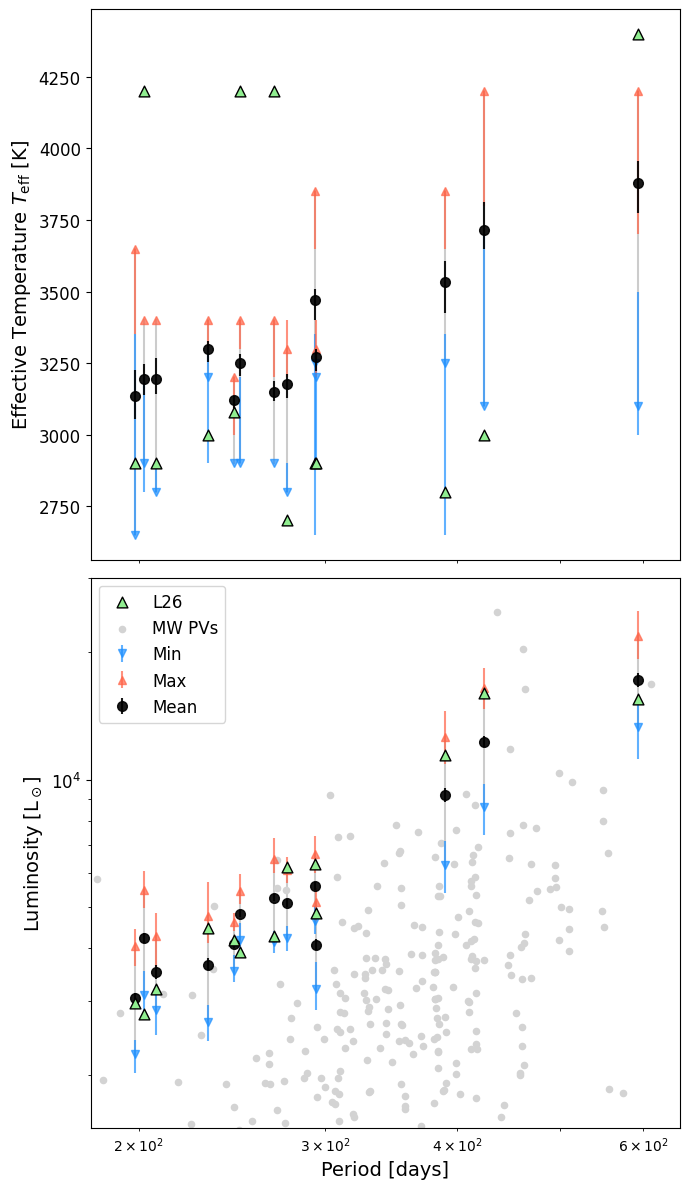}
    \caption{Top panel: Effective temperature values for each star as a function of pulsation period. Error bars represent the total $T_{\rm eff}$ amplitude spanned during the entire cycle. Black dots indicate median $T_{\rm eff}$ values across the cycle, while green triangles represent the static $T_{\rm eff}$ derived from the time-averaged SED fit by \citetalias{liberatori2026ggscs}. Bottom panel: Luminosity values for each star at different period regimes. Symbols for the analyzed SMC targets are the same as in the top panel. Grey points represent C-rich LPVs in the MW from \citet{GaiaCollaborationandTrabucchi2023lpvs} crossmatched with the catalogue from \citetalias{liberatori2026ggscs}.}
    \label{fig:Period_Teff_Luminosity_BestFit}
\end{figure}

Figure~\ref{fig:Period_Teff_Luminosity_BestFit} (top panel) displays the $T_{\rm eff}$ amplitude for each star as a function of its pulsation period. From a theoretical standpoint, as AGB stars evolve and their pulsation periods lengthen, they become more extended and cooler, leading to an expected anti-correlation between $T_{\rm eff}$ and period. 

However, our phase-resolved results exhibit higher median $T_{\rm eff}$ values for the longest-period targets. 
This apparent trend can driven by the degeneracy between effective temperature and circumstellar dust. 
As shown in Sect.~\ref{subsect:optical_depth}, longer-period stars possess significantly higher optical depths ($\tau_{0.55} \ge 1$). 
In this heavily obscured regime, the broad-band SED is dominated by dust reprocessing, severely reducing the fit's sensitivity to the underlying photospheric temperature. 
Because we strictly fixed the stellar mass and C/O ratio across the pulsation cycle, the minimization algorithm could reproduced the massive bolometric amplitudes of these obscured sources by selecting higher $T_{\rm eff}$ models and compensating with drastically increased dust optical depths. 

Nevertheless, when these strict phase-resolved constraints are relaxed, this degeneracy can be mitigated: the unconstrained, time-averaged fits by \citetalias{liberatori2026ggscs} for these exact same long-period stars (green triangles in Fig.~\ref{fig:Period_Teff_Luminosity_BestFit}) yield $T_{\rm eff}$ values of $\sim 2800-3000$~K, aligning with theoretical expectations. 
The only exception is the longest-period star, which is also the most dust-obscured, for which the high amount of dust around the stellar surface likely prevents us to actually infer accurate effective temperature values.
Derived temperatures in heavily dust-enshrouded regimes should therefore be interpreted with caution.

\begin{table*}[htbp]
\small
    \centering
    \renewcommand{\arraystretch}{1.3} 
    \setlength{\tabcolsep}{4pt}
    \caption{Properties and derived parameters of the selected stars.}
    \label{tab:results_table}
    \begin{tabular}{l c c c c c c c c c c c c c c}
            \hline\hline
        $Gaia$ DR3 id & Period & C/O & M & $T_{\rm eff, min}$ & $T_{\rm eff, max}$ & \dots & $L_{\rm min}$ & $L_{\rm max}$ & $\tau_{\rm min}$ & $\tau_{\rm max}$ & $\tau_{\rm p16, min}$ & $\tau_{\rm p84, min}$ & $\tau_{\rm p16, max}$ & $\tau_{\rm p84, max}$ \\
         & (days) & & ($M_\odot$) & (K) & (K) & \dots & ($L_\odot$) & ($L_\odot$) & & & & & & \\
        \hline
        4685915083569253120 & 202 & 1.40 & 3.0 & 2900 & 3400 & \dots & 3099 & 5483 & 0.02 & 0.55 & 0.0002 & 0.0916 & 0.3 & 0.9 \\
        4687404887428798720 & 294 & 1.40 & 3.0 & 3200 & 3300 & \dots & 3204 & 5125 & 0.3 & 0.95 & 0.2 & 0.5 & 0.8 & 1.5 \\
        4685488919677678080 & 269 & 1.40 & 3.0 & 2900 & 3400 & \dots & 4129 & 6499 & 0.02 & 0.5 & 0.0001 & 0.3 & 0.4 & 0.6 \\
        4688954717829223936 & 390 & 2.00 & 1.0 & 3250 & 3850 & \dots & 6281 & 12579 & 0.2 & 0.8 & 0.0784 & 0.3 & 0.5 & 1.0 \\
        4685883129008411264 & 293 & 2.00 & 1.0 & 3250 & 3850 & \dots & 4635 & 6685 & 0.3 & 0.9 & 0.3 & 0.4 & 0.7 & 1.0 \\
        4685776167110162944 & 246 & 1.05 & 2.0 & 2900 & 3200 & \dots & 3529 & 4617 & 0.0001 & 0.3 & 0.0001 & 0.02 & 0.1 & 0.4 \\
        4688871601593755008 & 198 & 2.00 & 1.0 & 2650 & 3650 & \dots & 2251 & 4055 & 0.0001 & 0.5 & 0.0001 & 0.04 & 0.2 & 0.6 \\
        4688894107214534144 & 593 & 2.00 & 3.0 & 3100 & 4200 & \dots & 13297 & 21779 & 1.0 & 3.5 & 0.9 & 1.5 & 3.5 & 4.0 \\
        4685776167120724992 & 233 & 1.40 & 3.0 & 3200 & 3400 & \dots & 2669 & 4751 & 0.0095 & 0.6 & 0.0001 & 0.1 & 0.5 & 0.8 \\
        4685949546407892608 & 424 & 2.00 & 2.0 & 3100 & 4200 & \dots & 8619 & 16449 & 0.09 & 1.0 & 0.04 & 0.5 & 0.7 & 1.5 \\
        4685649620189877248 & 249 & 1.40 & 3.0 & 2900 & 3400 & \dots & 4181 & 5449 & 0.2 & 0.5 & 0.1 & 0.3 & 0.4 & 0.6 \\
        4685898006715794688 & 276 & 1.40 & 3.0 & 2800 & 3300 & \dots & 4214 & 6099 & 0.3 & 1.0 & 0.03 & 0.4 & 0.8 & 1.5 \\
        4689068208025791232 & 207 & 1.10 & 2.0 & 2800 & 3400 & \dots & 2860 & 4271 & 0.0001 & 0.3 & 0.0001 & 0.01 & 0.0001 & 0.5 \\
        \hline

    \end{tabular}
    \vspace{1mm}
    \tablefoot{
        \tablefoottext{a}{This table is available in its entirety at the CDS. The full version includes additional columns providing the derived photometric pulsation parameters (mean magnitudes, amplitudes, phases, and their respective errors) for each individual band, the OGLE cross-match IDs, and the asymmetric $1\sigma$ uncertainties (16th and 84th percentiles) for the derived physical parameters at both minimum and maximum light.}
    }
\end{table*}

\subsection{Luminosity}
The bolometric luminosity at each phase was derived empirically via spline interpolation of the multi-band photometry, as done in \citetalias{liberatori2026ggscs}. We chose this approach to keep input luminosities strictly independent of theoretical atmospheric models.

Throughout the pulsation cycle, the bolometric luminosity undergoes substantial variations, often exceeding several tens of percent relative to the mean value. 
Figure~\ref{fig:diagnostic} presents, as an example, how the luminosity of the star \textit{Gaia} DR3 4685883129008411264 spans a wide range of values, going from $4635~L_\odot$ to $6685~L_\odot$.

Figure~\ref{fig:Period_Teff_Luminosity_BestFit} (bottom panel) demonstrates a strong correlation between average luminosity and pulsation period, clearly recovering the classical Period-Luminosity (PL) relation expected for Mira variables \citep[e.g.,~][]{feast1989plrmiras, huges1990lpvs, Whitelock2006plmiras}.

\subsection{Optical Depth}
\label{subsect:optical_depth}
All analyzed stars exhibit a significant variation in optical depth (at 0.55$\rm \mu m$, hereafter defined as $\tau_{0.55}$) throughout their pulsation cycle. 
Most notably, almost every star shows a strong anti-correlation between $\tau_{0.55}$ and luminosity: minimum optical depths systematically coincide with maximum luminosity and effective temperature.

This behavior is physically expected. 
In a simplified scenario of a stationary wind or a fixed shell, the enhanced radiation field at maximum brightness (minimum radius and maximum temperature) pushes the inner dust condensation radius outward. 
While the actual circumstellar dynamics is more complex (with pulsation-driven shocks playing a role in dust formation), the intense radiation field hinders dust condensation and survival in both scenarios.
This effectively reduces the total dust column density, yielding a lower optical depth. 
Our empirical results closely trace the analytical relation $\tau_{0.55} \propto L^{-1/2}$ derived by \cite{bressan1998}.

Overall, the sample has a mean optical depth of $\overline{\tau_{0.55}} \sim 0.53$ and an average phase-dependent excursion of $\overline{\Delta \tau_{0.55}} \approx 0.68$.
As with $T_{\rm eff}$, resolving $\tau_{0.55}$ variations provides a much more coherent physical picture of the circumstellar envelope than static fits.

\begin{figure}
    \centering
    \includegraphics[width=1\linewidth]{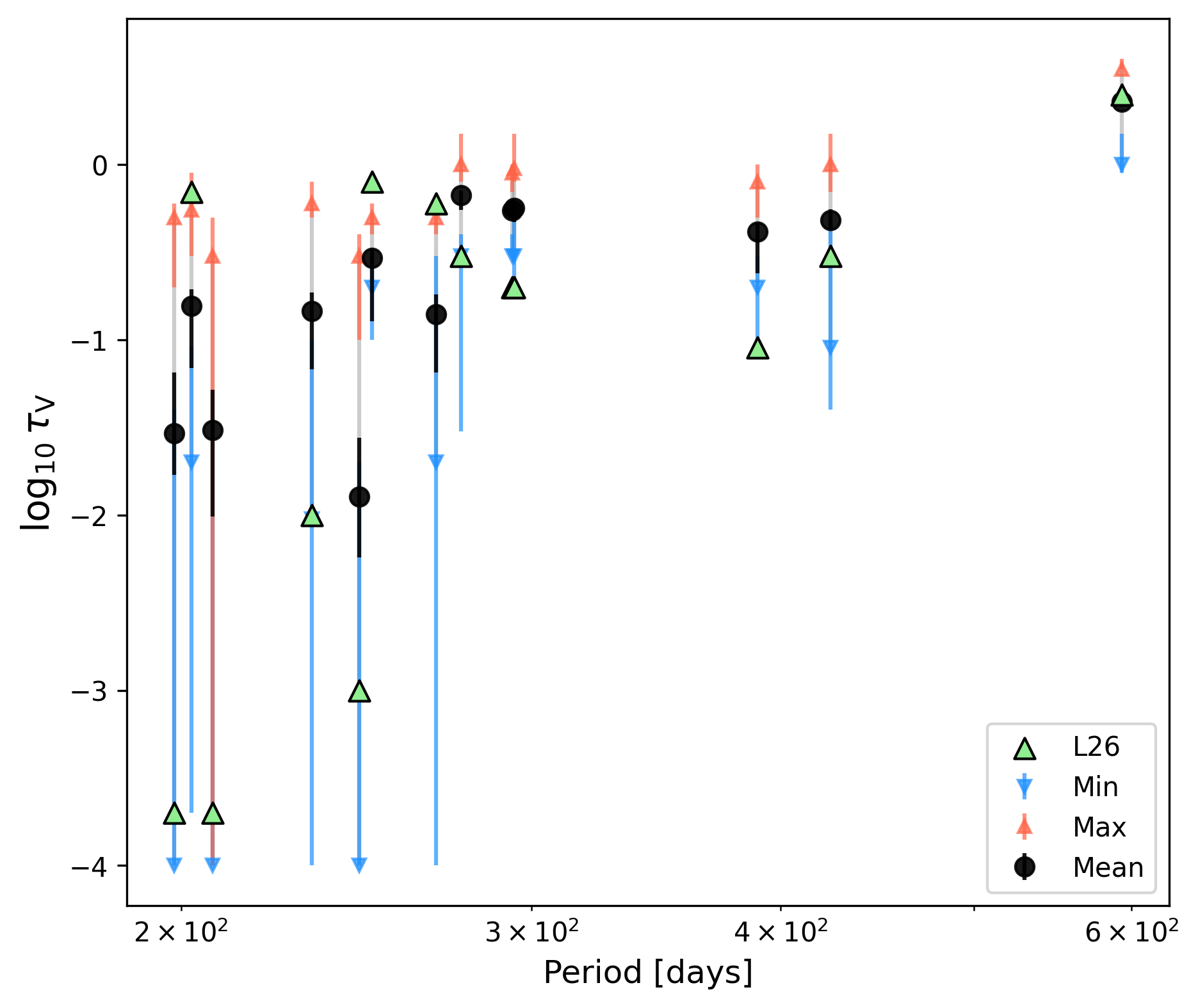}
    \caption{Optical depth values for each star as a function of pulsation period. Color labels are the same as in Fig.~\ref{fig:Period_Teff_Luminosity_BestFit}.}
    \label{fig:Period_Tau_BestFit}
\end{figure}

Finally, Fig.~\ref{fig:Period_Tau_BestFit} shows that the average $\tau_{0.55}$ mildly correlates with the pulsation period, tending to be higher in stars with longer periods. 
We also highlight that the most obscured star in our sample exhibits the longest pulsation period (593~days).

\subsection{Mass and C/O ratio}
Stellar mass and surface C/O ratio for each source were jointly fixed to unique values for each star, in order to avoid a nonphysical variation of this values across the pulsation cycle.

Out of the 13 stars in our sample, 7 have a mass of $3~M_\odot$, 3 stars have a mass of $1~M_\odot$ and 3 stars are characterized by a mass of $2~M_\odot$. 
The apparent restriction to coarse integer masses is a direct consequence of the grid's sampling density, as intermediate masses lack sufficient $T_{\rm eff}$ coverage to fit a full pulsation cycle.
The values of current mass are not necessarily consistent with evolutionary models. However, it is well established that broad-band photometric SED fitting is weakly sensitive to the assumed stellar mass. 
As explored in detail by \citet{aringer2009crichgiantsphotometry}, while mass and surface gravity can introduce complex variations in specific molecular absorption features, their overall influence on photometric colors remains marginal. 
These mass-induced variations are, in fact, significantly smaller than the uncertainties introduced by circumstellar dust reddening and atmospheric dynamics. This aspect is examined in more detail in Sect.~\ref{sec:Discussion}.

Regarding the surface chemistry, the derived C/O ratios span a range from 1.05 to 2.00. We find 6 stars being characterized by a C/O of 1.4, 5 stars by a C/O=2.0, and 2 stars with values of 1.05 and 1.10, respectively. Similar to the stellar mass, the apparent bunching of the solutions around C/O values of 1.4 and 2.0 is an artifact of the grid's sampling density, as these are the only chemical compositions providing a continuous and dense $T_{\rm eff}$ coverage in the COMARCS models.

It is worth noting that for carbon stars in a low-metallicity environment such as the SMC, one might physically expect C/O values exceeding 2.0. However, our determinations are artificially truncated by the upper boundary of the available model grid. Theoretically, this truncation at C/O $\le$ 2.0 introduces specific degeneracies. 
If a star possesses a true C/O $> 2.0$, its SED will be intrinsically redder in the optical due to enhanced molecular blanketing. To compensate for the missing molecular opacity, the SED fitting algorithm might artificially alter the remaining free parameters.

However, we emphasize that our specific sample consists of \textit{Gaia}-selected targets, which are inherently optically bright carbon stars. Because they are not deeply dust-enshrouded, their optical emission is well-constrained, and the derived optical depths are naturally low. If the fitting algorithm were severely compensating for a truncated C/O ratio by artificially inflating the dust opacity, we would expect systematically higher $\tau_{0.55}$ values. Consequently, while the C/O upper limit might severely bias the parameters of heavily obscured, extreme carbon stars, its impact on our specific optically-bright sample is likely marginal. Therefore, we consider our derived stellar and circumstellar parameters to be robust, although deriving completely unique C/O ratios from broad-band photometry remains intrinsically degenerate \citep{aringer2009crichgiantsphotometry}.

\subsection{Radial Pulsation and Velocity}
By tracking the phase-resolved $T_{\rm eff}$ and luminosity, we traced the continuous stellar radius variation via the Stefan-Boltzmann law. Our analysis of the 13 stars yields an average minimum radius of 186 $R_\odot$ and an average maximum radius of 287 $R_\odot$, which corresponds to a radial expansion of $\sim$54\% from minimum to maximum. 

\begin{figure}[h]
    \centering
    \includegraphics[width=\linewidth]{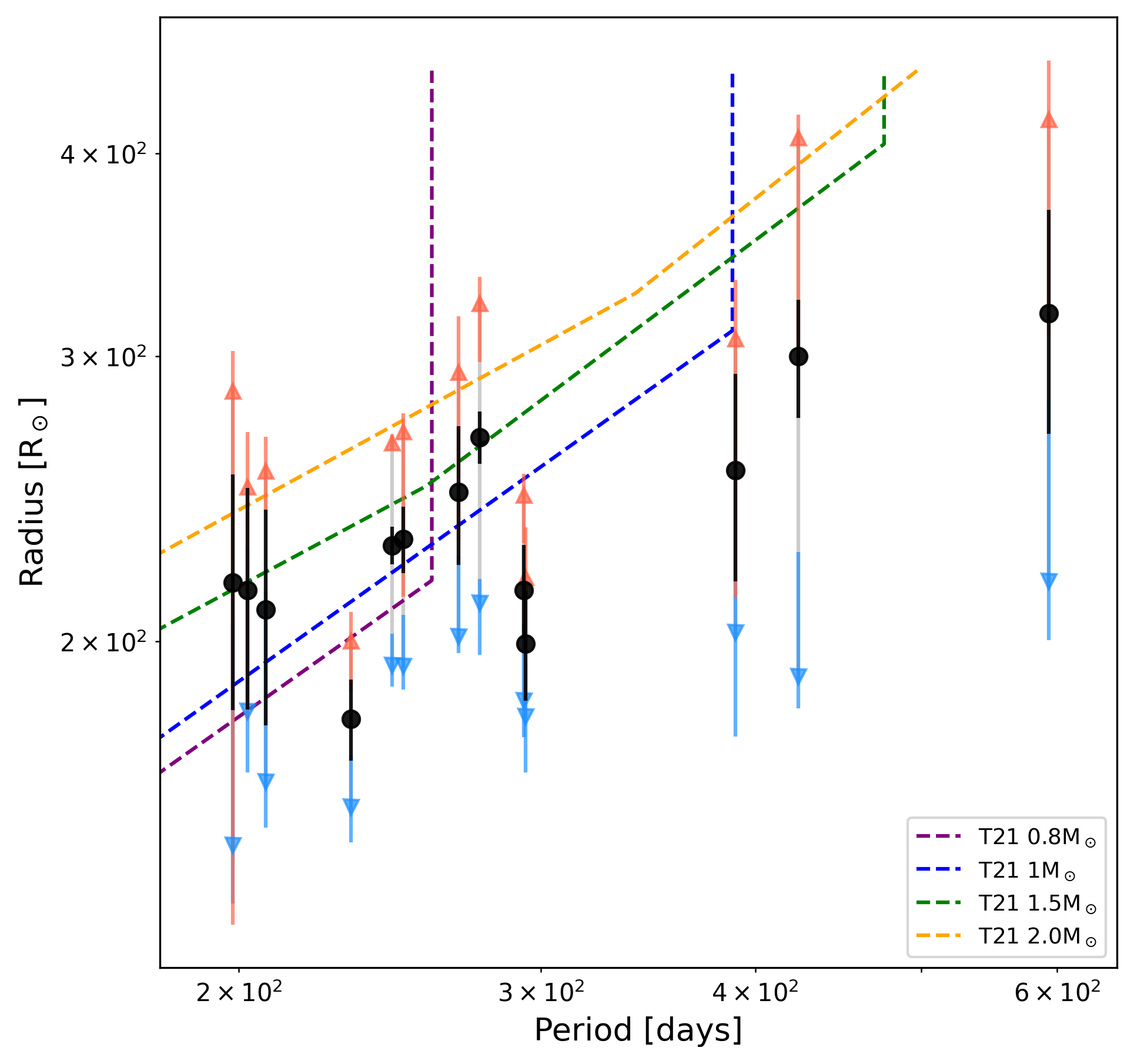}
    \caption{Derived stellar radii ranges as a function of pulsation period. Black dots represent the median radius across the cycle, while blue downward and red upward triangles indicate the minimum and maximum radii, respectively. Dashed lines represent the theoretical non-linear PMR relations from \citet{trabucchi2021longperiods} for different stellar masses (0.8, 1.0, 1.5, and 2.0 $M_\odot$).}
    \label{fig:Period_Radius_T21}
\end{figure}

To validate our results, we compared the derived radii with the theoretical non-linear Period-Mass-Radius (PMR) relations by \citet{trabucchi2021longperiods} (Fig.~\ref{fig:Period_Radius_T21}). Our targets populate the region between the 0.8 and 2.0 $M_\odot$ tracks, consistent with the expected progenitors of SMC C-rich AGB stars. 
This comparison is physically robust, as the structural PMR relation is largely insensitive to the specific atmospheric chemistry or metallicity adopted in the theoretical models. 
The only notable outlier is our longest-period, most heavily dust-enshrouded target. 
In a such highly dusty regime, the dense circumstellar envelope severely obscures the true photosphere (making the empirical radius derivation highly uncertain), while simultaneously challenging the applicability of the theoretical models themselves.

Although these pulsational masses ($0.8 - 2.0~M_\odot$) are systematically lower than our unconstrained photometric masses, this discrepancy is an artifact of the grid's coarse mass sampling. 
As demonstrated in Appendix~\ref{asec:comparison-models-priors}, introducing the period as an active prior naturally drives the SED fit toward these lower, theoretically expected masses, further confirming the soundness of our geometric derivations.

Furthermore, by fitting a sinusoidal function to these radius variations and computing its time derivative, we determined the radial pulsation velocity. We obtained an average pulsation semi-amplitude (defined as the zero-to-peak velocity) of $\Delta V_\mathrm{puls} \approx 5.3$~km~s$^{-1}$.

This derivation, however, requires a smoothly sampled $T_{\rm eff}$ evolution. For stars where discrete grid steps cause abrupt temperature jumps, the non-continuous radius curve prevents reliable velocity estimation.

\begin{figure}[!h]
    \centering
    \includegraphics[width=0.95\linewidth]{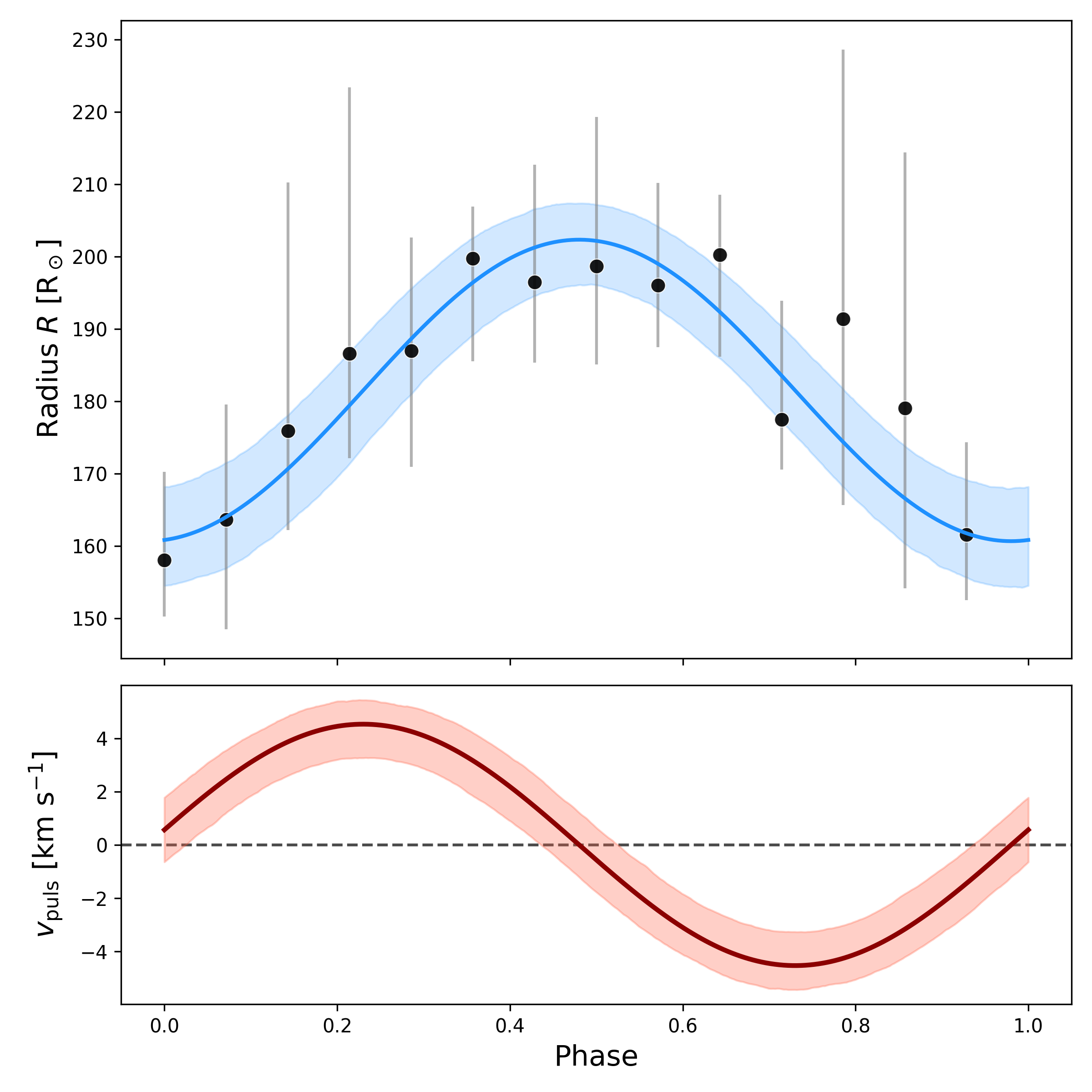}
    \caption{Radius evolution of a single star across the pulsation cycle. Points indicate values derived by inverting the Stefan-Boltzmann law; the solid line is the best-fit sinusoidal model. Below is the evolution of the velocity amplitude of radial pulsation with phase.}
    \label{fig:radial-velocity-evolution}
\end{figure}

Figure~\ref{fig:radial-velocity-evolution} shows a representative example for \textit{Gaia} DR3 4685776167120724992. The pulsation velocity semi-amplitude for this source is $\Delta V_\mathrm{puls} \approx 5$~km~s$^{-1}$. 
The values found for radial pulsation and pulsation velocity, presented in Table~\ref{tab:table_velocity}, are consistent with both dynamical models and recent kinematic measurements of expanding envelopes in AGB pulsators. From a theoretical standpoint, for instance, \cite{Eriksson2014synthetic} simulated the expansion of the stellar surface in their dynamical models using a piston at the inner boundary just below the photosphere, with the velocity prescribed as a sinusoidal variation with semi-amplitudes set to 2, 4, and 6 $\rm km \ s^{-1}$, values that are absolutely compatible with our findings.

From an observational perspective, although the \textit{Gaia} Focused Product Release (FPR) provides epoch radial velocities for thousands of Long Period Variables \citep{GaiaCollaborationandTrabucchi2023lpvs}, direct radial velocity time-series are not available in the literature for our specific targets, as they are not included in this kinematic sample. 
Nevertheless, our photometrically-derived velocities are strongly supported by both individual high-resolution spectroscopic measurements and large-scale statistical samples.

As demonstrated by \cite{Nowotny2010lineformation}, the kinematic behavior of the inner pulsating layers is largely universal among Mira variables, exhibiting consistent radial velocity amplitudes and structural variations regardless of their specific atmospheric chemistry. 
Relying on this universal behavior, we can robustly benchmark our derived velocities against well-studied Galactic counterparts. 
For instance, time-series spectroscopy of near-infrared molecular lines in the carbon-rich Mira S~Cep reveals a radial velocity semi-amplitude of $\approx 7$~km~s$^{-1}$ \citep{Nowotny2005dynamics}, which is in excellent agreement with our typical derived values.

\begin{figure}[!h]
    \centering
    \includegraphics[width=0.95\linewidth]{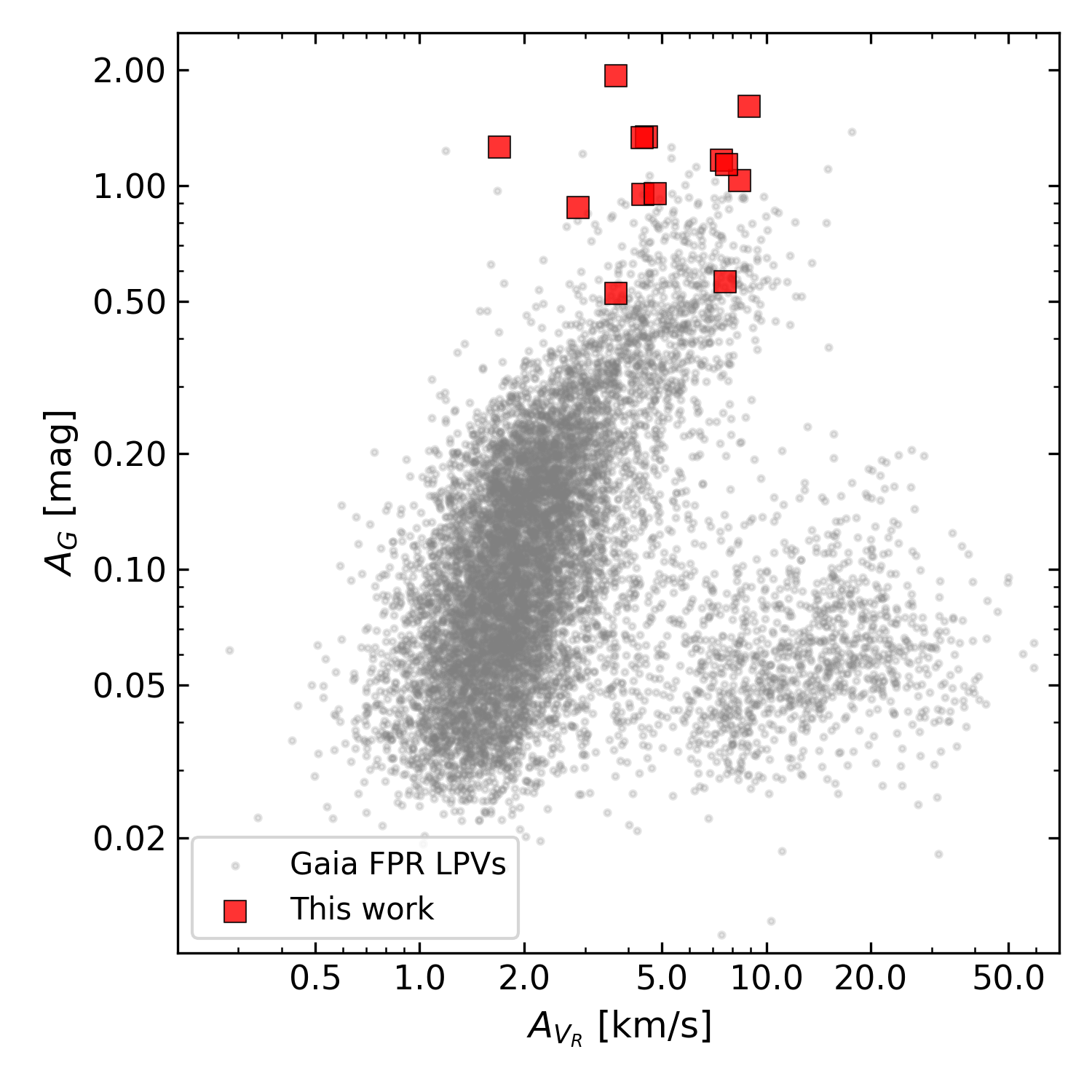}
    \caption{Pulsation amplitude in the $G$ band ($A_G$) as a function of the radial velocity semi-amplitude ($A_{V_R}$). The grey dots represent the population of Long Period Variables from the \textit{Gaia} FPR \citep{GaiaCollaborationandTrabucchi2023lpvs}, while the red squares denote the stars analyzed in this work. Our targets naturally populate the kinematic and photometric locus expected for large-amplitude Mira variables.}
    \label{fig:amplitude_vs_velocity}
\end{figure}

Furthermore, we can statistically validate our results by comparing the properties of our targets against the entire \textit{Gaia} FPR LPV population. Figure~\ref{fig:amplitude_vs_velocity} illustrates the $G$-band pulsation amplitude as a function of the radial velocity semi-amplitude ($A_{V_R}$). 
Our sample is consistent with the upper locus of the \textit{Gaia} FPR distribution, populating the region expected for large-amplitude, carbon-rich Mira variables. 

We note, however, a fundamental physical distinction regarding our derived velocities. Our photometrically derived radii trace the stellar photosphere via the Stefan-Boltzmann law, rather than the physical mass elements of the pulsating envelope. 
As the expanding envelope becomes more diffuse, the photosphere naturally lags behind the faster-moving outward material. 
Consequently, our derived expansion velocities must be strictly interpreted as lower limits to the true kinematic velocity of the gas. As a result, the true physical velocities of our targets could be larger, which would shift our sample further to the right in Figure~\ref{fig:amplitude_vs_velocity}, following the natural extension of the \textit{Gaia} sequence. Nevertheless, the consistency between the order of magnitude of our photometric lower limits and the values from large-scale kinematic surveys provides a robust validation of our phase-resolved fitting methodology.

\begin{table}[htbp]
\small
    \centering
    \renewcommand{\arraystretch}{1.3} 
    \setlength{\tabcolsep}{4pt}
    \caption{Stellar radius ranges and pulsation velocity semi-amplitudes for all the stars in the sample.}
    \label{tab:table_velocity}
    \begin{tabular}{l c c c}
        \hline\hline
        $Gaia$ DR3 id & $\rm R_{min}$ & $\rm R_{max}$ & $\rm \Delta V_{puls}$ \\
         & ($\rm R_\odot$)  & ($\rm R_\odot$) & (km/s) \\
        \hline
        4685898006715794688 & 211 & 323 & 5 \\
        4688954717829223936 & 203 & 308 & 4 \\
        4688871601593755008 & 150 & 285 & 8 \\
        4685776167110162944 & 193 & 265 & 8 \\
        4685776167120724992 & 158 & 200 & 4 \\
        4685488919677678080 & 201 & 293 & 8 \\
        4685883129008411264 & 184 & 246 & 2 \\
        4685915083569253120 & 181 & 249 & 5 \\
        4689068208025791232 & 164 & 255 & 7 \\
        4688894107214534144 & 218 & 420 & 4 \\
        4685949546407892608 & 190 & 409 & 9 \\
        4687404887428798720 & 180 & 219 & 3 \\
        4685649620189877248 & 193 & 269 & 4 \\
        \hline
    \end{tabular}
    \vspace{1mm}
\end{table}

\section{Discussion}
\label{sec:Discussion}
The fundamental merit of our phase-resolved approach lies in its ability to reconstruct a coherent, time-dependent physical picture of the pulsating carbon star, which traditional time-averaged SED fitting methods inherently fail to capture. Although the underlying atmospheric (COMARCS) and dust (DUSTY) models are structurally static, our methodology allows for the temporal variation of these static conditions across the entire pulsation cycle. By tracking the fundamental parameters in this manner, we can directly observe the dynamic connection between the star and its surrounding envelope, overcoming the limitations of standard mean-photometry fits.

Specifically, our results trace a clear thermodynamic and structural sequence. When the star reaches its maximum brightness, it is in a contracted, hot state, characterized by a minimum radius and maximum $T_{\rm eff}$, which drives a peak in bolometric luminosity. 
This intense radiation field pushes the dust condensation front to larger distances. 
Consequently, the total column density of the circumstellar dust decreases, naturally explaining the systematic anti-correlation we observe between the optical depth ($\tau_{0.55}$) and the luminosity. Conversely, as the star expands and cools toward minimum light, the weakened radiation field allows new dust to condense closer to the photosphere, rapidly driving the optical depth to its peak.

This photometric and thermodynamic picture is strongly supported by our kinematic derivations. The evolution of the stellar radius, derived directly from the $T_{\rm eff}$ and luminosity modulations, translates to macroscopic photospheric expansion velocities of $\Delta V_{\rm puls} \approx 5~{\rm km~s^{-1}}$.
The fact that this purely photometric derivation agrees with independent spectroscopic measurements of AGB atmospheres is a powerful validation of the method.

Ultimately, the ability to simultaneously recover the thermodynamic variations, the cyclic dust processing, and the correct radial kinematics highlights the main achievement of this pilot study: our phase-resolved empirical methodology is not merely fitting isolated data points, but is effectively tracing the pulsational engine and the coupled response of the mass-loss-driving envelope in real time.

In the following subsections, we evaluate our findings by comparing them to classical static approaches and extensively discuss the crucial caveats of our analysis, including the structural limits of the model grids and the conceptual limitations of applying hydrostatic atmospheres to highly dynamic environments.

\subsection{Comparison with time-averaged SED fitting methodologies}
A fundamental aspect of this work is evaluating how our phase-resolved approach compares to classical time-averaged SED fitting, which represents the standard methodology in the current literature. Since our procedure introduces a fundamentally different conceptual approach to deriving stellar and circumstellar parameters, a strict one-to-one comparison is inherently limited. Nevertheless, it is instructive to quantify the discrepancies arising from the two methods. 

To this end, we benchmark our findings against the results published by \citetalias{liberatori2026ggscs}. We restrict our comparison to this specific study, as the authors have already demonstrated that their derived parameters are fully consistent with the broader available literature within the uncertainties. The static parameters from \citetalias{liberatori2026ggscs} are overlaid on our dynamic ranges in Figs.~\ref{fig:Period_Teff_Luminosity_BestFit} and \ref{fig:Period_Tau_BestFit}.

A qualitative analysis reveals partial agreement. For 3 out of the 13 stars, the time-averaged $T_{\rm eff}$ from \citetalias{liberatori2026ggscs} falls within the dynamic range captured by our phase-resolved analysis. When accounting for the $1\sigma$ uncertainty boundaries, the time-averaged $T_{\rm eff}$ is consistent with our dynamic ranges for 7 out of the 13 stars.

Similarly, for the optical depth, 6 out of 13 static values fall within our derived limits. 

The most solid agreement is found for the bolometric luminosity, where the static values from \citetalias{liberatori2026ggscs} fall within our pulsation-driven variations for 10 stars. 
This strong consistency is expected, as both studies derive this parameter via multi-band integration—an entirely empirical method strictly independent of theoretical atmospheric models. Regarding the three remaining stars whose static luminosities fall outside our phase-resolved ranges, their values are nevertheless consistent when their $1\sigma$ uncertainties are considered.

To address the discrepancies in $T_{\rm eff}$ and $\tau_{0.55}$, it is crucial to note that they arise directly from fundamental methodological differences rather than observational inconsistencies. First, \citetalias{liberatori2026ggscs} utilized a different photometric dataset (including the full 2MASS and WISE suites), whereas this work relies on a tailored combination of Gaia, VMC $K_{\rm s}$, and WISE W1-W2 photometry. Second, and most importantly, \citetalias{liberatori2026ggscs} performed a blind fit of time-averaged photometry over the entire ATHENA-C grid without imposing prior constraints on mass and C/O ratio. In contrast, our approach extracts phase-specific SEDs and restricts the allowed parameter space by fixing the mass and C/O across the pulsation cycle. Because atmospheric opacities and dust condensation are non-linear processes, fitting a static model to a time-averaged SED does not yield the same physical parameters as the true time-averaged parameters of a dynamic envelope. This underscores the necessity of time-resolved methodologies for a physically accurate characterization of large-amplitude variables.

\subsection{Methodological limitations}
A significant structural limitation affecting our phase-resolved SED fitting is the discrete parameter resolution of the underlying atmospheric models. By enforcing constant mass and C/O values throughout the pulsation cycle to ensure physical consistency, we inherently restrict the available parameter space. For a specific mass and C/O combination, the COMARCS grid provides a sparsely populated sequence of $T_{\rm eff}$ values. Consequently, the derived temperature variations often exhibit abrupt jumps rather than a smooth, continuous evolution (as discussed in Sect.~\ref{sec:Results} regarding radial velocities). 
This is an intrinsic limitation of the COMARCS model grid utilized and cannot be circumvented with the existing models.

Conversely, parameters governing the circumstellar envelope, such as the optical depth $\tau_{0.55}$, are much more homogeneously covered. As detailed by \citetalias{liberatori2026ggscs}, the optical depth was computed for the ATHENA-C grid via DUSTY radiative transfer modeling across every combination of mass, C/O, and temperature, allowing for a much finer and smoother continuous variation in our fits. \\

We note that in this study, we restricted our mid-infrared analysis to the WISE W1 and W2 bands. Incorporating time-series photometry beyond 5~$\mu$m would theoretically provide tighter constraints on the coldest dust in the outermost regions of the circumstellar envelope, thereby reducing the absolute uncertainties on the optical depth. However, long-wavelength data with a temporal resolution comparable to the WISE W1 and W2 bands are not currently available for these stars. Furthermore, extensive tests conducted in \citetalias{liberatori2026ggscs} using time-averaged photometry out to 22~$\mu$m demonstrated that omitting the longer-wavelength bands does not systematically bias the derived median stellar and dust parameters, but merely increases their formal uncertainties. We are therefore confident that the omission of $>5\,\mu$m data does not compromise the robustness of our phase-resolved results.

\subsubsection{Sensitivity of the SED fitting to the mass}
As mentioned in Sect.~\ref{sec:Results}, purely photometric SED fitting is notoriously insensitive to stellar mass. Consequently, the mass values derived from our standard minimum $\chi^2$ approach tend to be overestimated when compared to evolutionary models of AGB stars in the SMC, which predict initial masses strictly below $3\,M_\odot$ for carbon stars \citep[e.g.,][and references therein]{pastorelli2019MNRAS}.

To break this degeneracy and assess the reliability of our mass estimates, we explored an alternative parameter-selection criterion that incorporates stellar variability (detailed in Appendix~\ref{asec:comparison-models-priors}). Since the pulsation period is coupled to mass and radius via theoretical period-mass-radius relations \citep{trabucchi2021longperiods}, we can infer a dynamical mass using the observed period alongside the radius independently derived from $T_{\rm eff}$ and luminosity via the Stefan-Boltzmann law.

Applying this modified, period-constrained method forces the derived current stellar masses to peak strongly around $1\,M_\odot$. 
While this lower mass regime is evolutionarily more accurate than the one based purely on SED fitting, enforcing this constraint exposes the structural limitations of the models grid. 
The resulting $T_{\rm eff}$ and $\tau_{0.55}$ curves become significantly less smooth, and potentially less realistic, than those derived from the classical fit. 
Therefore, while incorporating variability yields a more realistic mass scale, the exact determination remains inherently imprecise due to grid discretisation.

\subsubsection{Adequacy of hydrostatic models and opacities for dynamic atmospheres}
A profound conceptual limitation intrinsic to this class of analyses lies in the hydrostatic nature of the employed atmospheric models. 
The ATHENA-C grid relies on COMARCS and DUSTY to approximate the real atmospheres and dust shells of carbon stars. 
However, Mira variables are defined by intense, radial pulsations and strong shock waves propagating through their outer layers.

Modeling these highly dynamic systems using steady-state, hydrostatic approximations is therefore inherently limited. 
A more physically consistent approach requires modern radiation-hydrodynamics models, such as the 1D DARWIN models \citep{hofner2016darwin, bladh2019carbonmodels, eriksson2023synthphot}, or ultimately, rigorous 3D simulations \citep[e.g., the CO5BOLD models;][]{Freytag2002co5bold} coupled with non-equilibrium dust formation. 
Previous efforts, such as those by \citet{Nowotny2011synthlcurve}, have successfully investigated time-dependent variations for individual Galactic objects using dynamical models. Although significant computational advancements are being made, producing extensive dynamical grids for routine SED fitting remains out of reach for large observational samples. We plan to incorporate such advanced dynamical models in future extensions of this work.

Furthermore, the derived absolute parameters are sensitive to the adopted microphysical input data. On the atmospheric side, recent significant opacity updates are expected to alter the predicted colors of future COMARCS models \citep{sloan2026jwst}. 
Similarly, the choice of dust opacity constants, such as our adoption of the Zubko data, or the treatment of the grain size, introduce systematic uncertainties in the circumstellar parameters. 
Therefore, while hydrostatic models provide the most practical tool currently available to trace physical trends, the absolute derived parameters should be treated with caution, as they merely approximate a highly dynamic environment.

\section{Conclusions}
\label{sec:Conclusions}

\noindent In this work, we presented a phase-resolved SED fitting analysis of a selected sample of 13 carbon-rich Mira variables in the Small Magellanic Cloud. By combining time-series photometry from \textit{Gaia}, WISE, VMC, 2MASS, and OGLE, we reconstructed simultaneous multi-band light curves and extracted independent SEDs at 14 distinct phases across the pulsation cycle.

This dynamic approach allowed us to coherently trace the variations of effective temperature, optical depth, and bolometric luminosity. We found typical peak-to-peak $T_{\rm eff}$ variations ranging from $\sim 100$~K up to over $1000$~K across the cycle. Moreover, we observed a strong anti-correlation between the circumstellar optical depth $\tau_{0.55}$ and the bolometric luminosity, which is physically consistent with the dust condensation zone being pushed outward during phases of maximum radiation. We also successfully recovered the classical Period-Luminosity relation and derived radial expansion velocities ($\Delta V_{\rm puls} \approx 5.3~{\rm km~s^{-1}}$), in agreement with independent spectroscopic measurements of expanding AGB envelopes.

A comparison with classical time-averaged SED fitting results, such as those from \citetalias{liberatori2026ggscs}, demonstrates the inherent limitations of static approaches. While bolometric luminosities remain consistent because they are derived empirically, static fits often fail to capture the true dynamic ranges of $T_{\rm eff}$ and $\tau_{0.55}$. This discrepancy highlights the necessity of time-resolved methodologies to correctly characterize large-amplitude variables. 

As emphasized in Sect.~\ref{sec:Discussion}, the determination of these parameters inevitably depends on the underlying models. The discrete parameter resolution of the hydrostatic grid introduces non-continuous variations in parameters such as $T_{\rm eff}$. Furthermore, the use of steady-state hydrostatic atmospheric models to describe inherently dynamic, pulsating environments represents a fundamental conceptual limitation. 

Despite these caveats, this study represents a significant step forward from traditional static SED modeling. Directly accounting for the time-dependent behavior of these AGB stars enables a more complete and physically coherent picture of the interplay between stellar pulsation and circumstellar dust formation. 

Future progress will benefit from extending this phase-resolved fitting procedure to larger and more diverse samples across different environments. Ultimately, the integration of fully consistent 3D radiation-hydrodynamics models coupled with non-equilibrium dust formation will be essential to overcome current hydrostatic limitations and to accurately characterize the complex physics of mass loss in pulsating carbon stars.

\begin{acknowledgements}
This project has received funding from the European Union’s Horizon 2020 research and innovation programme under the Marie Skłodowska-Curie grant agreement No 101072454, as part of the Milky Way-$Gaia$ Doctoral Network (\url{https://www.mwgaiadn.eu/}).
The contribution by MT and GP is funded by the European Union – NextGenerationEU and by the University of Padua under the 2023 STARS Grants@Unipd programme (project ``CONVERGENCE'').
The OGLE project has received funding from the Polish National Science
Centre grant OPUS-28 2024/55/B/ST9/00447 awarded to AU.

This project has made use of data products from the Two Micron All Sky Survey, which is a joint project of the University of Massachusetts and the Infrared Processing and Analysis Center/California Institute of Technology, funded by the National Aeronautics and Space Administration and the National Science Foundation;
data products from the Wide-field Infrared Survey Explorer, which is a joint project of the University of California, Los Angeles, and the Jet Propulsion Laboratory/California Institute of Technology, and NEOWISE, which is a project of the Jet Propulsion Laboratory/California Institute of Technology. WISE and NEOWISE are funded by the National Aeronautics and Space Administration;
data from the European Space Agency (ESA) mission {\it Gaia} (\href{https://www.cosmos.esa.int/gaia}{https://www.cosmos.esa.int/gaia}), processed by the {\it Gaia} Data Processing and Analysis Consortium (DPAC, \href{https://www.cosmos.esa.int/web/gaia/dpac/consortium}{https://www.cosmos.esa.int/web/gaia/dpac/consortium}). Funding for the DPAC has been provided by national institutions, in particular the institutions participating in the {\it Gaia} Multilateral Agreement;
This publication makes use of the following free/open source software and/or libraries: the Starlink Tables Infrastructure Library \citep[STILTS and Topcat][]{stilts}; IPython \citep{ipython} and Jupyter \citep{jupyter} notebooks; the \textsc{Python} libraries \textsc{NumPy} \citep{numpy}, \textsc{SciPy} \citep{scipy}, \textsc{matplotlib} \citep[a \textsc{Python} library for publication quality graphics][]{matplotlib}, and \textsc{Astropy} \citep[a community-developed core \textsc{Python} package for Astronomy][]{astropy2013,astropy2018,astropy2022}.
This research has made use of NASA's Astrophysics Data System Bibliographic Services, and of the following services provided by CDS, Strasbourg: the SIMBAD data base, the VizieR catalogue access tool \citep[DOI: 10.26093/cds/vizier,][]{vizier}, the ``Aladin sky atlas'' \citep{aladin}, and the cross-match service \citep{cds2012,cds2020}.
\end{acknowledgements}

\bibliographystyle{aa} 
\bibliography{references,acknowledgements/acknowledgements}

\appendix

\section{Simulating the CMD Pulsational Loops}
\label{appendix:cmd_simulation}

\begin{figure}[h]
    \centering
    \includegraphics[width=0.9\linewidth]{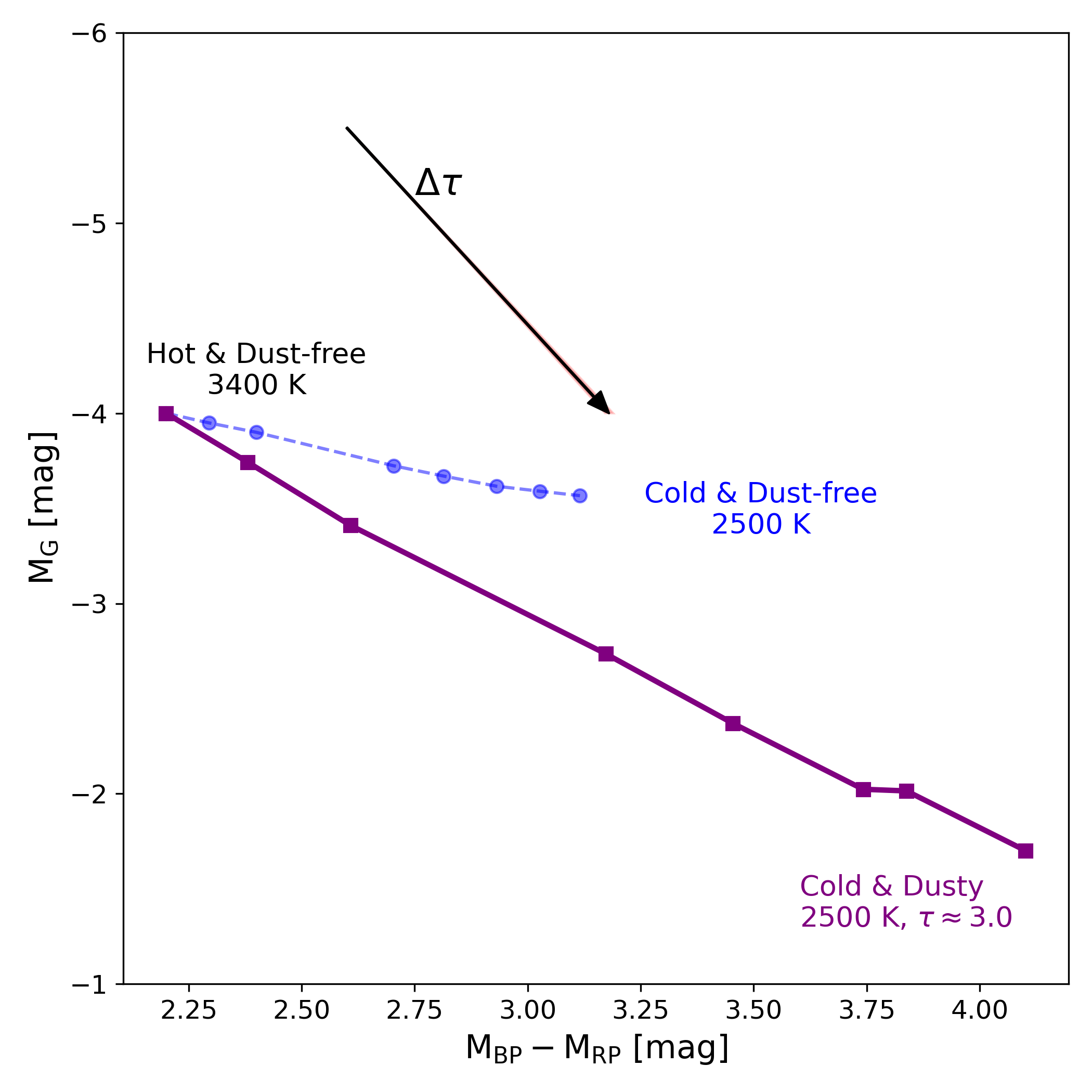}
    \caption{Simulation of a pulsation cycle in the CMD for a model with $M = 3.0\,M_\odot$ and C/O $\sim 1.40$. The blue dashed line traces a bare photosphere cooling from 3400~K to 2500~K with no dust. The purple solid line simulates a dynamically coupled envelope where dust condenses (up to $\tau_{0.55} \approx 3.0$) as the star cools, aligning perfectly with the reddening vector (black arrow).}
    \label{fig:appendix_cmd_sim}
\end{figure}

To explicitly demonstrate the physical drivers behind the elongated loops observed in the \textit{Gaia} CMD (Sect.~\ref{sec:Results}), we simulated the color-magnitude excursion of a pulsating star using our ATHENA-C model grid. We selected a canonical configuration ($M = 3.0\,M_\odot$, C/O $\sim 1.40$) and modeled two distinct evolutionary paths, shown in Figure~\ref{fig:appendix_cmd_sim}.

The blue dashed line represents a static, "dust-free" track. Here, the star cools from 3400~K down to 2500~K, but the optical depth remains fixed at $\tau_{0.55} \approx 0$. As shown, the slope of this pure $T_{\rm eff}$ variation significantly differs from the slope of the circumstellar reddening vector. 

The purple solid line simulates a dusty pulsation cycle track (dynamic track). In this scenario, as the effective temperature decreases, the optical depth is simultaneously increased to simulate dust condensation, reaching $\tau_{0.55} \approx 3.0$ at 2500~K. This track deviates sharply from the dust-free sequence and naturally aligns with the reddening vector. 

While the exact elliptical shapes observed in Figure~\ref{fig:cmd_evolution} are generated by the phase lag between the photospheric thermal variations and the circumstellar dust response, these simulated lines effectively represent the major axes of those ellipses. This test conclusively proves that the alignment of a star's pulsational loop with the reddening vector is a direct indicator that its photometric variations are primarily driven by the periodic formation and destruction of circumstellar dust rather than simple temperature changes.

\section{Methodological comparison of parameter selection criteria}
\label{asec:comparison-models-priors}

As outlined in Sect.~\ref{subsect:phase-resolved_sedfit} and discussed in Sect.~\ref{sec:Discussion}, purely photometric SED fitting exhibits weak sensitivity to stellar mass. To address this, we explored two distinct criteria for selecting the constant stellar mass ($M$) and surface C/O ratio across the pulsation cycle:
\begin{itemize}
    \item Case 1: Minimum overall mean $\chi^2$. This purely statistical approach evaluates every available $M$ and C/O combination in the COMARCS grid, computing the mean $\chi^2$ averaged over all 14 phases. The combination yielding the lowest overall mean $\chi^2$ is adopted.
    \item Case 2: Pulsation period constraint. This physically motivated approach selects $M$ and C/O based on an independent physical observable: the pulsation period. For each $M$ and C/O combination in the grid, we perform the phase-resolved fit, derive the best-fit $T_\mathrm{eff}$ at each phase, and compute the corresponding stellar radius by inverting the Stefan-Boltzmann law using our empirically derived multi-band bolometric luminosity. We then calculate the analytical pulsation period from the \citetalias{trabucchi2021longperiods} period–mass–radius relation, evaluated at the median radius and the fixed test mass. The $M$ and C/O pair predicting a period closest to our observed \texttt{LombScargle} period ($P_{\rm LC}$) is selected. This self-consistent approach links SED-derived atmospheric parameters directly to the star's pulsational properties.
\end{itemize}
For each star, figures representing the evolution of $T_{\rm eff}$, $\tau_{0.55}$ and bolometric luminosity with the pulsation phase by assuming results from both Case~1 and Case~2 are presented in Appendix~\ref{appendix:diagnostic_plots}.

Comparing the results from these two methods reveals that parameters strictly governing the atmospheric and circumstellar structure, the $T_{\rm eff}$ and the optical depth ($\tau_{0.55}$), exhibit minimal discrepancies. 
The overall shape of the observed multi-band SED strongly constrains these variables. 
Furthermore, because the bolometric luminosity is derived empirically from the observations (model-free) and $T_{\rm eff}$ remains similar, the resulting stellar radii are highly compatible in both scenarios.

The fundamental divergence between the two methods lies in the derived current stellar mass and, consequently, the analytically predicted period. Results based on pure $\chi^2$ minimization (Case~1) lead to a sample with a systematically higher mean current mass of approximately $2.3\,M_\odot$, with 7 out of 13 sources assuming the value of $3.0\,M_\odot$. 

Conversely, employing the period constraint (Case~2) effectively fine-tunes the SED fit. Because the radius is already tightly constrained by the SED, the method is forced to adjust the mass parameter to match $P_{\rm LC}$. This systematically selects lower mass values, shifting the sample mean to approximately $1.1\,M_\odot$. As a direct result of this mass fine-tuning, the periods predicted by Case~2 naturally align with the observed actual periods. The systematic discrepancies in both the predicted periods and the derived masses between the two methodologies are summarized in Table~\ref{tab:apx:results_table}.

From a physical perspective, the lower current masses ($\sim 1\,M_\odot$) derived via the period-constrained method (Case~2) represent a more realistic scale for the majority of our sample. As typical for these objects, their relatively short pulsation periods suggest correspondingly low luminosities and lower envelope masses. Conversely, significant mass-loss would inflate the stellar envelope, consequently lengthening the pulsation period. This physical scenario is perfectly exemplified by our most extreme source (Gaia DR3 4688894107214534144): it exhibits the longest period ($P=593$ days), the largest optical amplitude, and the highest optical depth, consistent with an inflated envelope undergoing severe mass-loss.

However, the magnitude of these discrepancies and the quality of the resulting parameter curves are intrinsically tied to the discrete nature of the ATHENA-C and COMARCS grids. As discussed in Sect.~\ref{sec:Discussion}, forcing the mass constraint in a sparsely sampled grid often limits the availability of optimal $T_{\rm eff}$ models, leading to less smooth parameter variations. A fully homogeneous grid with finer, continuous mass sampling would likely resolve this tension entirely.
\begin{table*}[!ht]
\small
    \centering
    \renewcommand{\arraystretch}{1.3} 
    \setlength{\tabcolsep}{4pt}
    \caption{Mass, radius and theoretical period derived by utilizing the theoretical relation from \cite{trabucchi2021longperiods} with the results of the SED fitting derived in Case~1 and Case~2. Last column represents the real observed period from our light curve fit.}
    \label{tab:apx:results_table}
    \begin{tabular}{l c c c c c c c c c c c c c c}
            \hline\hline
$Gaia$ DR3 id & $\rm Mass_{\chi^2}$ & $\rm Radius_{\chi^2}$ & $\rm Period_{Trabucchi, \chi^2}$ & $\rm Mass_{T21}$ & $\rm Radius_{T21}$ & $\rm Period_{Trabucchi, T21}$ & $\rm Period_{LC}$ \\
 & ($M_\odot$) & ($R_\odot$) & (days) & ($M_\odot$) & ($R_\odot$) & (days) & (days) \\
\hline
4685915083569253120 & 3.0 & 212 & 126 & 1.0 & 203 & 220 & 202 \\
4687404887428798720 & 3.0 & 198 & 111 & 1.0 & 198 & 213 & 293 \\
4685488919677678080 & 3.0 & 244 & 160 & 1.0 & 246 & 285 & 268 \\
4688954717829223936 & 1.0 & 255 & 298 & 1.0 & 267 & 317 & 389 \\
4685883129008411264 & 1.0 & 207 & 227 & 1.0 & 211 & 231 & 293 \\
4685776167110162944 & 2.0 & 219 & 170 & 1.0 & 229 & 259 & 245 \\
4688871601593755008 & 1.0 & 192 & 205 & 1.0 & 192 & 205 & 198 \\
4688894107214534144 & 3.0 & 295 & 223 & 1.0 & 462 & 388 & 593 \\
4685776167120724992 & 3.0 & 184 & 98  & 1.0 & 196 & 211 & 232 \\
4685949546407892608 & 2.0 & 276 & 254 & 1.0 & 283 & 343 & 423 \\
4685649620189877248 & 3.0 & 219 & 133 & 1.0 & 213 & 235 & 249 \\
4685898006715794688 & 3.0 & 237 & 152 & 2.0 & 221 & 173 & 276 \\
4689068208025791232 & 2.0 & 196 & 140 & 1.0 & 206 & 224 & 207 \\
\hline
    \end{tabular}
    \vspace{1mm}
\end{table*}

\section{Diagnostic figures}
\label{appendix:diagnostic_plots}
\begin{figure*}
    \centering
    \includegraphics[width=0.3\linewidth]{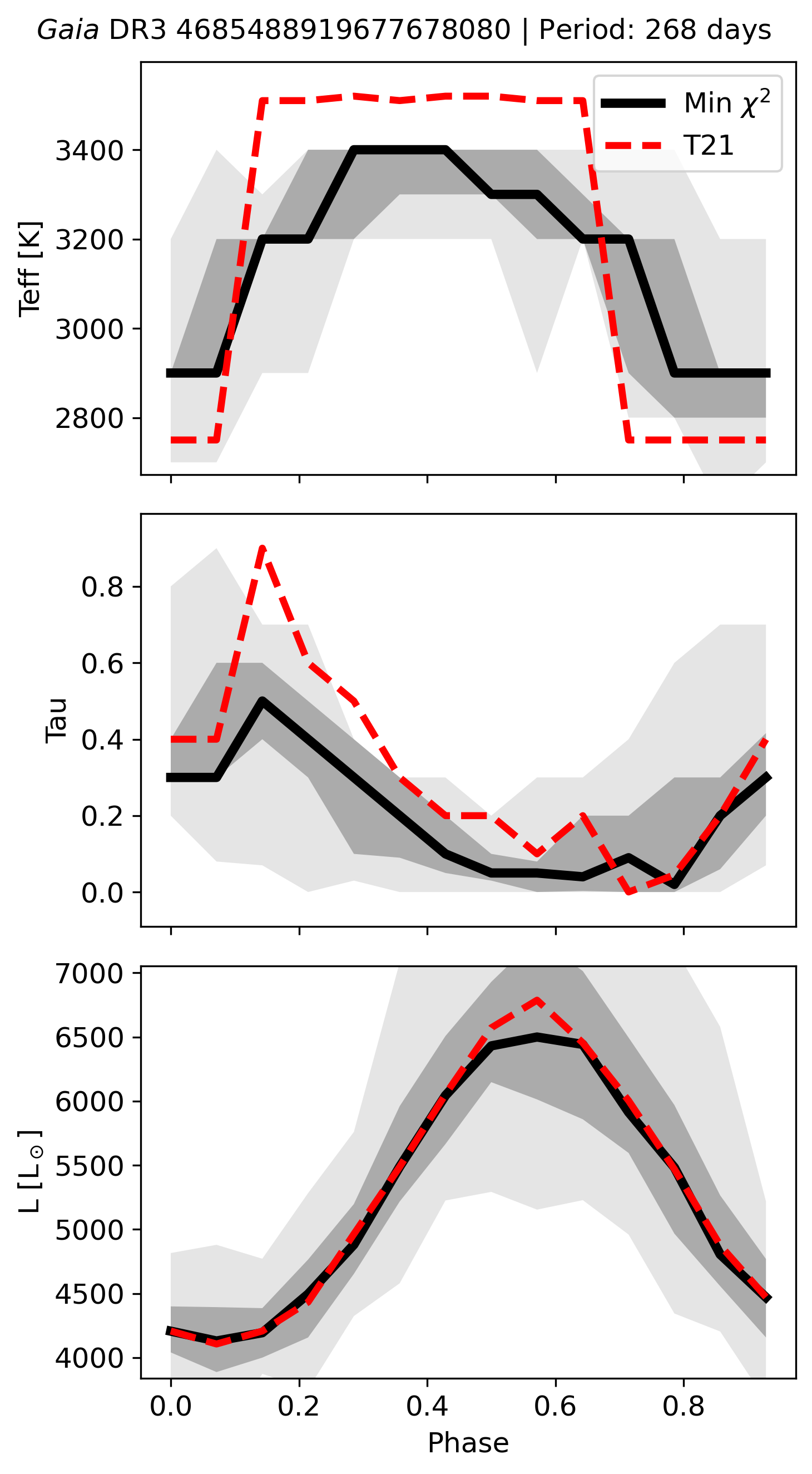}
    \includegraphics[width=0.3\linewidth]{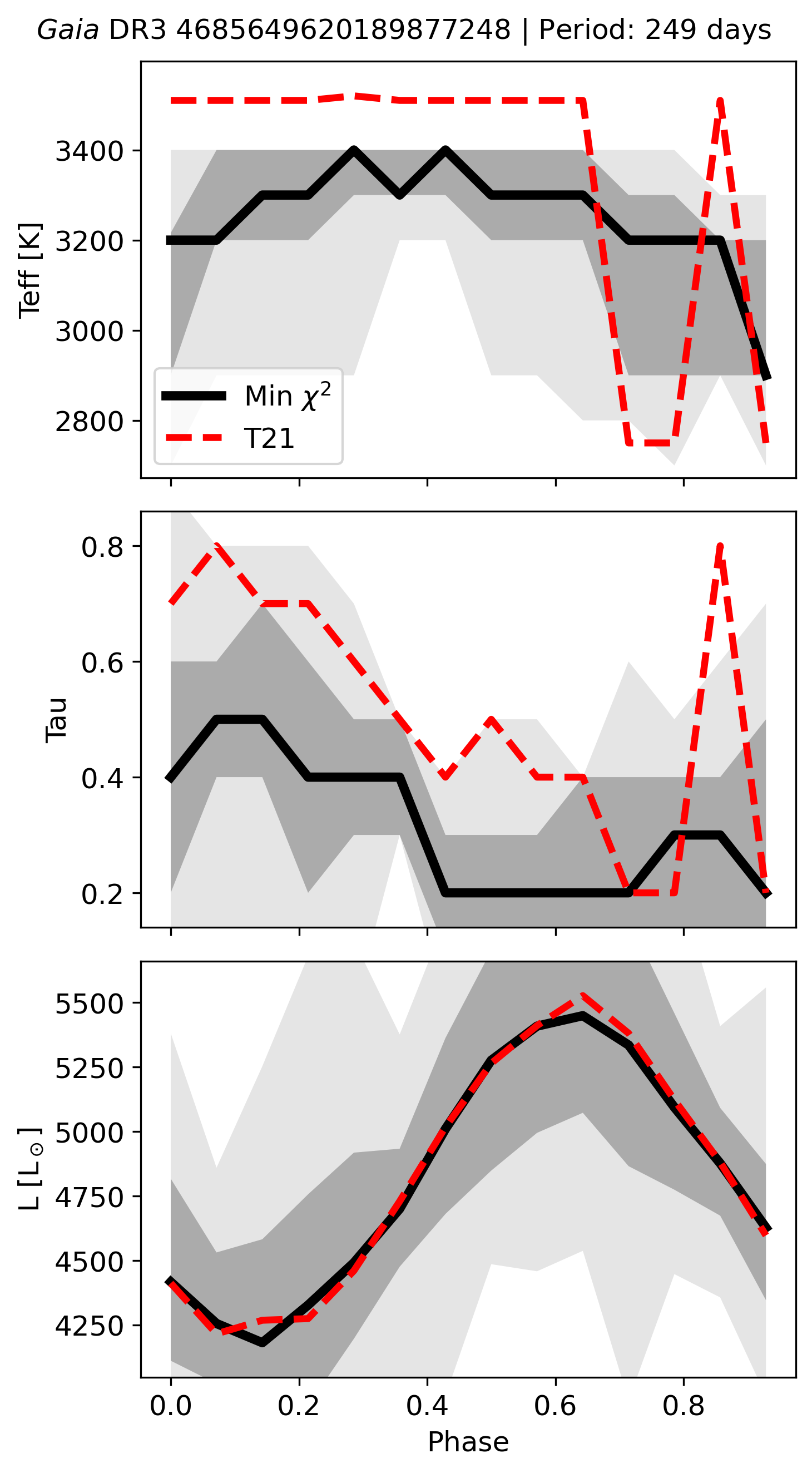}
    \includegraphics[width=0.3\linewidth]{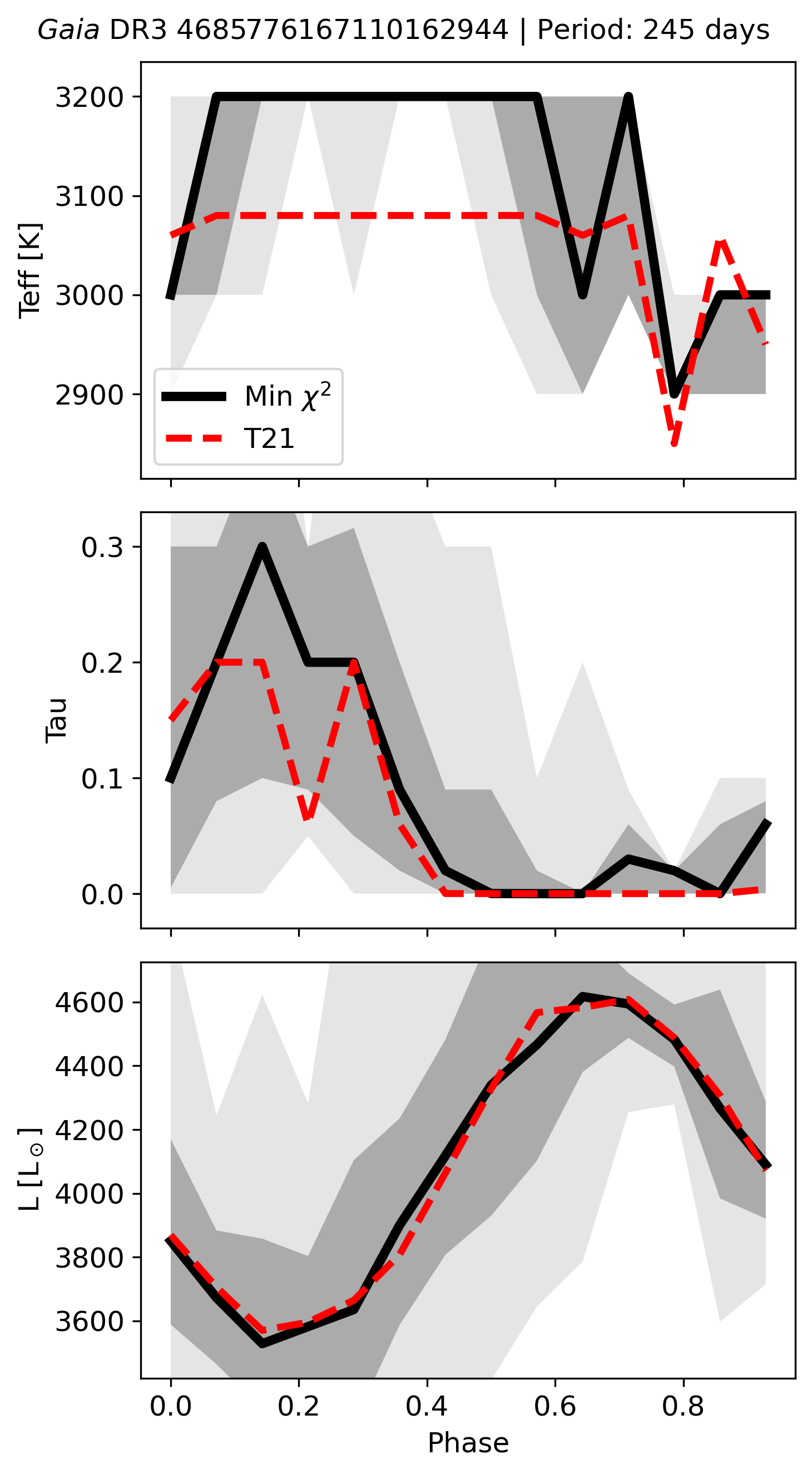}
    \caption{Diagnostic plots for additional stars showing the evolution of three parameters with pulsation phase. Grey lines and shaded areas are the same as Fig.~\ref{fig:diagnostic}, while red dashed lines represents results derived with fitting priors by Case~2 (see Appendix~\ref{asec:comparison-models-priors}).}
    \label{fig:appendix_diagnostics_1}

\end{figure*}

\begin{figure*}
    \centering
    \includegraphics[width=0.3\linewidth]{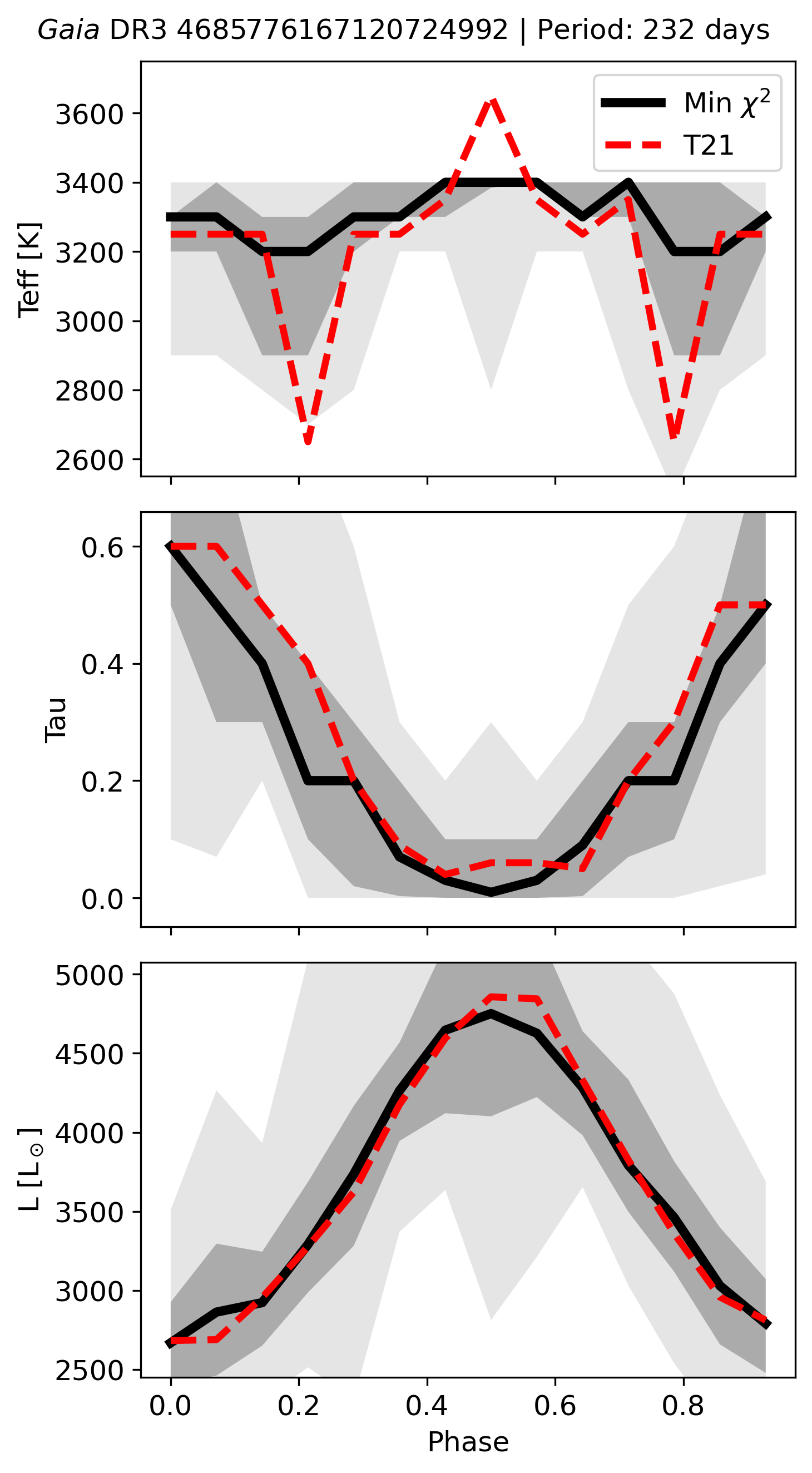}
    \includegraphics[width=0.3\linewidth]{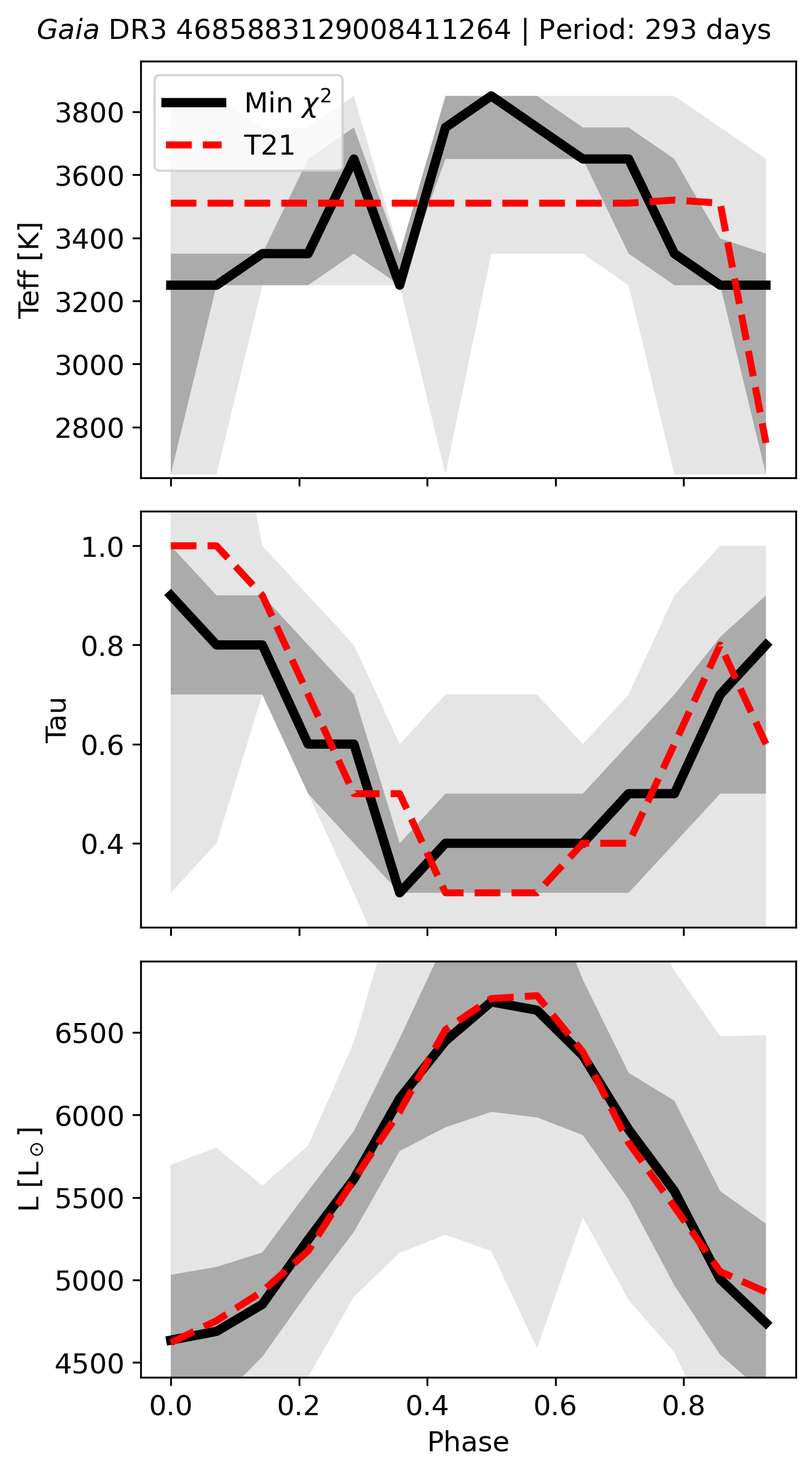}
    \includegraphics[width=0.3\linewidth]{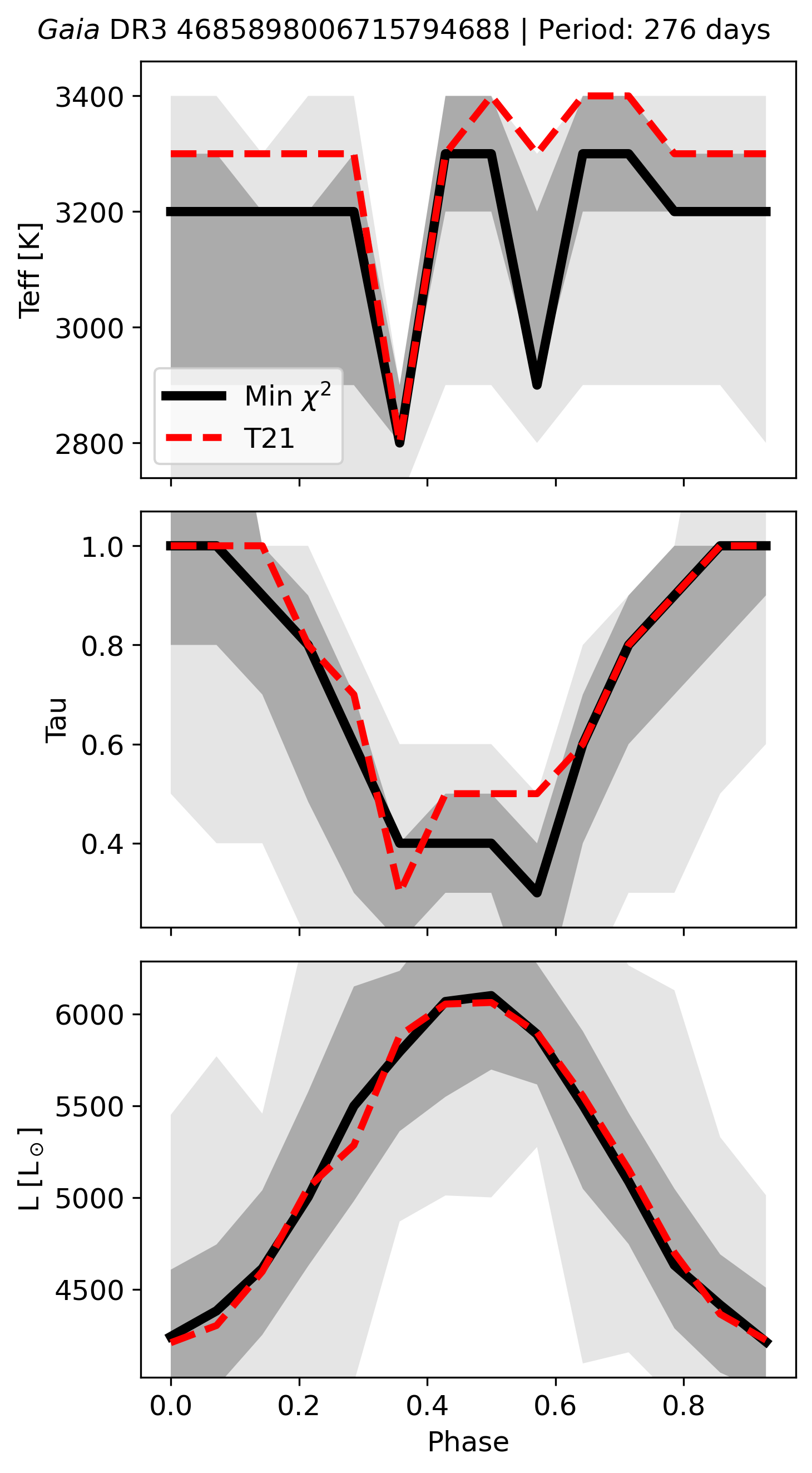}
    \includegraphics[width=0.3\linewidth]{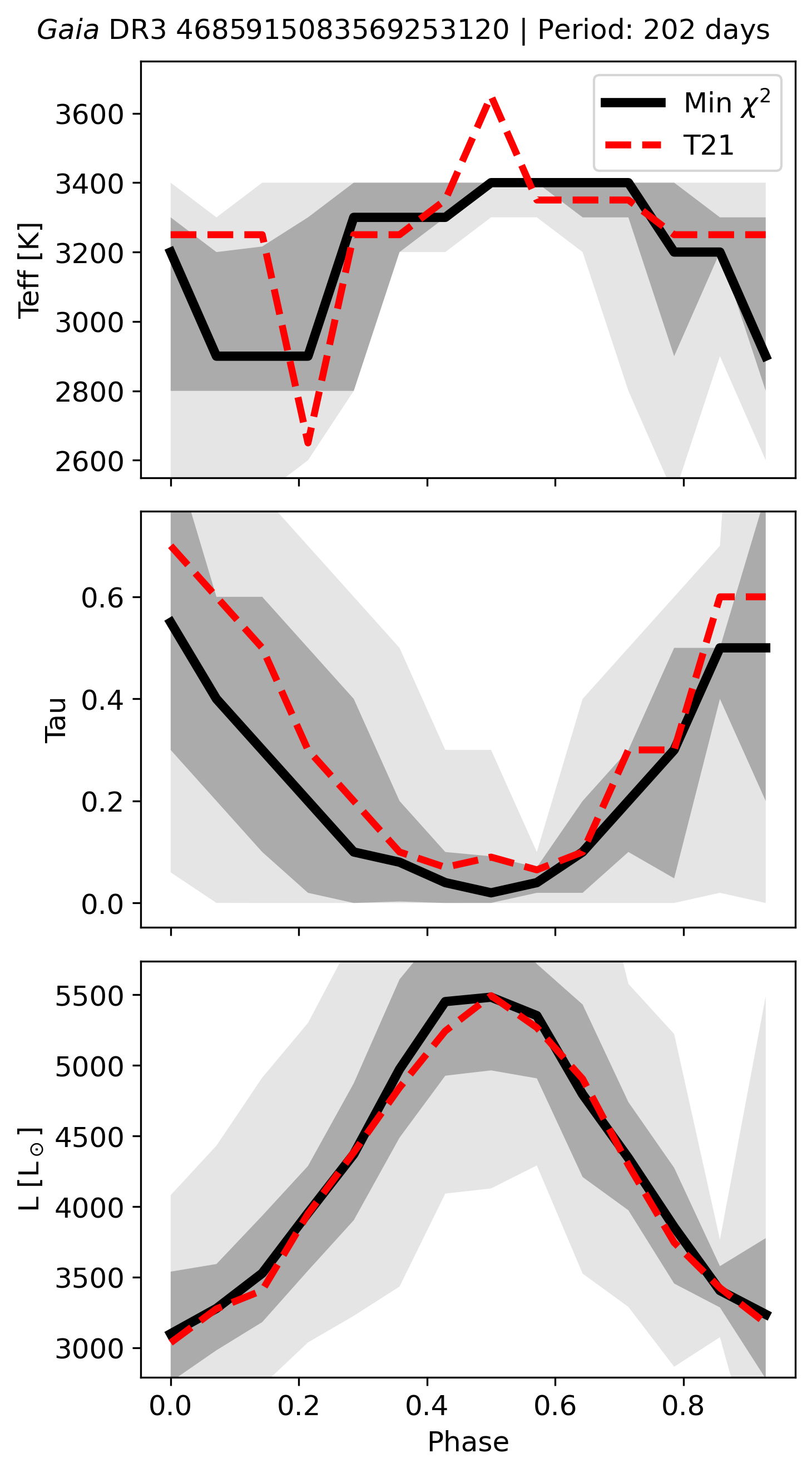}
    \includegraphics[width=0.3\linewidth]{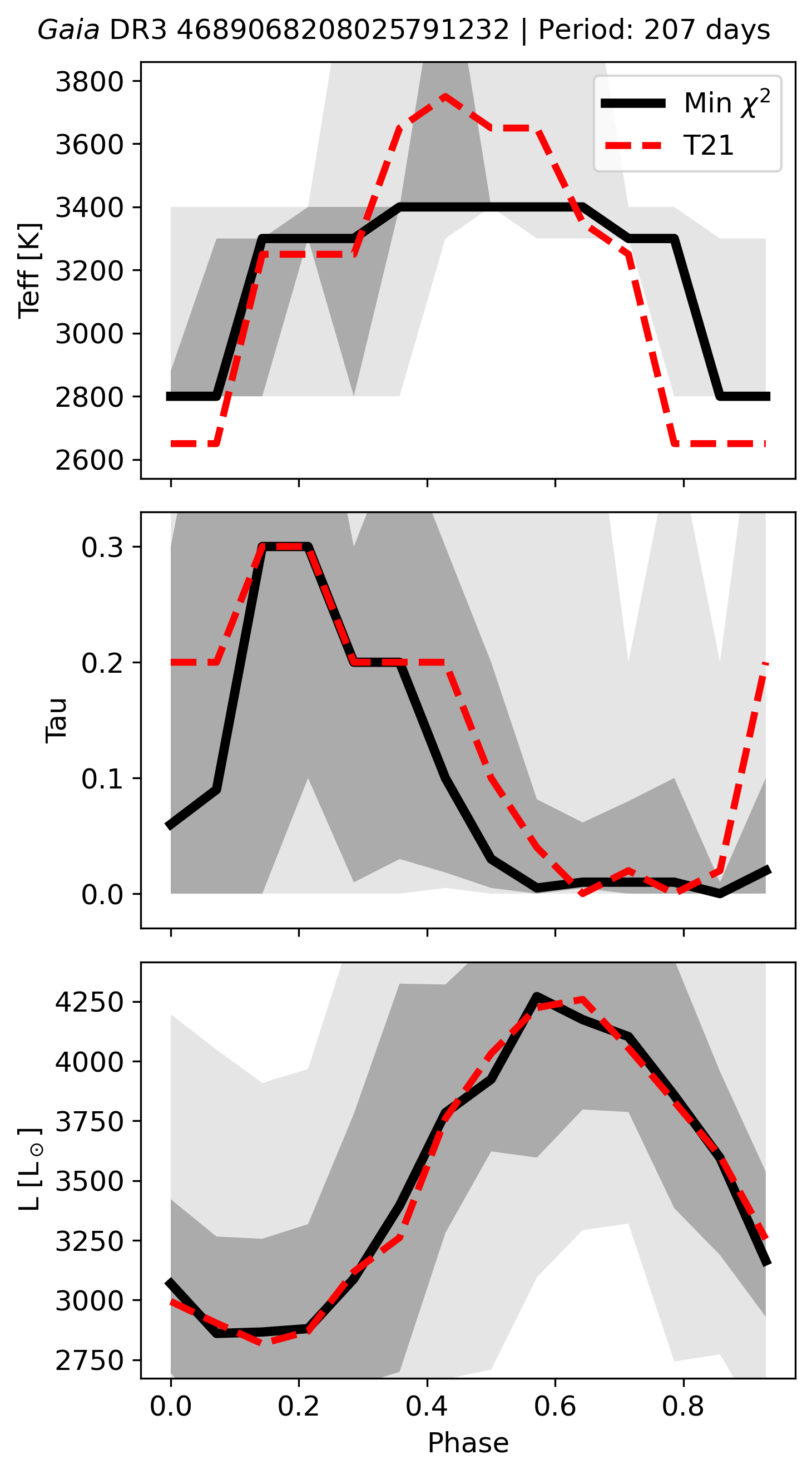}
    \includegraphics[width=0.3\linewidth]{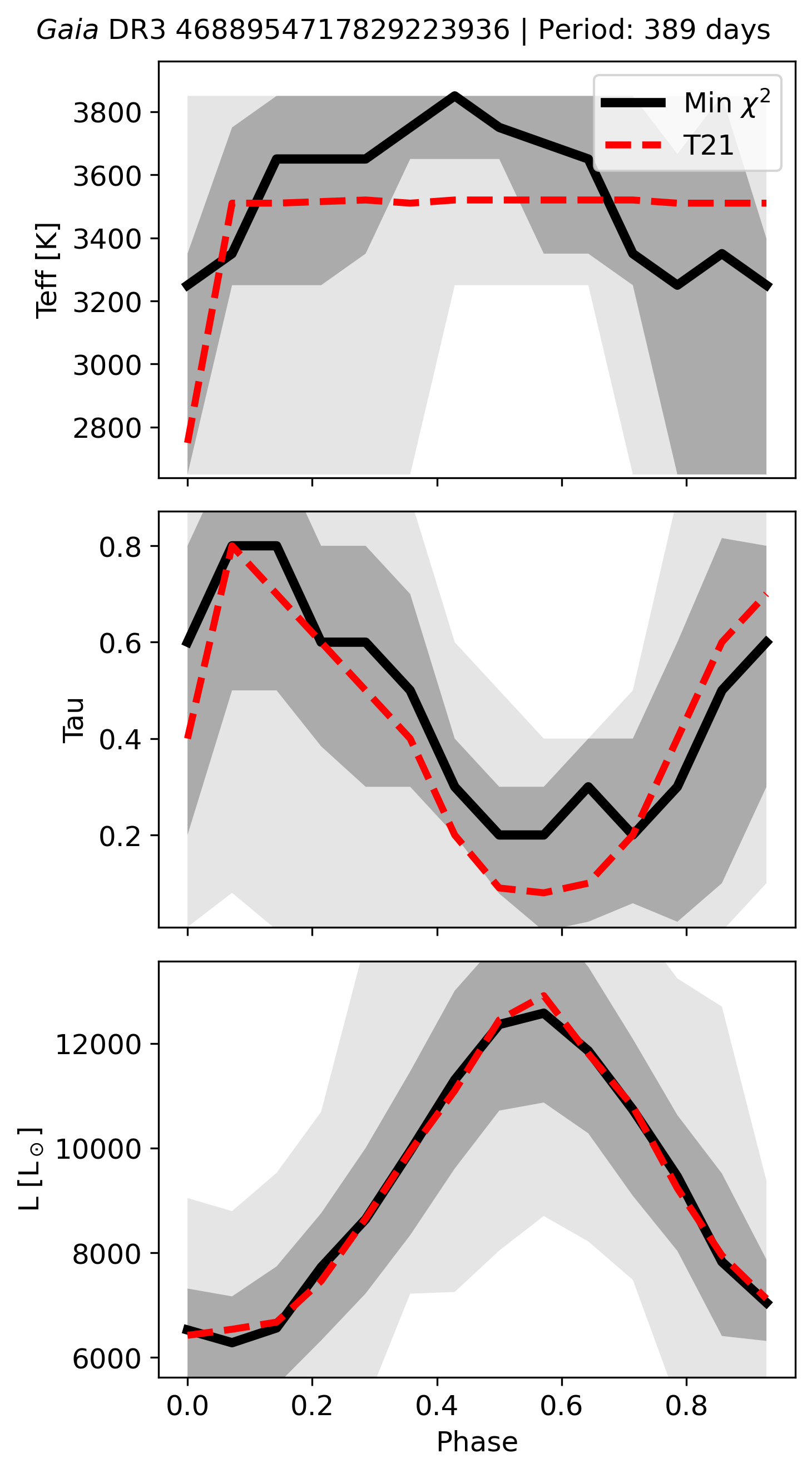}
    \caption{Diagnostic plots for additional stars showing the evolution of three parameters with pulsation phase. Lines and shaded areas are the same as Fig.~\ref{fig:appendix_diagnostics_1}.}
    \label{fig:appendix_diagnostics_2}
\end{figure*}

\begin{figure*}
    \centering
    \includegraphics[width=0.3\linewidth]{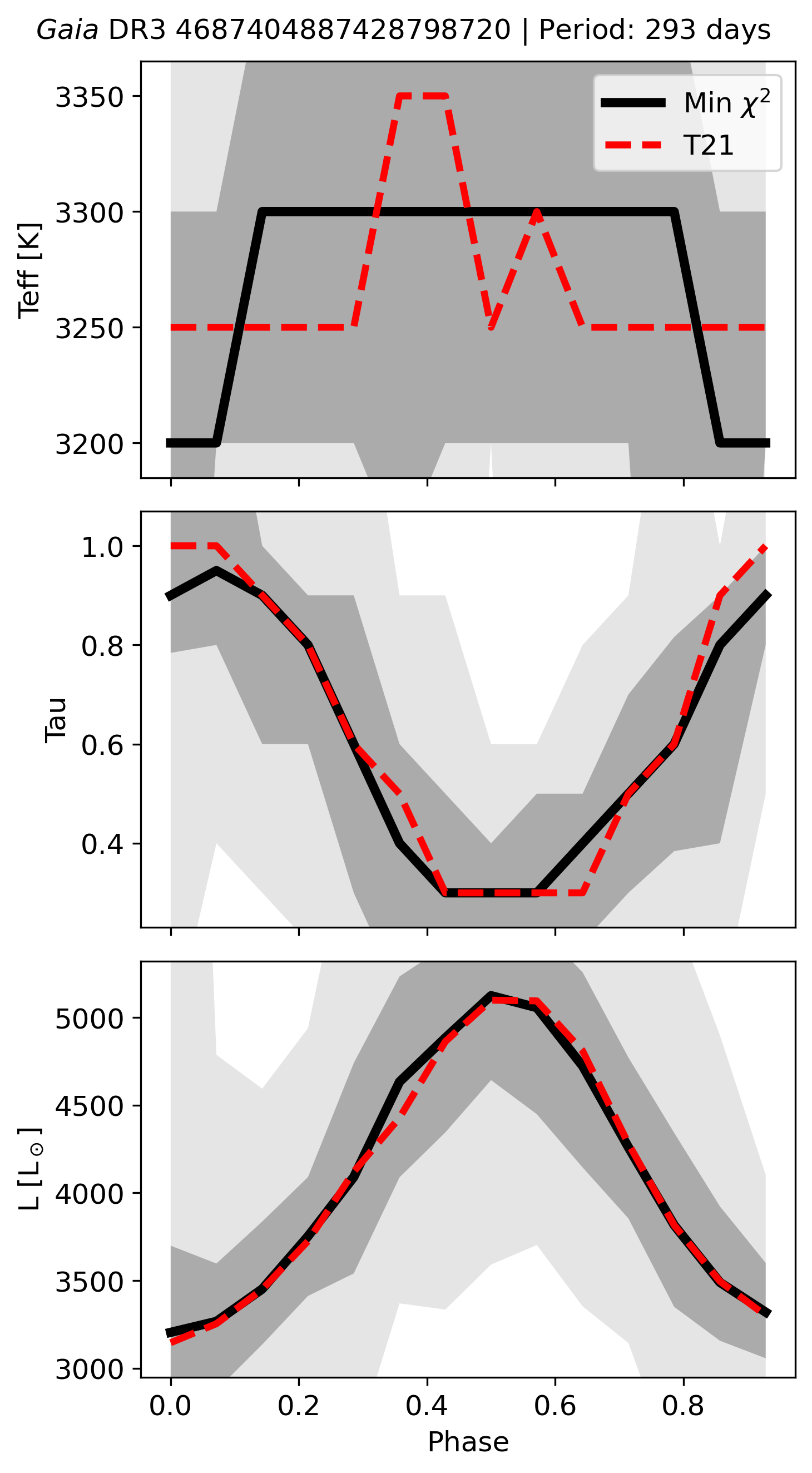}
    \includegraphics[width=0.3\linewidth]{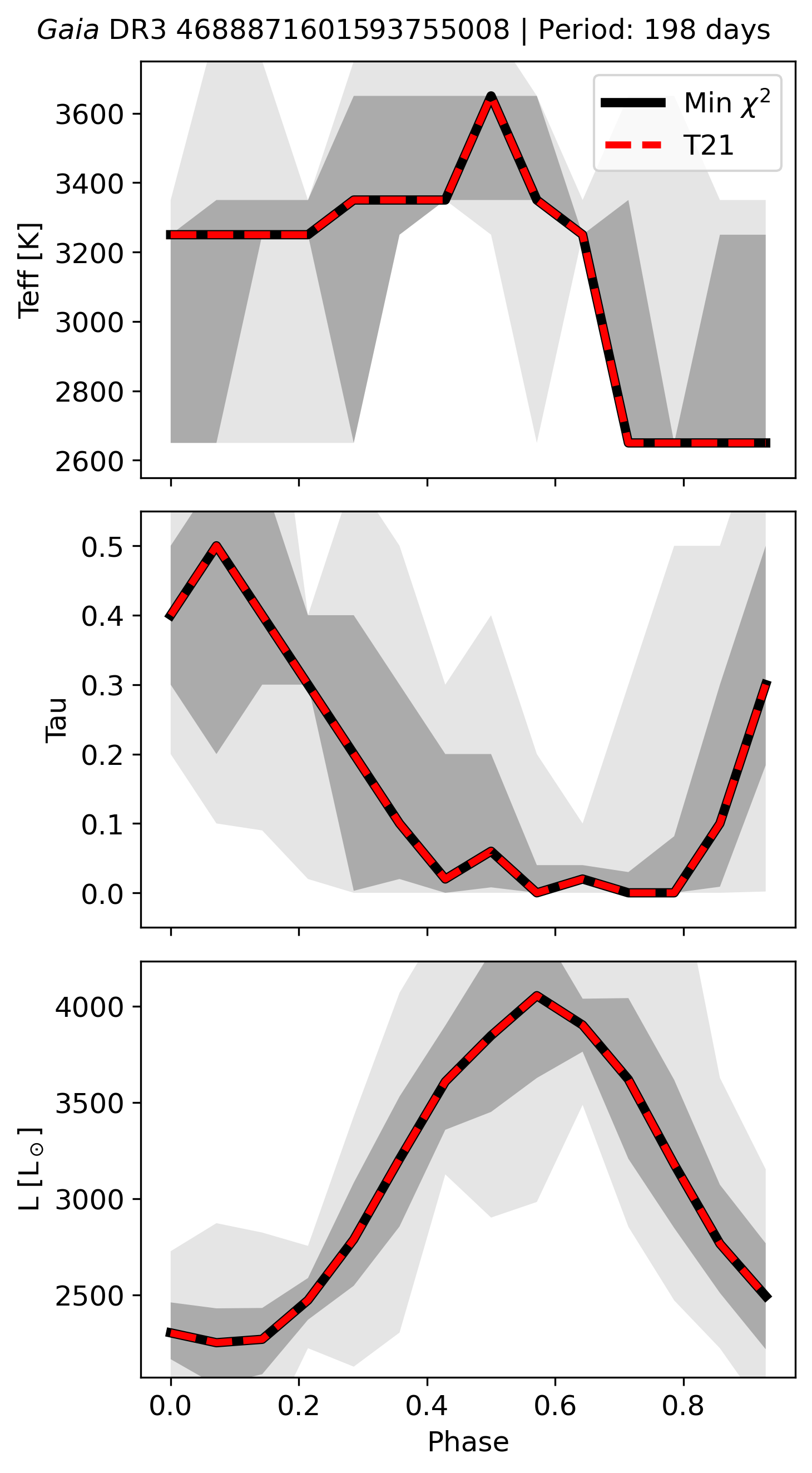}
    \includegraphics[width=0.3\linewidth]{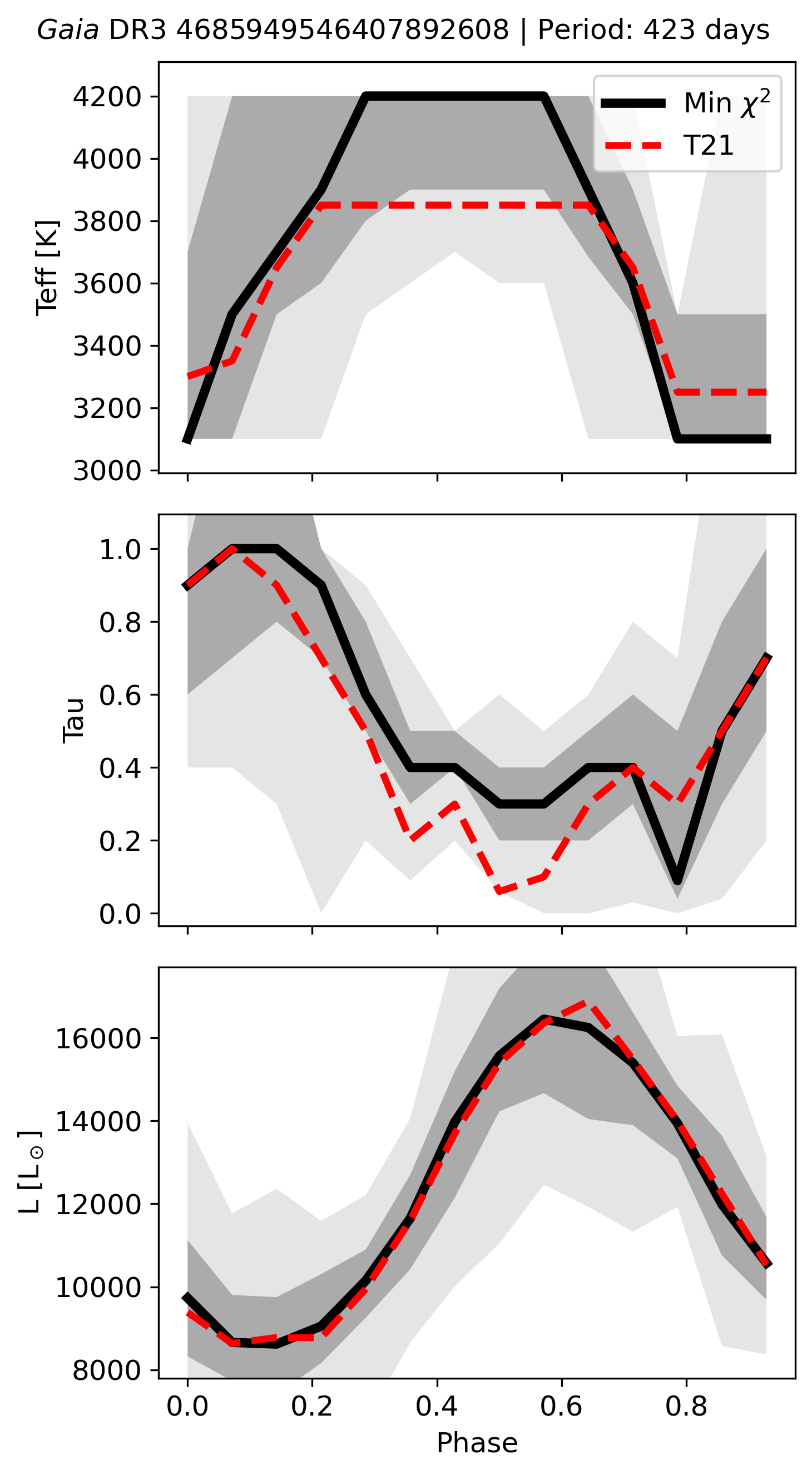}
    \includegraphics[width=0.3\linewidth]{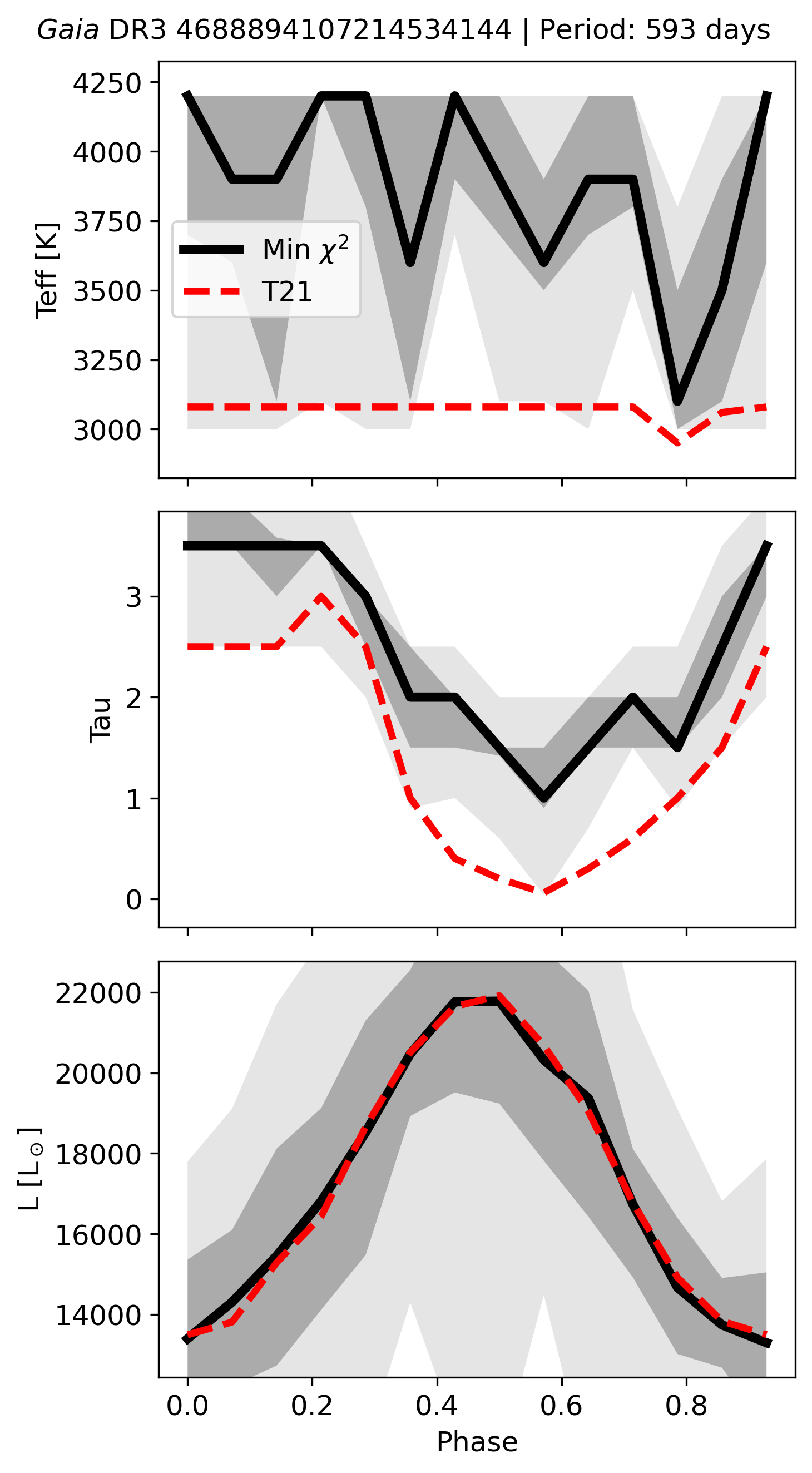}
    \caption{Diagnostic plots for additional stars showing the evolution of three parameters with pulsation phase. Lines and shaded areas are the same as Fig.~\ref{fig:appendix_diagnostics_1}.}
    \label{fig:appendix_diagnostics_3}
\end{figure*}

By following, the diagnostic Figures \ref{fig:appendix_diagnostics_1}, \ref{fig:appendix_diagnostics_2} and \ref{fig:appendix_diagnostics_3} present the evolution of  $T_{\rm eff}$, the optical depth, and the luminosity in the entire pulsation cycle for each star in our sample.
In these figures, both cases of results are presented: Case~1~results, presented and discussed in this manuscript, and Case~2~results, presented and discussed in Appendix \ref{asec:comparison-models-priors}.

\begin{figure*}
    \centering
    \includegraphics[width=0.33\linewidth]{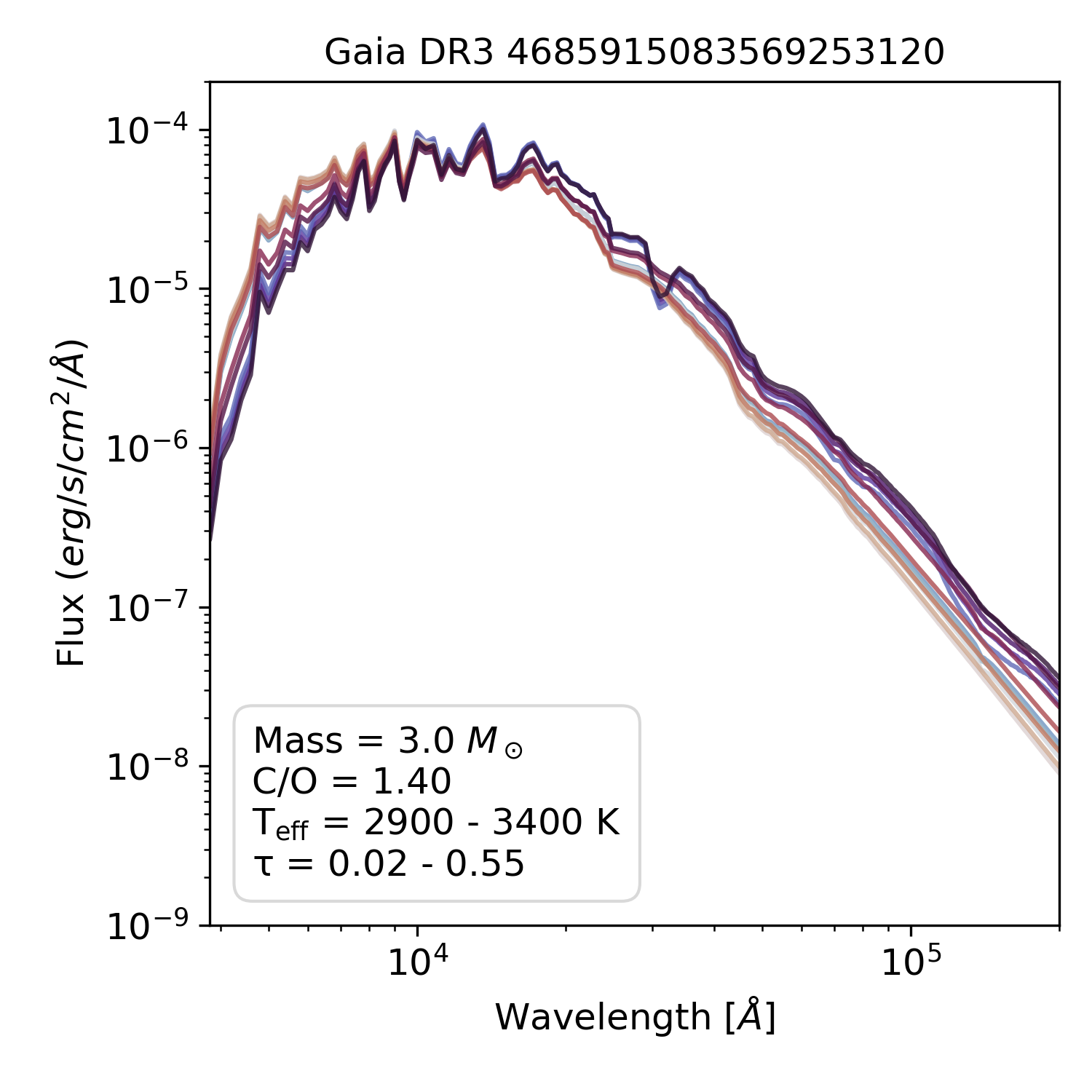}
    \includegraphics[width=0.33\linewidth]{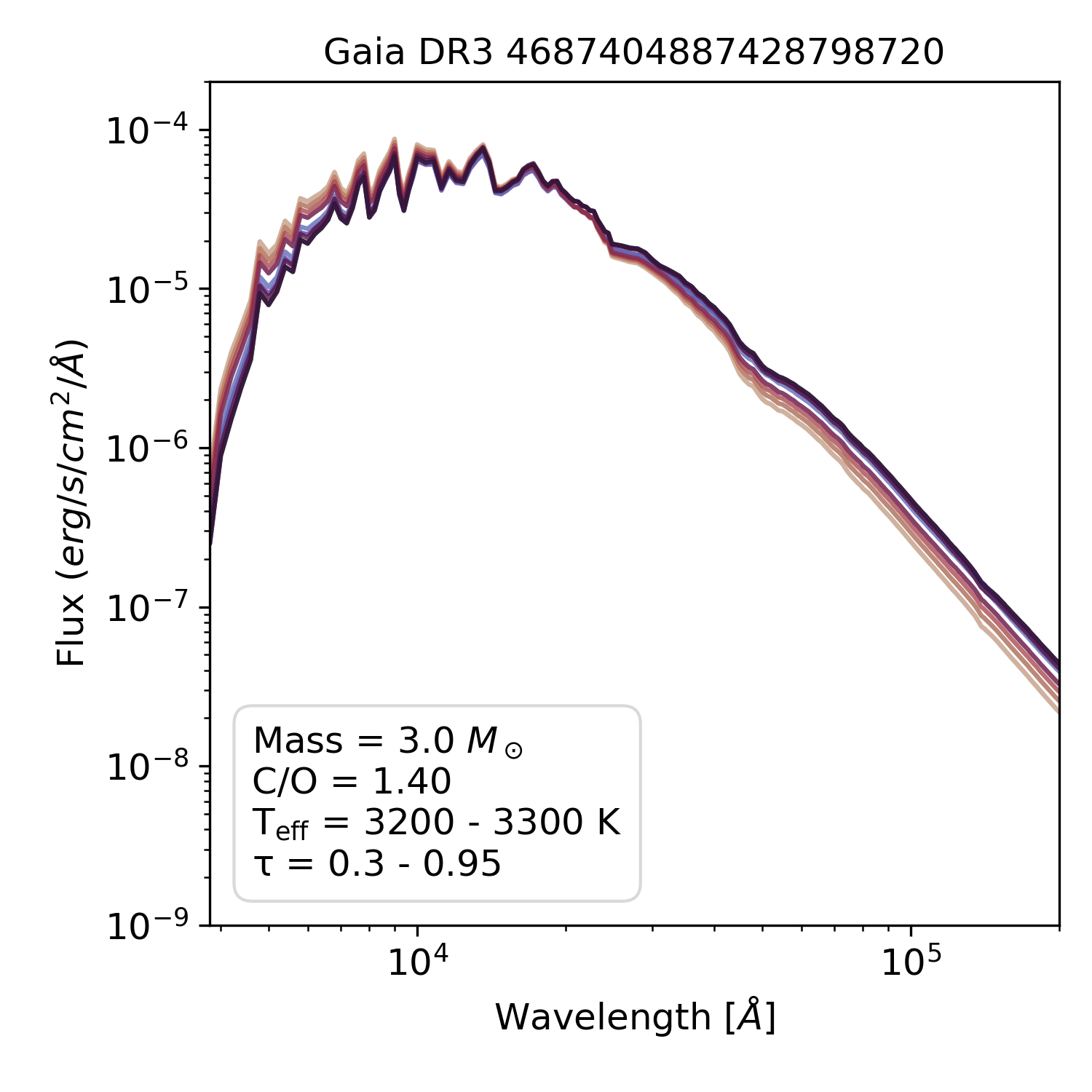}
    \includegraphics[width=0.33\linewidth]{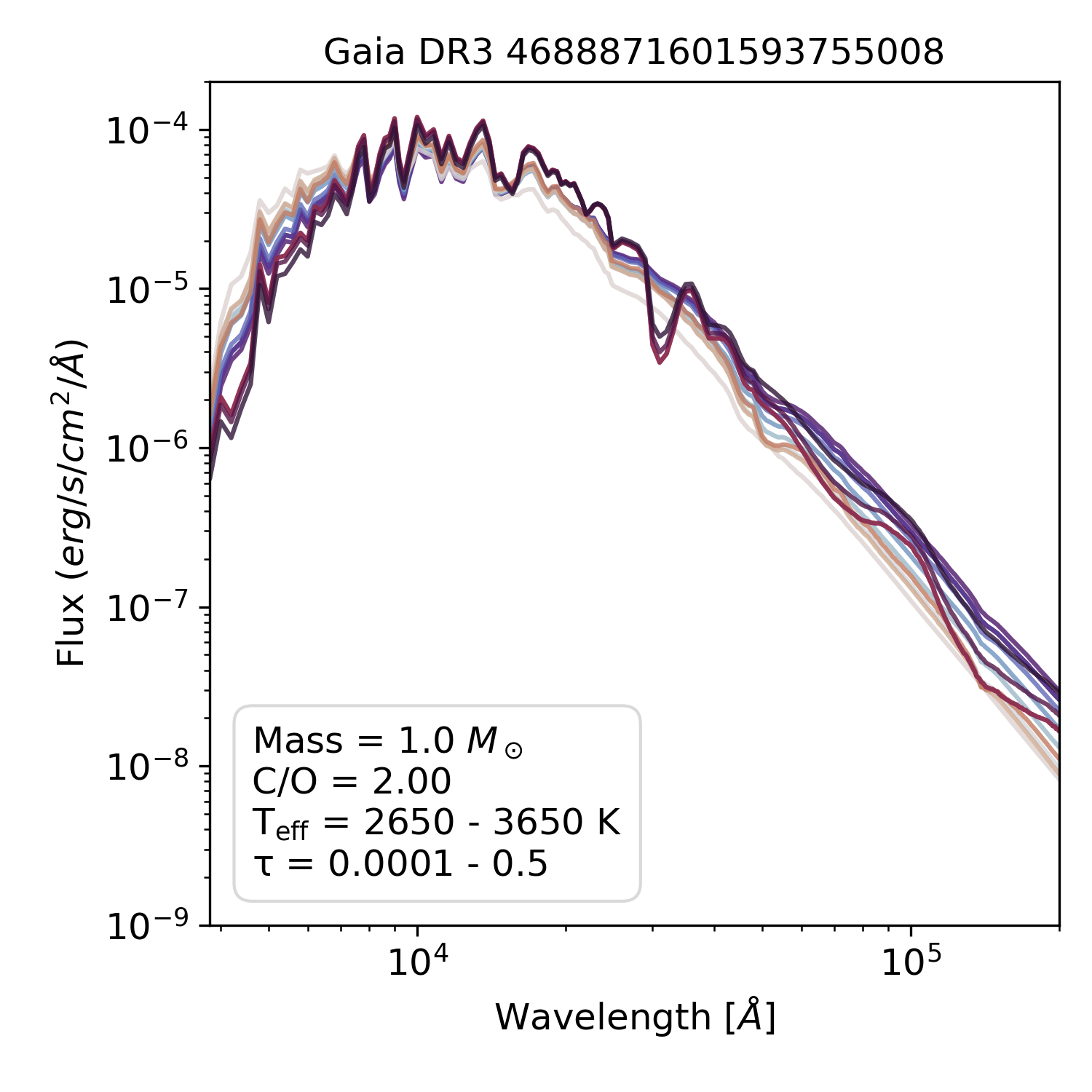}
    \includegraphics[width=0.33\linewidth]{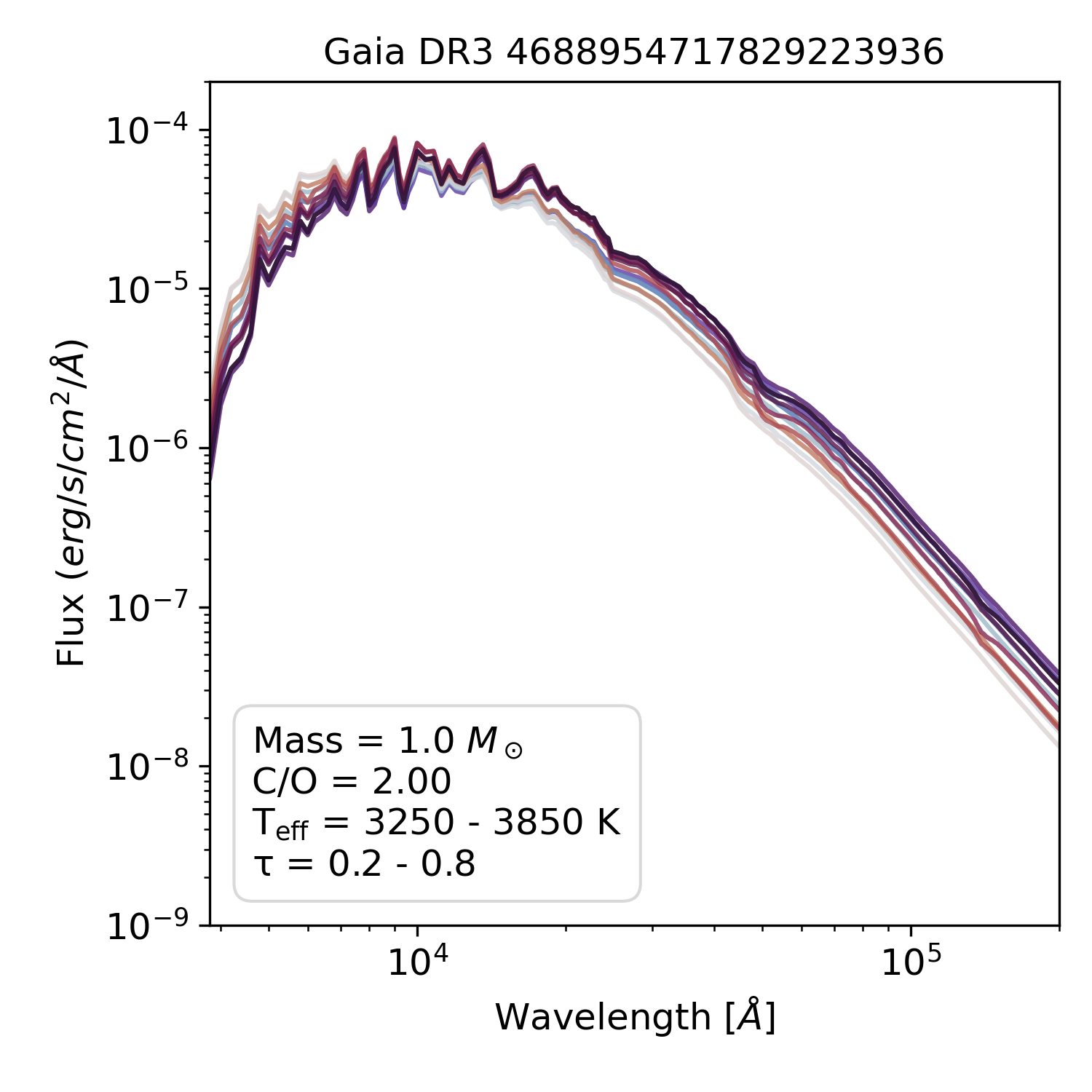}
    \includegraphics[width=0.33\linewidth]{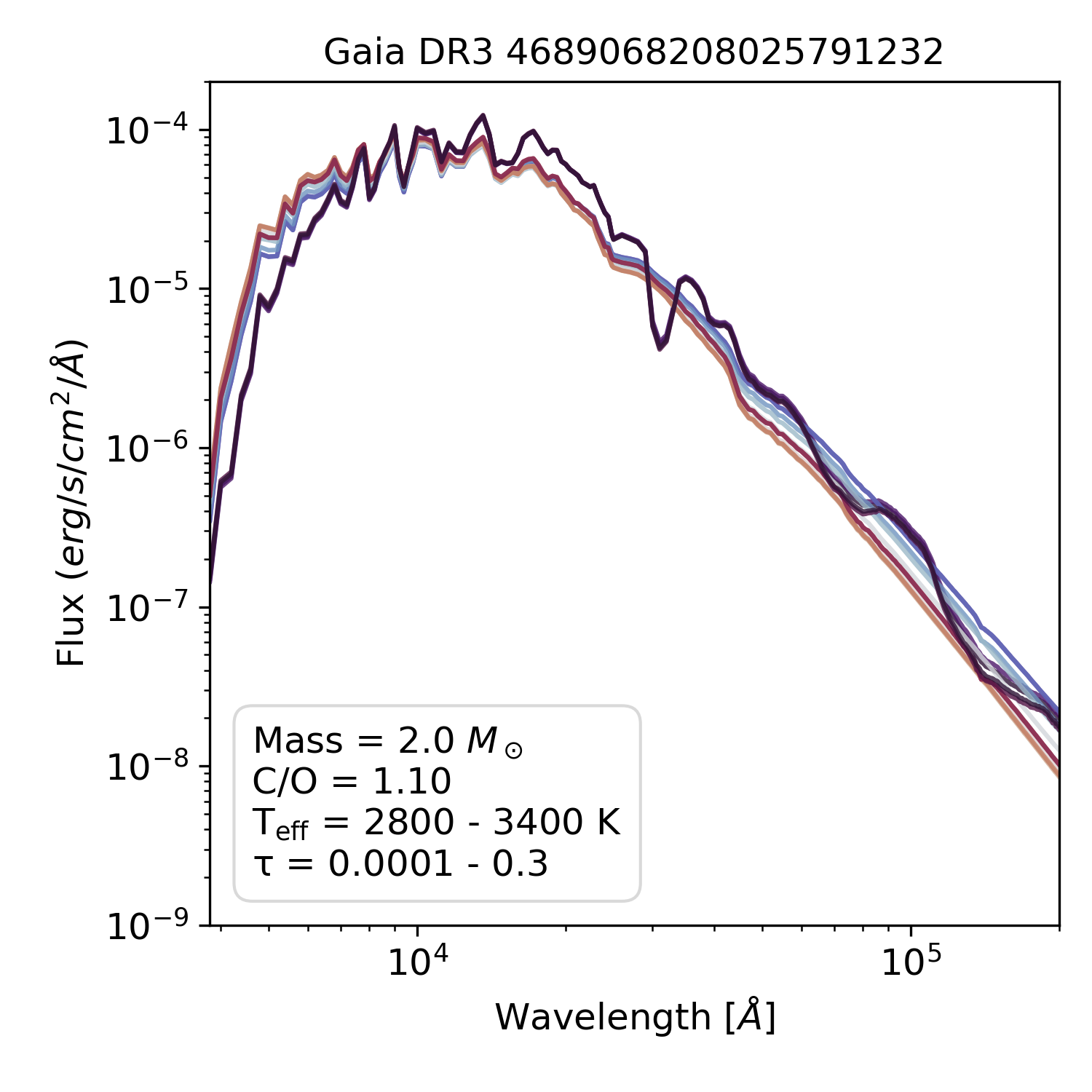}
    \includegraphics[width=0.33\linewidth]{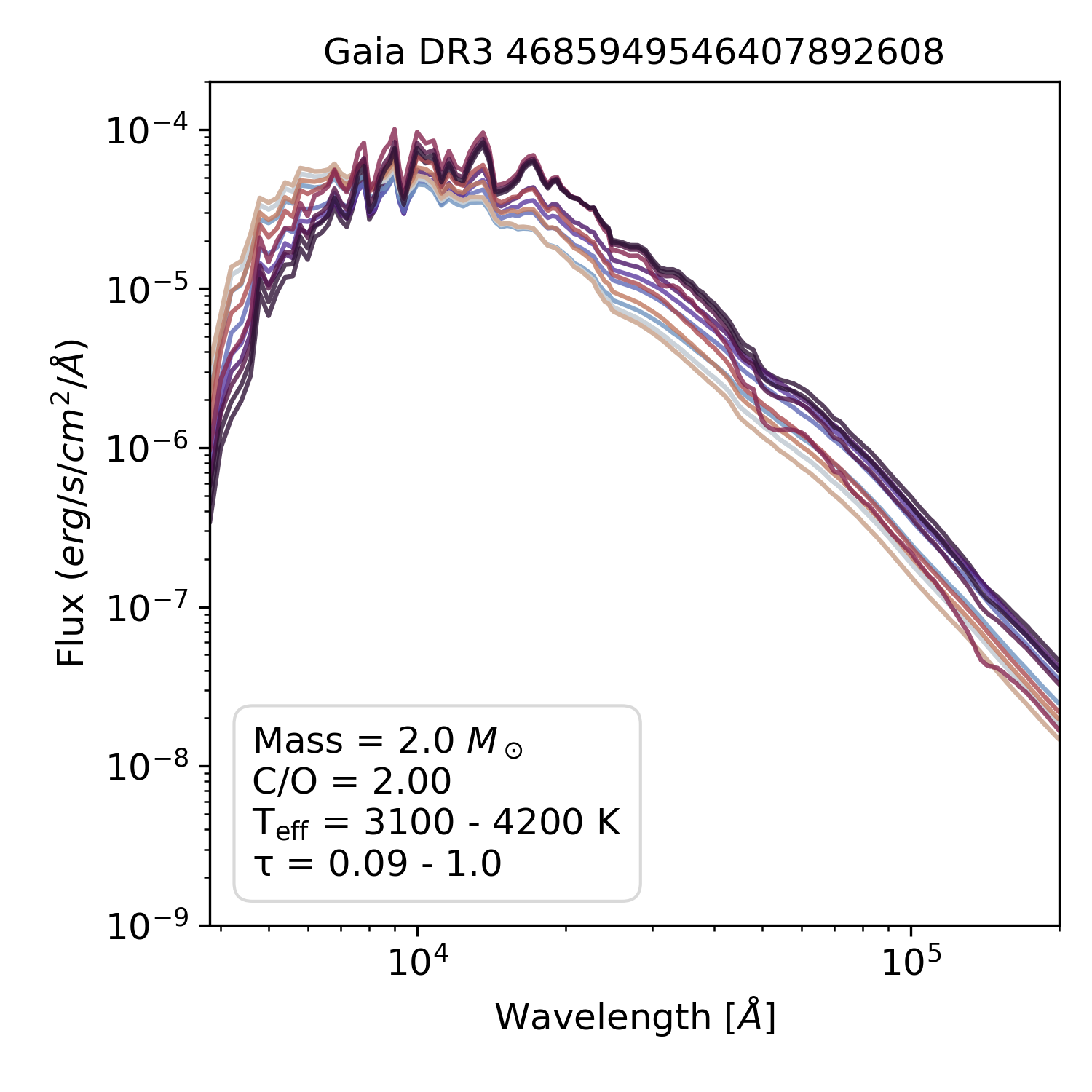}
    \includegraphics[width=0.33\linewidth]{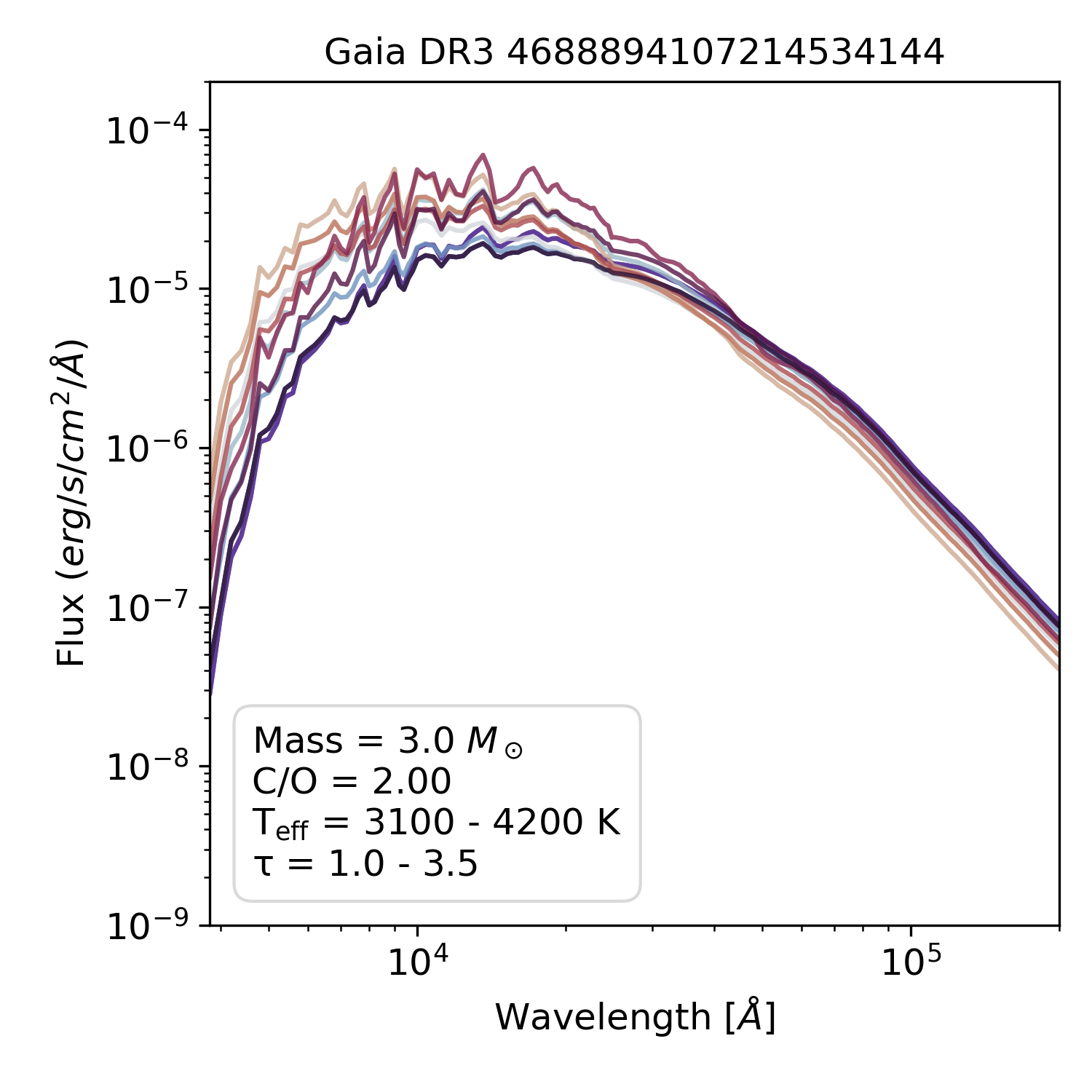}
    \caption{Best-fit SED models over the entire pulsation cycle. Colors indicate the pulsation phases. }
    \label{fig:cyclic_plot_apx_2}
\end{figure*}

Figures~\ref{fig:cyclic_plot_apx_1} and \ref{fig:cyclic_plot_apx_2} present the evolution of the best-fit ATHENA-C SED for each star along the pulsation cycle.

\begin{figure*}
    \centering
    \includegraphics[width=0.33\linewidth]{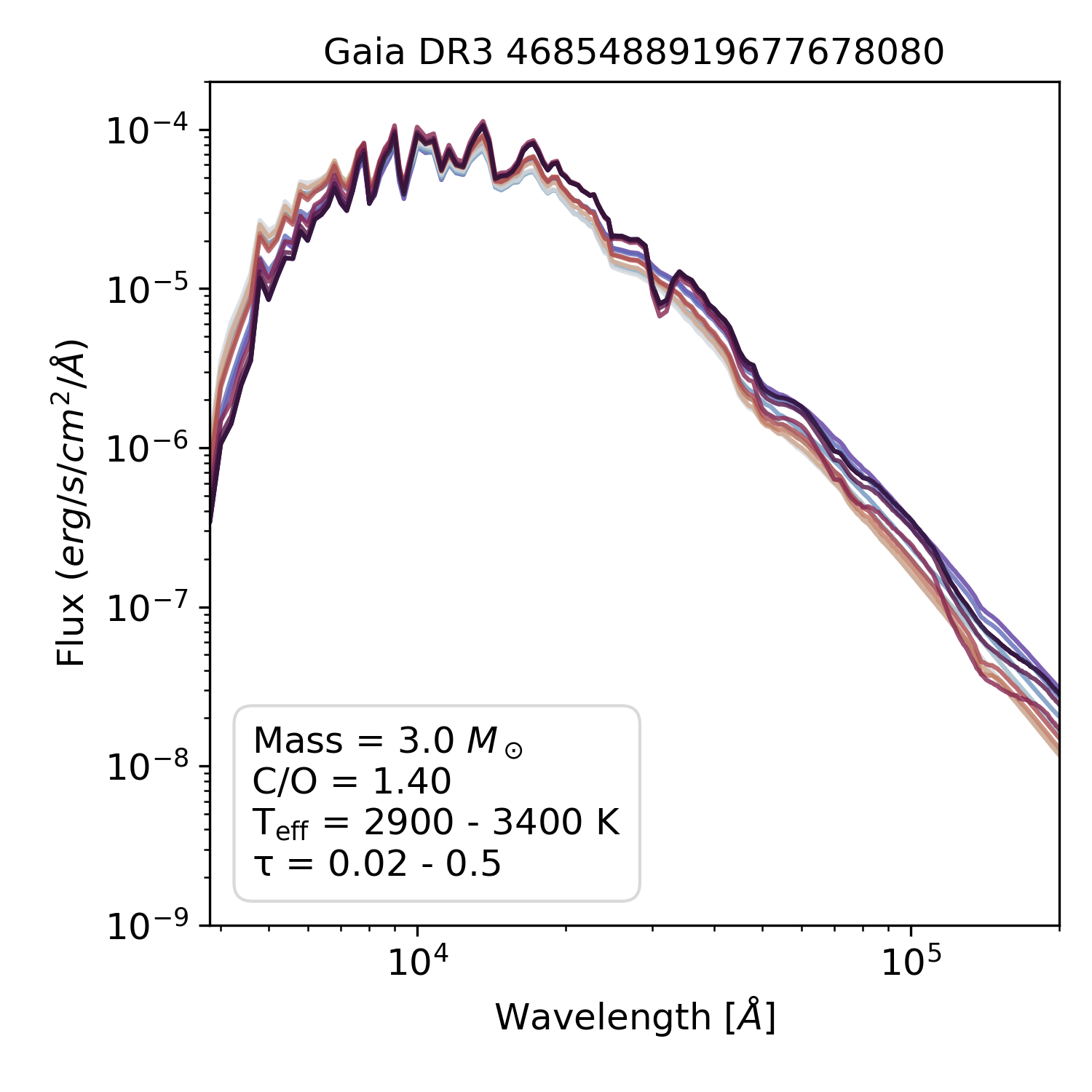}
    \includegraphics[width=0.33\linewidth]{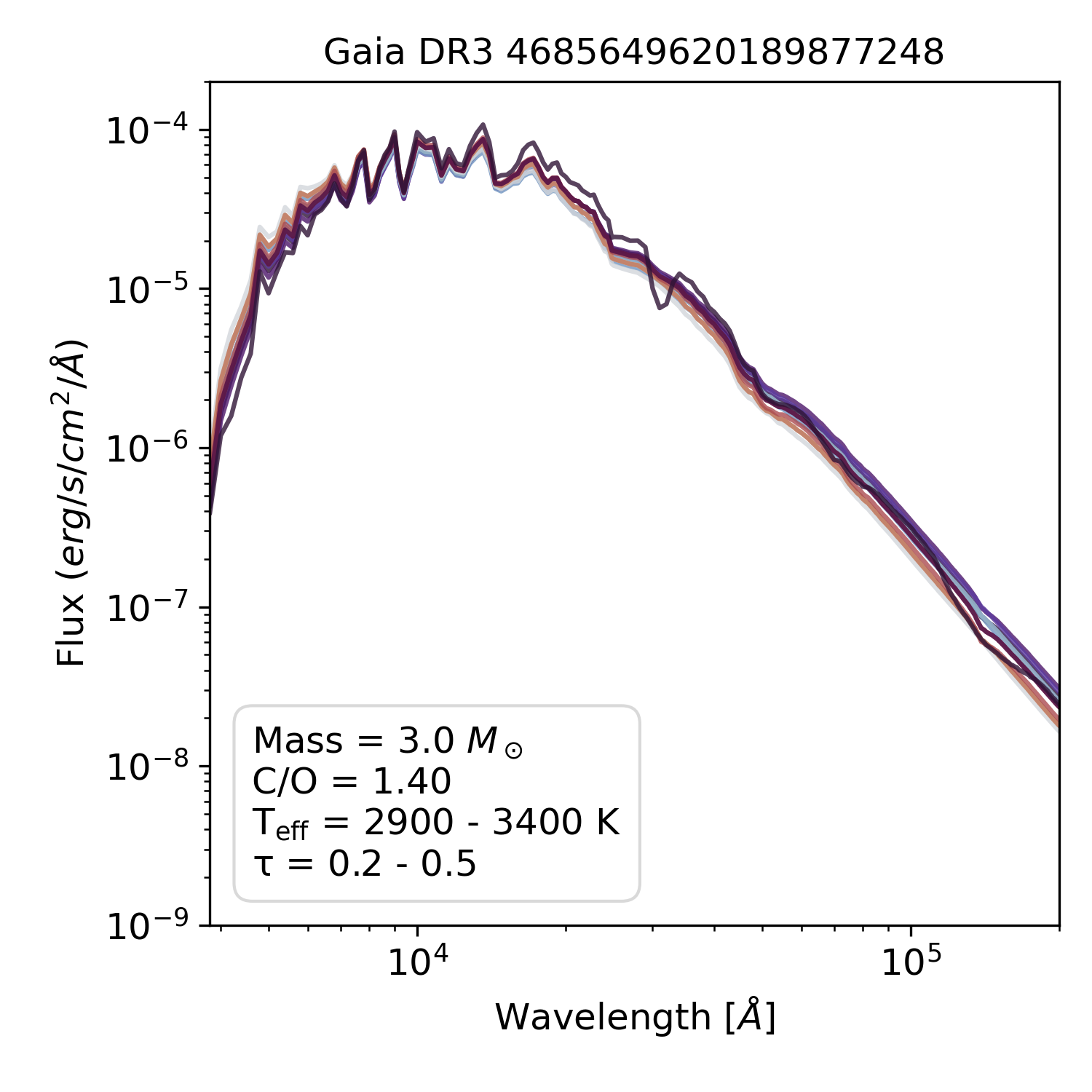}
    \includegraphics[width=0.33\linewidth]{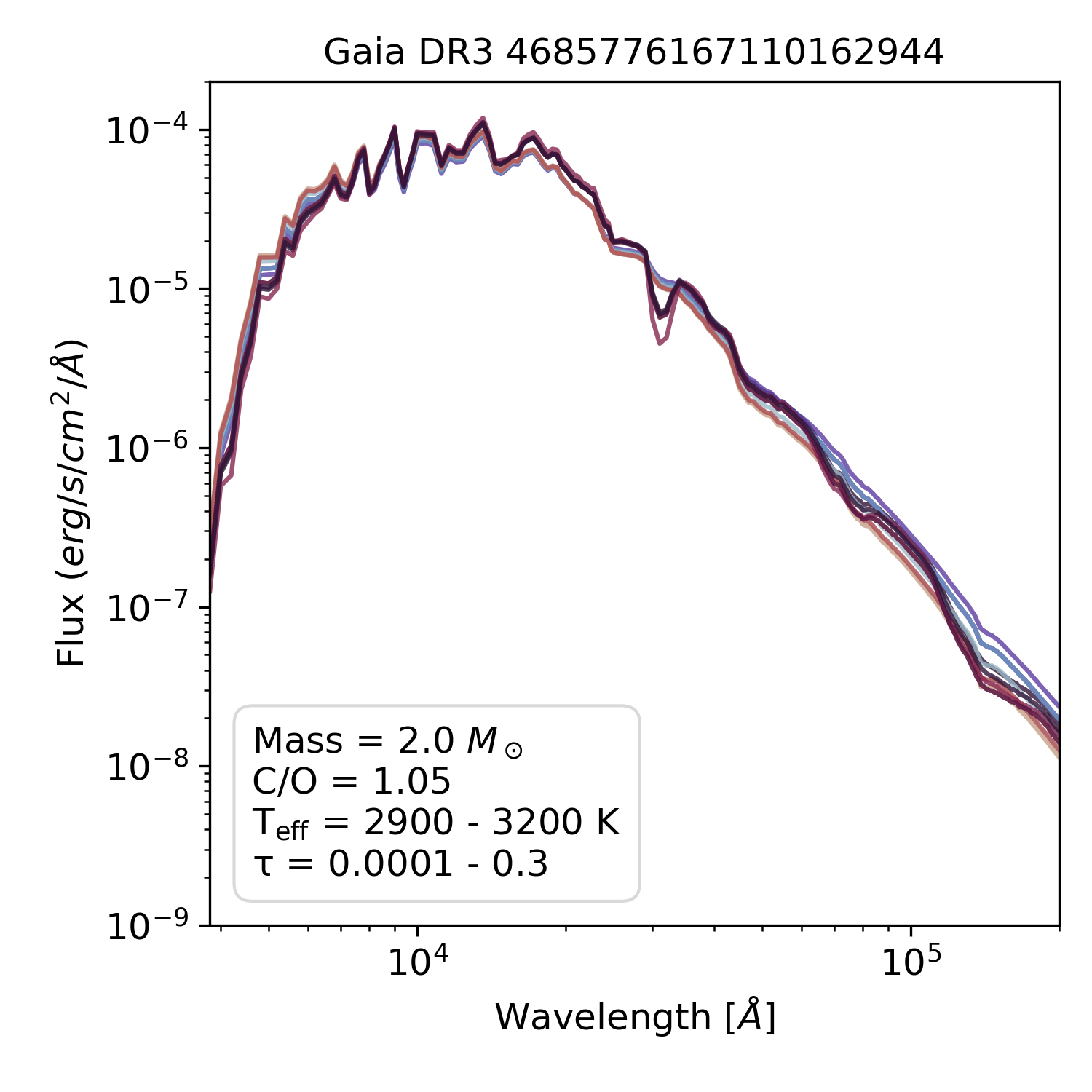}
    \includegraphics[width=0.33\linewidth]{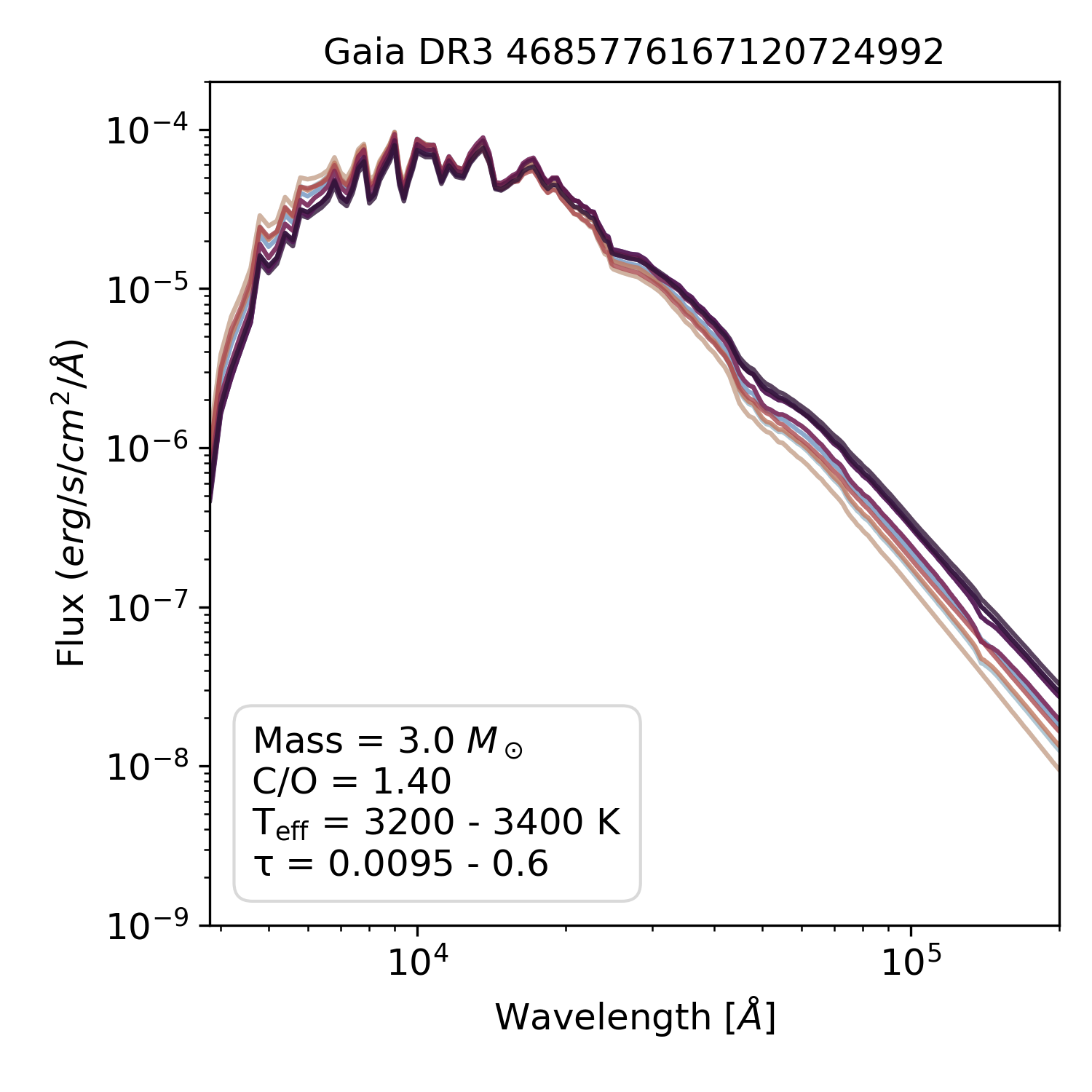}
    \includegraphics[width=0.33\linewidth]{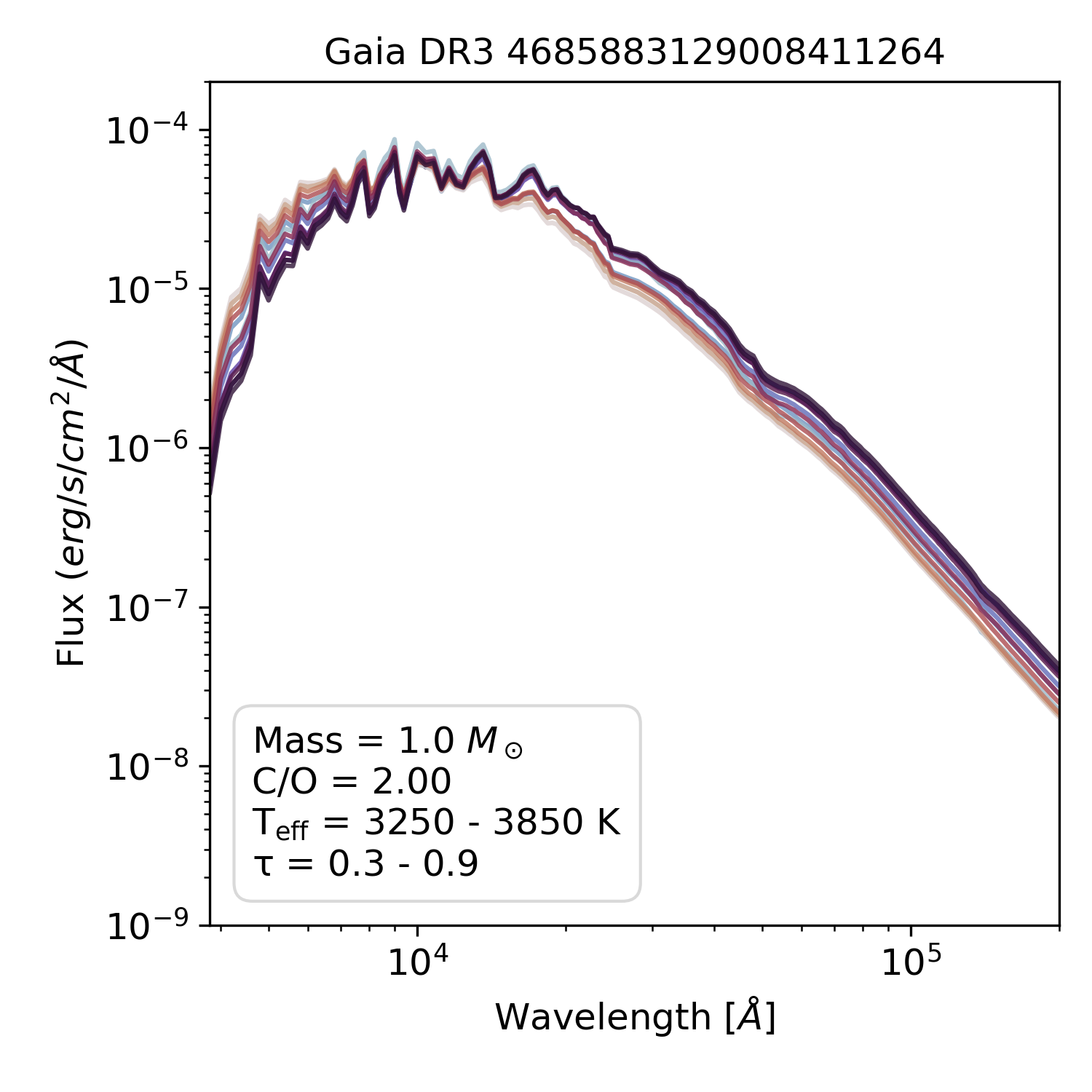}
    \includegraphics[width=0.33\linewidth]{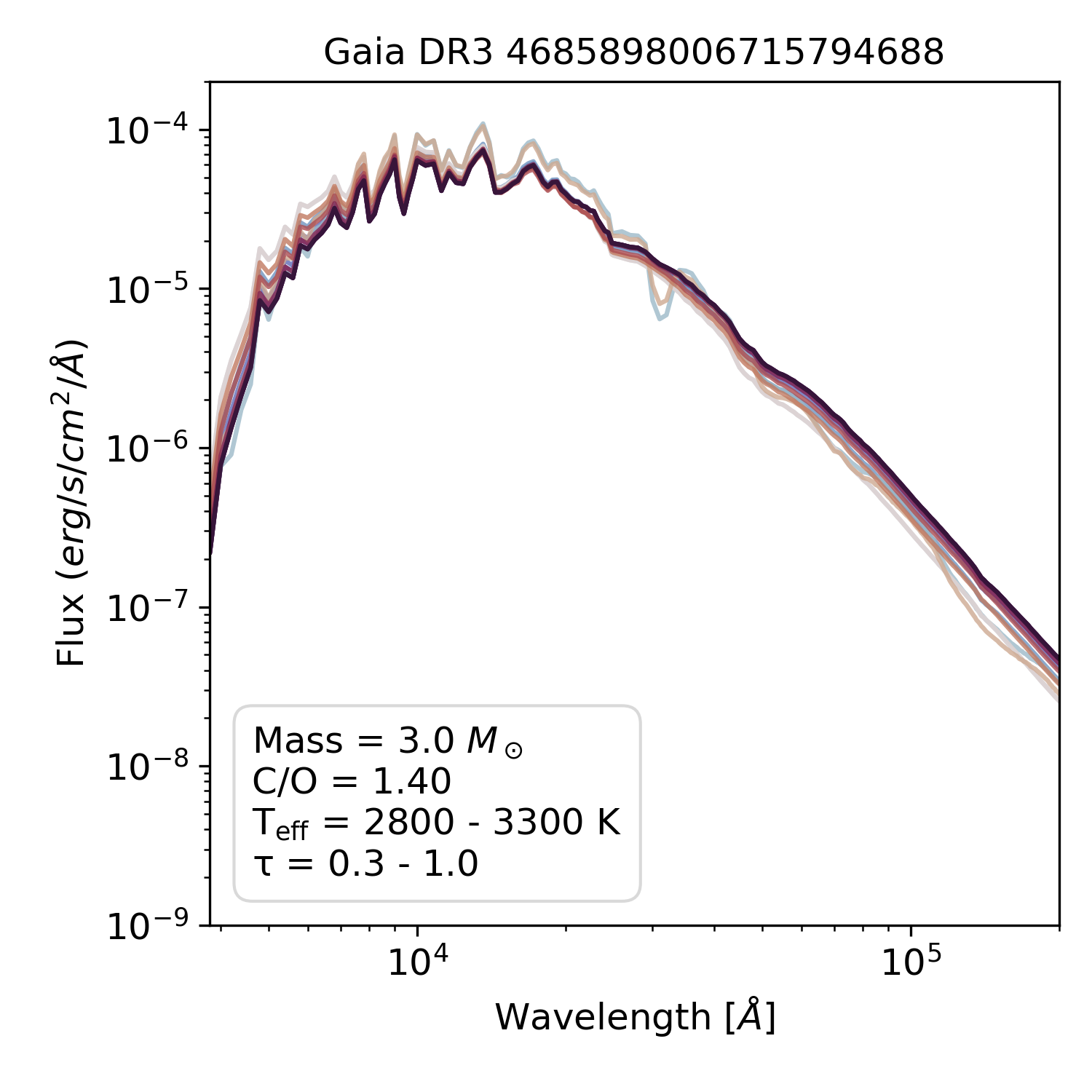}
    \caption{Best-fit SED models over the entire pulsation cycle. Colors indicate the pulsation phases. }
    \label{fig:cyclic_plot_apx_1}
\end{figure*}

\end{document}